\documentclass[fleqn,usenatbib]{mnras}

\usepackage{hyperref}
\usepackage[section]{placeins}
\usepackage{xcolor}
\usepackage[version=4]{mhchem}
\newcommand{\comment}[1]{}
\usepackage{multirow}

\usepackage[T1]{fontenc}

\DeclareRobustCommand{\VAN}[3]{#2}
\let\VANthebibliography\thebibliography
\def\thebibliography{\DeclareRobustCommand{\VAN}[3]{##3}\VANthebibliography}

\usepackage{graphicx}
\usepackage{rotating}
\usepackage{pdflscape}
\usepackage{xcolor}
\usepackage{aas_macros}

\definecolor{burgundy}{rgb}{0.50,0.00,0.13}

\usepackage{ulem}

\title[Modification of planetary heat budgets]{Modification of the radioactive heat budget of Earth-like exoplanets by the loss of primordial atmospheres}

\author[N. V. Erkaev  et al.]
{N.~V.~Erkaev$^{1,2,3} $, M.~Scherf$^{4,5,6}$, O.~Herbort$^{4,7,8,9,10}$ H.~Lammer$^4$, P.~Odert$^{6}$, D.~Kubyshkina$^{4}$, \newauthor M.~Leitzinger$^{6}$, P.~Woitke$^4$,
C. O'Neill$^{11}$
\\
$^1$Institute of Computational Modelling SB RAS, 660036, Krasnoyarsk, Russian Federation\\
$^2$Siberian Federal University, Krasnoyarsk, Russian Federation\\
$^3$Institute of Laser Physics, Siberian Branch of the Russian Academy of Sciences, Novosibirsk, Russian Federation\\
$^4$Space Research Institute, Austrian Academy of Sciences, Schmiedlstr. 6, A-8042, Graz, Austria\\
$^5$Institute of Geodesy, Technical University Graz, Austria\\
$^6$Institute of Physics/IGAM, University of Graz, Austria\\
$^7$St Andrews Centre for Exoplanet Science, University of St. Andrews, St. Andrews, UK\\
{$^8$SUPA, School of Physics \& Astronomy, University of St Andrews, North Haugh, St Andrews, UK}\\
{$^9$School of Earth \& Environmental Sciences, University of St Andrews, Irvine Building, St Andrews, UK}\\
{$^{10}$Institute for Astrophysics, University of Vienna, Türkenschanzstrasse 17, A-1180 Vienna, Austria}\\
$^{11}$Planetary Research Centre, Department of Earth and Environmental Science, Macquarie University, Sydney 2109, Australia}

\date{\today}

\pubyear{2021}

\pagerange{\pageref{firstpage}--\pageref{lastpage}} \pubyear{2021}

\def\LaTeX{L\kern-.36em\raise.3ex\hbox{a}\kern-.15em
    T\kern-.1667em\lower.7ex\hbox{E}\kern-.125emX}

\begin{document}

\label{firstpage}

\maketitle
\begin{abstract}
The initial abundance of radioactive heat producing isotopes in the interior of a terrestrial planet are important drivers of its thermal evolution and the related tectonics and possible evolution to an Earth-like habitat. The moderately volatile element K can be outgassed from a magma ocean into a H$_2$-dominated primordial atmospheres of protoplanets with assumed masses between 0.55-1.0$M_{\rm Earth}$ at the time when the gas disk evaporated. We estimate this outgassing and let these planets grow through impacts of depleted and non-depleted material that resembles the same $^{40}$K abundance of average carbonaceous chondrites until the growing protoplanets reach 1.0$M_{\rm Earth}$. we examine different atmospheric compositions and, as a function of pressure and temperature, calculate the proportion of K by Gibbs Free Energy minimisation using the GGChem code. We find that for H$_2$-envelopes and for magma ocean surface temperatures that are $\ge$ 2500 K, no K condensates are thermally stable, so that outgassed $^{40}$K can populate the atmosphere to a great extent. However, due to magma ocean turn-over time and the limited diffusion of $^{40}$K into the upper atmosphere, from the entire $^{40}$K in the magma ocean only a fraction may be available for escaping into space. The escape rates of the primordial atmospheres and the dragged $^{40}$K are simulated for different stellar EUV-activities with a multispecies hydrodynamic upper atmosphere evolution model. Our results show that one can expect that different initial abundances of heat producing elements will result in different thermal and tectonic histories of terrestrial planets and their habitability conditions.
\end{abstract}


\section{INTRODUCTION}\label{sec:intro}

Terrestrial planets accrete mass by numerous collisions between small rocky material, which accumulate to larger ones, thereby forming planetesimals, and Moon and Mars-mass planetary bodies that grow finally to protoplanets {and fully grown planets that might evolve towards Earth-like Habitats, {i.e., planets that allow for the evolution of life as we know it} \citep[e.g.,][]{lambrechts2012,Brasser2013,Johansen2015,Birnstiel2016,Johansen2021,Lammer2021}. Over its entire evolution, such a rocky planet can pass through a maximum of three different atmospheric stages. Depending on the mass of the growing protoplanet at the time when the disk gas dissipates, the first atmospheric stage will be a captured primordial H$_2$-dominated gaseous envelope \citep[e.g.,][]{Lammer2020b,Lammer2021} with a magma ocean below that can only solidify after the accretion halts and the overlying and escaping primordial atmosphere becomes thin enough to allow surface temperatures that are below the melting point of the underlying silicates. The second stage is constituted by the catastrophic outgassing of a steam {or carbon-rich} atmosphere during magma ocean solidification {\citep[][]{Sossi2020,Bower2022,Gaillard2022}}. Secondary atmospheres that built up after the condensation or loss of the steam atmosphere through further volcanic degassing define the third and last stage of atmospheric evolution  of rocky planets \citep[see, e.g.,][]{Lammer2018}.}

{As mentioned above, a primordial hydrogen-dominated atmosphere can only be {retained} in case that the growing protoplanet reaches a certain mass already within the disk phase since the accreted atmosphere would otherwise not be stable and immediately lost after nebula dissipation through the so-called boil-off \citep[e.g.,][]{Owen2020}; as simulations suggest, this critical mass might be around $\approx 0.5\,M_{\rm Earth}$ \citep{Johnstone2015}. For the Earth, there are isotopic indications that the planet accreted such a primordial atmosphere and must, therefore, already have grown to a substantial size within the solar nebula \citep[e.g.,][]{Harper1996a,Harper1996b,Jacobsen2008,Williams2019,Lammer2021}, {and must have lost any primordial atmosphere later-on \citep[][]{Lammer2020b}}. That such a fast accretion is indeed feasible was recently shown by \citet{Johansen2021} who were able to reconstruct the Earth via pebble accretion within just 5 million years. Moreover, \citet{Lammer2020b} were able to reconstruct the Earth's and Venus' atmospheric Ne and Ar isotopic compositions for an initial planetary mass of the time of the solar nebula dissipation of 0.55-0.60\,$M_{\rm Earth}$ and 0.80-1.0\,$M_{\rm Venus}$, respectively. It is, therefore, feasible to assume that protoplanets can accrete a sufficient mass to accrete a primordial atmosphere during the lifetime of the circumstellar disk, an assumption that is also confirmed by the findings of numerous low-mass exoplanets that are surrounded by a thick H$_2$-atmosphere \citep[e.g.,][]{Owen2020}.}

The bulk silicate composition of growing planets has long been linked to different planetary building blocks, such as ordinary and enstatite chondrites \citep[e.g.,][]{Dauphas2017}, or ureilites and carbonaceous chondrites \citep{Schiller2018}. Today's isotopic data further indicates that the Earth was assembled by $\approx$95\% dry or volatile depleted material with a few percent contribution of volatile rich carbonaceous chondrites \citep{Marty2012}{ as was recently also confirmed by the analysis of the Earth's Zn isotopes \citep{Steller2022,Savage2022}}. This is also in agreement with a recent study that reproduced the evolution of Earth's atmospheric isotopic ratios $^{20}$Ne/$^{22}$Ne, $^{36}$Ar/$^{38}$Ar, and $^{36}$Ar/$^{22}$Ne for proto-Earth masses of $\le$ 0.6$M_{\rm Earth}$ at the time when the gas disk dissipated \citep{Lammer2020b}. Depending on how fast planetary embryos grow to larger masses within the lifetime of the disk and at which orbital locations they originate or migrate to, these bodies will evolve to objects with different volatile and elemental abundances, which results in different elemental ratios compared to their initial ones \citep{Hin2017,Young2019,Sossi2019,Benedikt2020}.

\citet{Schaefer2007,Schaefer2010} and \citet{Fegley2016} performed chemical kinetic and equilibrium calculations for modeling the chemistry of the volatiles released by heating of different types of carbonaceous (CI, CM), ordinary (H, L, LL) and enstatite (EH, EL) chondrites as a function of temperature and pressure of steam atmospheres that are formed after magma ocean solidification of terrestrial planets. These authors found that major rock-forming (e.g., K, Al, Ca, Na, Fe, Si, Mg, S, P, Cl, F) and other moderately volatile elements, such as Rb and Zn, should also be outgassed and embedded in steam atmospheres at high temperatures (1000 - 2500 K). Besides steam atmospheres, moderately rock-forming elements will also outgas from an underlying magma ocean into primordial hydrogen-dominated atmospheres through equilibration between atmosphere and magma ocean before a first crust forms.

It is known that the heat producing elements inside present Earth are governing heat due to the radioactive decay of U (uranium), Th (thorium) and K (potassium) isotopes, which are $^{235}$U, $^{238}$U, $^{232}$Th and $^{40}$K. The early Earth inherited these isotopes together with $^{26}$Al, $^{60}$Fe and other short-lived nuclides from the interstellar medium during its growth phase \citep{Bizzarro2007,Makide2011,Schiller2015,ONeill2020}. These elements share the geochemical characteristic of being incompatible, i.e., staying preferentially within the melt with a ratio of concentration between melt and solid being close to 1. As a result U, Th and K become concentrated in the planet's continental crust and potentially also in deep reservoirs \citep{Labrosse2007}.

If one considers for instance enstatite chondritic (EC) material \citep{Dauphas2017} and/or ureilites \citep{Schiller2018} as the parent bodies of the Earth and other terrestrial planets, the initial ratio of K/U and K/Th is higher than the present ratio, indicating that some loss of the moderately volatile element K should have occurred during the protoplanet's accretion \citep{Lammer2020b}. {Besides, it is noteworthy that even though enstatite chondrites fit the isotopic composition of the Earth quite well, they have a composition that is too silicon-rich to represent direct precursors of the Earth \citep[][]{ONeill2008,Fegley2020}. In addition, the Earth generally shows a volatile depletion that is much stronger than for any enstatites.}

U and Th are heavy refractory lithophile elements, which belong to a group of elements that condense from the canonical solar nebula at higher temperatures than the main constituents of the rocky planets such as Mg, and Si. {The 50\% condensation temperatures $T_{\rm 50}$ of uranium and thorium are 1609\,K and 1630\,K, respectively,
while the much lighter potassium belongs to the moderately volatile elements and has a much lower {condensation temperature} of 993\,K \citep{Wood2019}.} $^{238}$U is the most abundant U-isotope with a half-life of $\tau_{1/2} = 4.468$ billion years (Gyr) \citep{Ruedas2017,ONeill2020}. The isotope $^{235}$U is much less abundant and decays more rapidly with $\tau_{1/2}= 0.704$\,Gyr, while $^{232}$Th has a long half-life of $\tau_{1/2}=14.05$\,Gyr \citep{ONeill2016}. Finally, the moderately volatile radioactive potassium isotope $^{40}$K decays with $\tau_{1/2}=1.248$\,Gyr of which 89\% transform into stable $^{40}$Ca by beta decay, and 11\% into $^{40}$Ar.

During the first Gyr after the formation of the solar system the radioactive decay of $^{40}$K produced more heat in early Earth's interior than all of the before mentioned radioactive U and Th isotopes together \citep{Turcotte2002,ONeill2020}. Today the relative contribution of radioactive nuclides to Earth's internal heating rate is, however, determined by $^{238}$U and $^{232}$Th, while $^{40}$K plays a minor role.

The initial heat production budget of a planet exerts a first order control on its thermal evolution, tectonics, and is hence very important to evolve a terrestrial planet into an Earth-like habitat \citep{ONeill2020}. The radioactive decay of $^{40}$K contributes to the long-term temperature evolution of a planet's interior which is important for driving a long-lived magnetic dynamo, \citep{Turcotte2002,Murthy2003,Nimmo2015,ONeill2020}. A right amount of such heat producing elements further indirectly support the development of secondary atmospheres via a correlation between mantle heating and volcanic outgassing \citep{Padhi2012,Johnson2018}. Investigating through which processes the amount of heat producing elements can be altered during the early evolutionary states of planet formation, is, therefore, very important to understand the subsequent evolution of a rocky exoplanet into a potential habitat.

In the following section we will briefly discuss our motivation, the overarching model structure and important differences to our previous study, \citet{Benedikt2020}. In Sect.~\ref{sec:model} and its related Appendix~\ref{App:Model}, we describe the multi-species hydrodynamic upper atmosphere escape model. In Sect.~\ref{sec:inputs}, we describe the input parameters of the study, including the applied EUV-evolution tracks of young G-stars, captured masses of primordial atmospheres by protoplanets with different masses at the time when the gas disk dissipated, the initial potassium and $^{40}$K abundances in the assumed magma ocean volumes, the K-fraction in the gas phase embedded in an H$_2$-dominated primordial atmosphere as a function of surface pressure and temperature (see also Appendix~\ref{App:Kfraction}), and impactors that deliver potassium to the growing protoplanet. Our results are presented and discussed in Sect.~\ref{sec:results}, which also includes a discussion of our result's consequences on planetary heat production and the evolution and habitability of Earth-like exoplanets. Finally, Sect.~\ref{sec:conclusion} concludes the work.

\section{MOTIVATION AND MODEL STRUCTURE}\label{ref:motivation}

{\citet{Benedikt2020} recently demonstrated that atmospheric escape from small planetary embryos with masses between $1\,M_{\rm Moon}$ and $1\,M_{\rm Mars}$ can efficiently lose moderately volatile elements such as $^{40}$K that are outgassed from the magma ocean either via dragging through the escape of the catastrophically outgassed steam atmosphere {(with a mean molar mass of $>22$ excluding outgassed moderately volatile elements}) or for the smaller ones ($\leq1\,M_{\rm Moon}$) directly through outgassing from the magma ocean . This study, therefore, showed that atmospheric escape can deplete small magmatic embryos that serve as building blocks for more massive growing protoplanets during the so-called giant impact phase of accretion. However, no study so-far investigated whether moderately elements can also be lost from heavier protoplanets with masses sufficient to accrete a primordial hydrogen-dominated atmosphere during the nebula phase. This initial situation diverges from the steam atmosphere case as studied by \citet{Benedikt2020} with respect to the initial mass of the planetary bodies but also in atmospheric composition and structure.}

It is, therefore, very important to investigate the effect of an escaping primordial H$_2$-dominated atmosphere and how this process can modify and influence the initial amount of $^{40}$K on terrestrial planets, {and how important this pathway might be compared to the loss of $^{40}$K from planetesimals and smaller-mass planetary embryos as mentioned above and, e.g., investigated by \citet{Hin2017} and our earlier study}. Since different heat budgets in planetary interiors might lead to different tectonic regimes and, therewith connected, to potentially failed carbon-silicate and nitrogen cycles, alterations in $^{40}$K abundances have important implications for the habitability of such planets. Venus and Mars, for instance, have different potassium bulk abundances than Earth {\citep[e.g.,][]{Basilevsky1997a,Abdrakhimov2002,Lammer2020b,Khan2022}} and do presently not show either plate tectonics or an active magnetic dynamo.

Here, we study for the first time the loss of the radioactive and moderately volatile element $^{40}$K through the dragging of a hydrodynamically escaping primordial atmosphere, in which $^{40}$K originates from accumulated H$_2$-dominated nebular gas, outgassing from an underlying global magma ocean and from impactors in case that the assumed protoplanet did not already accrete to its final mass within the nebula. {This present work, therefore, differs by several aspects from our previous study,} \citet{Benedikt2020}. These are:
\begin{itemize}
  \item {We investigate protoplanets with $M \geq 0.55\,M_{\rm Earth}$ that already accumulated a primordial atmosphere from the evaporating disk and have a magma ocean below it, whereas \citet{Benedikt2020} focus on smaller mass bodies at which steam atmospheres outgassed from the solidifying magma ocean.}
  \item {We let these protoplanets grow via impacts to a final mass of 1\,$M_{\rm Earth}$ since we are interested on the diversity of heat producing elements within Earth-like planets that evolved through different atmospheric escape and impact delivery scenarios.}
  \item {We study, for the first time, atmospheric evolution and the dragging of heavier trace elements by applying a fully hydrodynamic upper atmosphere model, while studies such as \citet{Kasting1983}, \citet{Zahnle1986}, \citet{Odert2018}, and \citet{Benedikt2020} only applied a modification of the energy-limited approach \citet{Watson1981} that includes dragging.}
  \item {Additionally, we apply equilibrium condensation models with the {\sc GGchem} code to evaluate {what fraction of the K is present in gaseous form and what K-bearing solid condensates are stable, if any}. However, we do not apply {\sc GGchem} self-consistently to avoid more complexity at the present stage of research but we use its results to contextualize our evolutionary outcomes.}
  \item {We calculate the final heat production of our different outcomes and compare these with a hypothetical planet that did not experience atmospheric escape of heat producing elements.}
\end{itemize}
{One should keep in mind that this model approach bears several simplifications and is, therefore, only a first step in studying the effect of atmospheric escape on the modification of the thermal heat budget during the accretion of an Earth-like planet. These include}
\begin{itemize}
  \item {We do not couple {{\sc GGchem}} with the overlying escaping atmosphere but assume different simple static and independent equilibrium values between the dissolved ratio of $^{40}$K in the atmosphere and within the magma ocean.}
  \item {There is neither ingassing of the atmosphere nor outgassing from the magma ocean (except for the static value of $^{40}$K) included in the model but a simple prescribed H-atmosphere that does not interact with the underlying magma ocean. We, therefore, neither include H$_2$ solubility in the magma, nor the catastrophic outgassing of the steam atmosphere at the time when the magma ocean solidifies or any other degassing that might replenish the atmosphere.}
  \item {For our simulations, we only assume one composition for protoplanets and impactors. We also do not take into account different redox states of the mantle.}
  \item {We assume a constant magma ocean with a depth of 1000\,km and a surface temperature of 2000\,K for all our scenarios.}
  \item {For the growth of the protoplanets below $1\,M_{\rm Earth}$, our impactors are just simple {Moon- or Mars-sized} mass additions to the planetary embryo with the same carbonaceous chondritic composition that is also assumed for the initial planetary embryos. The impactors do further not erode the atmosphere but only deliver mass and $^{40}$K.}
  \item {We do not simulate the loss of $^{40}$K from the impacting bodies but only assume some predefined depletion.}
  \item {Any other processes that might additionally alter the amount of heat producing elements within the final planet are omitted.}
  \item {Our resulting planets resemble Earth-mass planets that start from the same heat budget. However, the variety of exoplanets within the galaxy will be much more diverse, having different initial heat budgets, various final masses, and diverse orbits around stars other than solar-like ones with completely different radiation and plasma environments.}
\end{itemize}
Since we are mainly interested in hypothetical {Earth-mass} exoplanets, {will let our protoplanets always grow until a final mass of} 1.0\,$M_{\rm Earth}$ and orbit at 1\,AU inside the habitable zone of young G-stars with EUV evolution tracks that correspond to slow, moderate and fast rotating young G-type host stars.

\section{MULTISPECIES HYDRODYNAMIC UPPER ATMOSPHERE MODEL}\label{sec:model}

To model the evolution of the primordial hydrogen-dominated atmosphere, we use a multispecies hydrodynamic upper atmosphere model that simulates snapshots of the respective upper atmospheres {from the photospheric radius $r_0$ (i.e., the specific radius at which the atmosphere is opaque below and transparent above for the visible part of the spectrum) along the radial distance $r$ as long as the expanding thermosphere is still in the collisional regime. We simulate this atmosphere} for different time steps from 5\,Myr until 1\,Gyr until the atmospheres are lost. The mass conservation equations for the  hydrogen molecules, atoms and ions (H$_2$, H, H$_2^+$, H$^+$), and also $^{40}$K can be written as follows,
\begin{eqnarray}\label{eq:HConst1}
\frac{\partial \left(n_{\rm H}\right)}{\partial t} + \frac{1}{r^2}\frac{\partial \left(n_{\rm H} V r^2\right)}{\partial r}= - \nu_{\rm H} n_{\rm H} + \alpha_{\rm H} n_{e}n_{\rm H^+} + \nonumber \\
2\alpha_{\rm H_2}n_{\rm e}n_{\rm H_2^+} -  2\gamma_{\rm H} n n_{\rm H}^2, \\
\frac{\partial \left(n_{\rm H^+}\right)}{\partial t} + \frac{1}{r^2}\frac{\partial \left(n_{\rm H^+}V r^2\right)}{\partial r}=\nu_{\rm H} n_{\rm H} - \alpha_{\rm H} n_{e}n_{\rm H^+}, \\\label{eq:HConst4}\label{eq:HConst2}
\frac{\partial \left(n_{\rm H_2}\right)}{\partial t} + \frac{1}{r^2}\frac{\partial \left(n_{\rm H_2}V r^2\right)}{\partial r}=-\nu_{\rm H_2} n_{\rm H_2} + \gamma_{\rm H} n n_{\rm H}^2, \\\label{eq:HConst3}
\frac{\partial \left(n_{\rm H_2^+}\right)}{\partial t} + \frac{1}{r^2}\frac{\partial \left(n_{\rm H_2^+}V r^2\right)}{\partial r}=\nu_{\rm H_2} n_{\rm H_2} - \alpha_{\rm H_2} n_{e}n_{\rm H_2^+}, \\\label{eq:HConst4}
\frac{\partial n_K}{\partial t} + \frac{1}{r^2}\frac{\partial \left(n_K V_K r^2\right)}{\partial r}= 0. \label{Rho_K}
\end{eqnarray}
Here, $V$ and $V_{\rm k}$ are the velocities of the hydrogen and potassium species, $n_{\rm H}, n_{\rm H_2}, n_{\rm H^+}, n_{\rm H_2^+}$ are the number densities of the hydrogen atoms, molecules  and ions, respectively, $ n_{\rm K}$ is the potassium number density, $n$ the total number density, and $n_{\rm e}$ is the electron number density determined from the quasi-neutrality condition
\begin{equation}
n_{\rm e} = (n_{\rm H^+}+ n_{\rm H_2^+}), \label{el}
\end{equation}
where $\alpha_{\rm H}$ is the recombination rate related to the reaction \mbox{H$^{+}+$e$\rightarrow$H} of $4\times 10^{-12} (300/T)^{0.64}$ cm$^{3}$ s$^{-1}$, $\alpha_{\rm H_2}$ the dissociation rate (\mbox{H$_2$+e$\rightarrow$H + H}) = $2.3\times 10^{-8} (300/T)^{0.4}$ cm$^{3}$ s$^{-1}$, $\nu_{\rm H}$ the hydrogen ionization rate, and $\nu_{H_2}$ is the ionization rate of molecular hydrogen \citep{Yelle2004,Storey1995}. The ionization rates are scaled proportionally to the EUV flux which depends on the particular orbit location, i.e.,
\begin{equation}\label{eq:phiEuv}
\nu_{\rm H} = 2.58\cdot 10^{-8} \phi_{\rm EUV} , \quad  \nu_{\rm H_2} = 1.45 \cdot 10^{-8} \phi_{\rm EUV} ,
\end{equation}
where the ionization rates $\nu_{\rm H}$ and $\nu_{\rm H_2}$ in units of $1/\rm s$ are proportional to the EUV flux in units of $\rm erg /cm^2 /s$ \citep{Yelle2004}, and $\phi_{\rm EUV}$ is the function describing the EUV flux absorption in the atmosphere through
\begin{equation}
\phi_{\rm EUV}=\frac{1}{4\pi}\int_0^{\pi/2+\arccos(1/r)} J_{\rm EUV}(r,\theta)
2\pi\sin(\theta)d\theta.
\end{equation}
Here, $J(r,\theta)$  is the function of spherical coordinates that describes the spatial variation of the  EUV flux  due to the atmospheric absorption \citep{Erkaev2015}, i.e.,
\begin{eqnarray}
J(r,\theta) = J_\infty \exp(-\tau(r,\theta)), \\
\tau(r,\theta) = \int_{r}^{\infty}{\sigma_i n(s)\frac{s}{\sqrt{s^2-r^2 \sin(\theta)^2}}} d s ,
\end{eqnarray}
where $J_\infty$ is the intensity of incoming EUV radiation far away from the planet.

For the hydrogen constituents (see Equations~\ref{eq:HConst1},\ref{eq:HConst2},\ref{eq:HConst3},\ref{eq:HConst4}, and \ref{Rho_K}), we introduce a total mass density $\rho$ as a sum of the partial mass densities which is governed by the mass conservation equation,
\begin{eqnarray}
\frac{\partial \rho}{\partial t} + \frac{1}{r^2}\frac{\partial \left(\rho V r^2\right)}{\partial r}= 0. \label{Rho}
\end{eqnarray}
Velocities $V$ and $V_{\rm K}$ are  determined by the Euler momentum equations,
\begin{eqnarray}
\rho \frac{d  V}{d t}  +\nabla P= g\rho+
 \rho_{\rm K} \frac{1}{1+m_{\rm K}/m_{\rm H} }\nu_c (V_K-V)  , \\
 \rho_{\rm K} \frac{d  V_{\rm K}}{d t}  + \nabla P_{\rm K}= g\rho_{\rm K}+
 \frac{1}{1+m_{\rm K}/m_{\rm H} }\rho_{\rm K} \nu_c (V-V_{\rm K})  ,
 \end{eqnarray}
where $\rho$ is the total mass density of all hydrogen species,
\begin{equation}\label{eq:rhoH}
  \rho = m_{\rm H} (n_{\rm H} + n_{\rm H^+}) + m_{\rm H_2} ( n_{\rm H_2} + n_{\rm H_2^+}),
\end{equation}
and $P$ is the total pressure of hydrogen and electrons,
\begin{equation}\label{eq:PHe}
P=\left(n_{\rm H}+n_{\rm H^+}+n_{\rm H_2}+n_{\rm H_2^+}+n_{\rm e}\right)k_{\rm B} T,
\end{equation}
where $T$ is the upper atmosphere temperature and $k_{\rm B}$ is the  Boltzmann constant. Moreover, $\nu_c$ is the mean collision frequency,
\begin{equation}
\nu_c =   b_{\rm K,H} (n_{\rm H} +n_{\rm H^+}) +
   b_{\rm K,H_2} \frac{(1 +m_{\rm K}/m_{\rm H})}{(1 +  m_{\rm K}/m_{\rm H_2})} (n_{\rm H_2}+n_{\rm H_2^+}) ,
\end{equation}
where $m_{\rm H}$ and $m_{\rm K}$ are the hydrogen and potassium atomic masses, respectively. Further, $b_{\rm K,H}$ and $b_{\rm K,H_2}$ are the parameters related to collisions between potassium and hydrogen species, i.e.,
\begin{eqnarray}
  b_{\rm K,H} = \sigma_{\rm K,H} \sqrt{\frac{8 k_B T (m_{\rm K}+m_{\rm H})}{\pi m_{\rm K} m_{\rm H}}}, \\
  b_{\rm K,H_2} = \sigma_{\rm K,H_2} \sqrt{\frac{8 k_B T (m_{\rm K}+m_{\rm H_2})}{\pi m_{\rm K} m_{\rm H_2}}},
\end{eqnarray}
where $\sigma_{\rm K,H} $ and $\sigma_{\rm K,H_2} $ are the collision cross sections, and $g$ = -$\rho\frac{\partial U}{\partial r}$ with $U$ as the gravitational potential,
\begin{equation}
U = -\frac{G M_{\rm pl}}{r}.
\end{equation}
{The term $g$ is gravitational acceleration which decreases by increasing radius, i.e., with the expansion of the EUV heated upper atmosphere. Consequently, atmospheres that show a higher expansion result in more efficient escape.}

Since we assume that all species have the same temperature this is governed by the energy conservation equation,
\begin{equation}
\rho \frac{d E }{d t} = Q_{\rm EUV}  +   \frac{\partial }{r^2\partial r}\left(r^2 \chi \frac{\partial T}{\partial r}\right)
- P \frac{1}{r^2}\frac{\partial Vr^2}{\partial r} =0,
  \end{equation}
where $E$ is the thermal energy,
\begin{equation}
E = \left[ \frac{3}{2} (n_{\rm H} + n_{\rm H^+} + n_{\rm e} +n_{\rm K}) + \frac{5}{2}( n_{\rm H_2} + n_{\rm H_2^+}) \right] k_{\rm B} T ,
\end{equation}
and $\chi$ is the thermal conductivity {of the H gas in units of [$\rm erg\,cm\,s^{-1} K^{-1}$]}  \citep{Watson1981},
 \begin{equation}
 \chi = 4.45\cdot 10^4 \left(\frac{T}{1000}\right)^{0.7}.
 \end{equation}
The number densities of the species are to be expressed through the mass densities by dividing the corresponding masses of the particles. The stellar EUV volume heating rate, $Q_{\rm EUV}$, depends on the stellar EUV flux at the orbital distance of the test planets and on the atmospheric density. $Q_{\rm EUV}$ can then be written as
\begin{equation}
Q_{\rm EUV} = \eta \sigma_{\rm EUV}n_{\rm H}\phi_{\rm EUV},
\end{equation}
where $\eta$ is the ratio of the net local heating rate to the rate of the stellar radiative absorption which is typically $\approx15\%$ in a hydrogen atmosphere \citep{Shematovich2014}. As in \citet{Murray-Clay2009},\citet{Erkaev2013},\citet{Lammer2013} and \citet{Lammer2014}, we assume a single wavelength for all photons
and use the average EUV photoabsorption cross sections $\sigma_{\rm EUV}$ of $5\times10^{-18}$ cm$^{2}$ and $3\times10^{-18}$ cm$^{2}$ for H atoms and H$_2$ molecules, respectively. At the exobase we set zero conditions for the radial derivatives of the density and temperature (so called free boundary conditions).

Further details of the multispecies hydrodynamic upper atmosphere evolution model are described in Appendix~B. The lower boundary conditions, i.e., the photospheric radius $r_0$, the equilibrium temperature $T_0$, and the atmospheric pressure $p_0$ at $r_0$ are described in Section~\ref{sec:parAtm}.


\section{INPUT PARAMETERS}\label{sec:inputs}

For the input values as given in Table~\ref{tab:params} we use the same methods as described in detail in \citet{Lammer2020b}. Here, we summarize briefly the input parameters and assumptions regarding the EUV-evolution tracks of solar-like G-stars, initial values of primordial atmospheres around various protoplanetary masses at the time when the disk dissipated, the calculation of initial elemental abundances in the atmosphere, the underlaying magma ocean and the chosen impactor composition.

\begin{table*}
  \begin{center}
    \caption{Assumed planetary parameters and initial $^{40}$K amount within the magma ocean$^a$.}
    \label{tab:params}
\begin{tabular}{l|c|c|c|c|c|c|r}
  \hline
   & $M_{\rm p}/M_{\rm Earth}$ & $R_{0}/R_{\rm Earth}$ &$ M_{\rm at}/M_{\rm p}$  &mass impact $\Delta M /M_{\rm Earth}$ & $M_{\rm ^{40}K,mo}$[g] &$ M_{\rm ^{40}K,neb}$[g] & $M_{\rm ^{40}K,imp}$[g]\\
   \hline
  Case 1 & 0.55 & 0.989 & 1.39e-4  &  0.45 ( $t \leq $20 Myr)& 9.195e20 & 1.058e14 & 1.595e21\\
  Case 2  & 0.6 &1.132 & 0.9952e-3& 0.4 ($t\leq$ 20 Myr) & 9.801e20 & 7.929e14 & 1.418e21\\
 Case 3 & 0.7 & 1.418 & 0.602e-2   & 0.3 ($t\leq$ 20 Myr)& 1.097e21 & 5.830e15 & 1.063e21\\
  Case 4 & 0.8 & 1.704 & 1.564e-2 &  0.2 ($t\leq$ 20 Myr)& 1.209e21 & 1.721e16 & 7.088e20\\
  Case 5 &1.0 & 2.275 & 0.046  & no impact & 1.421e21 & 6.364e16 & no impact\\
  \hline
\end{tabular}
\end{center}
{$^a$Without atmospheric escape all 5 cases would result in a final planet with a $^{40}$K abundance equivalent to the abundance of average carbonaceous chondrites, i.e., $a_{\rm ^{40}K}$\,=\, 0.5934\,$\mu \rm g/g$. The amount of $^{40}$K accreted from the nebula is negligible.}
\end{table*}
\subsection{EUV-evolution of young G-stars}\label{sec:parStars}

Although the bolometric luminosity of solar-like stars is increasing over time \citep{Sackmann2003,Guedel2007}, the fluxes of X-rays and EUV decrease over their whole main sequence lifetime \citep{Ribas2005,Guedel2007,Claire2012,Tu2015,Johnstone2021}. These fluxes are initially saturated at very high levels before they start to continuously decay towards lower values. The magnetic and EUV luminosity evolution of G stars is further strongly dependent on their initial rotation rate \citep{Johnstone2015,Tu2015}. Fast rotators are in the saturation phase $t_{\rm sat}$ for up to $\approx$225 Myr and their EUV flux afterwards decline faster over time than in the case of slow or moderate rotators which leave saturation after $\approx$5, and $\approx$25 Myr, respectively. We use for each of the below described model cases 3 different EUV flux evolution power laws that correspond to slowly, moderately and fast rotating young G-stars \citep{Tu2015,Lammer2020b}, i.e.,
\begin{eqnarray}\label{eq:rotators}
\quad{\rm EUV}_{\rm slow} &=& 5.75\times 10^{31} t^{-0.93} \mbox{erg/s} \\  t \ge t_{\rm sat} &=& 5\, \rm Myr \nonumber ,\\
\quad {\rm EUV}_{\rm mod} &=& 4.7\times 10^{32} t^{-1.18}  \mbox{erg/s} \\   t \ge t_{\rm sat} &=& 25\, \rm Myr , \nonumber  \\
\quad{\rm EUV}_{\rm fast} &=& 1.2\times 10^{36} t^{-2.15}  \mbox{erg/s} \\   t \ge t_{\rm sat} &=& 225\, \rm Myr \nonumber .
\end{eqnarray}
The EUV surface flux at the orbit location of the studied planets can be found as the ratio
\begin{equation}
I_{\rm EUV} = {\rm EUV}_{\rm law}/(4\pi d_{\rm p}^2),
\end{equation}
where $d_{\rm p}$ is the orbital distance of the planet, which we assume to be 1\,AU.

\subsection{Captured primordial atmospheres}\label{sec:parAtm}

If protoplanets grew to masses that are $>$0.5$M_{\rm Earth}$ within the protoplanetary nebula {(with a characteristic lifetime of $\sim 2-5$\,Myr \citep{Mamajek2009})}, they accumulate H$_2$-dominated primordial atmospheres by accreting gas from the circumstellar disk \citep{Sekiya1980,Johnstone2015,Lee2015,Stoekl2015,Stoekl2016,Lammer2014,Lammer2020b}{; a scenario that was likely followed by the Earth and can be seen through the observation numerous exoplanets that still host an H-rich envelope (see Section~\ref{sec:intro} for details)}. In this study we assume 5 different cases with planetary masses at the end of the disk phase of 0.55, 0.6, 0.7, 0.8 and 1.0\,$M_{\rm Earth}$. The equilibrium temperature $T_0$ at 1\,AU at the time of disk evaporation is assumed for all 5 cases to be 250\,K {at the photospheric radius $r_0$. A greenhouse effect may lead to slightly higher equilibrium temperatures at $r_0$ by a few degrees but this has negligible effects on atmospheric structure and escape rates \citep[e.g.,][]{Mordasini2015}.}

During the disk dissipation, the protoplanet's photospheric radius $r_0$ of the H$_2$-dominated gas envelope shrinks fast by ``boil-off''-driven thermal escape \citep{Lammer2016,Owen2016,Lammer2018} until the escape flux becomes first equal to and then smaller than the maximum possible EUV-driven escape flux \citep{Fossati2017}. Here, $r_0$ corresponds to the level where the tangential optical depth $\tau = 1$, which is approximately at an atmospheric pressure $p_0$ of 100\,mbar \citep{Brown2001,Lammer2016,Lopez2014}. During this extreme thermal atmospheric escape phase, the bulk atmosphere flows upwards, even without additional external energy exposure, such as the stellar EUV flux \citep{Lammer2016, Owen2016,Fossati2017,Lammer2018}. The remaining gravitationally-attracted H$_2$-envelope of the protoplanet after the ``boil-off'' phase depends on the equilibrium temperature $T_0$ at the orbital location and the protoplanet's mass $M_{\rm p}$. After disk dispersal, the remaining H$_2$-envelope is then lost or affected by an EUV-driven hydrodynamic hydrogen flow that can drag heavier trace elements such as $^{40}$K with it \citep{Lammer2020a,Lammer2020b}.

To estimate the photospheric radius $r_0$, and, hence, the atmospheric mass remaining at the planet at the end of the boil-off phase, \citet{Lammer2020b} employed the following procedure. For each protoplanetary mass considered in their study, these authors assigned a set of possible photospheric radii ($r_{0,i}$) in a way that average densities of the resulting planets correspond to the range between the atmosphereless rocky body and the planet, hosting an extended hydrogen atmosphere. For each pair of $M_{\rm p}$ and $r_{0,i}$ they applied a 1D hydrodynamical model following \citet{Erkaev2015} and \citet{Kubyshkina2018b,Kubyshkina2018a} to estimate the atmospheric mass loss rate and compare it to the one predicted by the energy-limited approximation. The respective $r_{0,i}$ at which the two mass loss rates become comparable (factor of 2), they adopt as the actual photospheric radius $r_0$ at the time when the ``boil-off'' regime changes to the EUV-driven escape regime (i.e., the beginning of the simulation in the present study). As the mass loss rates increase steeply with increasing $r_{0,i}$ (and, hence, the atmospheric mass $M_{\rm at}$), after this point any extra atmospheric mass would be removed within a period of time negligible in comparison to the protoplanetary disk lifetime.


The initial atmospheric mass fraction $f = M_{at}/M_{p}$ of the captured H$_2$-envelopes themselves that remained from the ``boil-off'' phase were estimated by \citet{Lammer2020b} by using an initial model integrator that is based on an adaptive, implicit RHD-code
\citep{Lammer2014,Johnstone2015,Stoekl2015,Stoekl2016}. This code provides a framework for the implicit solution of the equations of radiation hydrodynamics on adaptive grids \citep{Dorfi2006,Stoekl2007,Stoekl2008,Erkaev2014,Lammer2014}.
The relative atmospheric mass fraction $f$ as a function of $R_0$ {is based on \citet{Stoekl2015} and \citet{Stoekl2016} }and approximated through \citep{Lammer2020b},
\begin{equation}
f = \left \{ \left [ \log_{10}(r_0)+ 0.07151 \right] /1.14767 \right \}^{(1/0.32033)},
\end{equation}
where {$r_0$} is normalized to the radius of present Earth. When the protoplanet grows in mass, {$r_0$} decreases with time.  For our 5 different cases, the relevant values of $r_0$, and $M_{\rm at}$
are listed in Table~\ref{tab:params}.


\subsection{K abundances in primordial atmospheres, magma ocean and impactors}\label{sec:comp}

If the protoplanets accreted masses that could accumulate H$_2$-dominated primordial atmospheres within the disk lifetime, these captured gaseous envelopes initially contained also the elements in the abundances of their host stars. In the present study we consider the known K abundance $a_{\rm K,neb}$ of the Sun as given in \citet{Lodders2009} in a primordial atmosphere relative to the abundance of hydrogen $a_{\rm H,neb}$ by the simplified assumption that all K and its condensed species are distributed equally within the disk. Then we compute the total mass of each specific element within the captured H$_2$-envelope by
\begin{equation}
m_{\rm K,neb} = (a_{\rm K,neb}/a_{\rm H,neb})M_{at}A_{\rm K},
\end{equation}
where $M_{\rm at}$ is the mass of the accreted hydrogen atmosphere, and $A_{\rm K}$ is the atomic mass number of potassium.

The initial accreted mass ratios of potassium and hydrogen in the atmosphere is K/H\,=\, 1.351351361$\times 10^{-7}$ \citep{Lodders2009}, which is equivalent to the K/H ratio in the nebula. For $^{40}$K we retrieve a ratio of $^{40}$K/H\,=\, 2.3166$\times 10^{-10}$ by assuming that 0.147\% among all K isotopes \citep{Lodders2009} belong to $^{40}$K.

Below the captured H$_2$-atmospheres the temperature reaches 1500--3500\,K or even more, which is sufficient to form a global magma ocean \citep{Ikoma2006,Bouhifd2011,Stoekl2016,Lammer2020b}. Based on magma ocean depth estimates that were modeled by \citet{Lammer2020b}, we assume a {constant} average depth of a global magmatic layer of 1000\,km {for each of the different planets}, from which $^{40}$K isotopes will be outgassed into the surrounding H$_2$-dominated gaseous envelope. Since the convective velocities in magma oceans are in the order of 10 m s$^{-1}$ \citep{Solomatov2007}, one can assume that the potassium will be delivered from the bottom to the top within a timescale of days, thus a depletion of $^{40}$K near the surface-atmosphere interface through outgassing should not occur.

To estimate the mean abundance of K in the magma ocean, we assume {average} carbonaceous chondritic (CC) composition{, which is calculated based on the mean value of different types of carbonaceous chondrites} 
(CI, CM, CO, CV) as given in \citet{Wasson1988}, which results in $a_{\rm K,mo} = 403.7\,\mu \rm g/g$ for all potassium isotopes, and $a_{\rm ^{40}K}$\,=\, 0.5934\,$\mu \rm g/g$ for $^{40}$K. The $^{40}$K amount in the magma ocean, $M_{\rm ^{40}K,mo}$, is given in the last column of Table~\ref{tab:params}. In case of the impactors, we also assume {the same average} carbonaceous chondritic composition, i.e., $a_{\rm ^{40}K,imp}$\,=\, 0.5934\,$\mu \rm g/g$. To let the protoplanets grow from the time when the disk dissipated, we assume regular impacts related to mass additions within the first 20 Myr until each case reaches its final mass of 1.0\,$M_{\rm Earth}$.

In order to model {{provide estimates on} the amount of $^{40}$K in the atmosphere through outgassing from the} molten crust, we use the equilibrium condensation model {\textsc{GGchem} \citep{Woitke2018} and include a total of 18 elements (H C N O F Na Mg Al Si P S Cl K Ca Ti Cr Mn Fe). This results in the consideration of a total of 471 gas species and 208 condensates in the calculation. {This equilibrium chemistry model takes into account the condensation of all condensates, which are saturated. Although the condensate phase for some pressure-temperature conditions considered in this work result in a partially molten crust, the dissolution of gas species into the melt are not taken into account.}

{Calculations with the equilibrium chemistry solver \textsc{GGchem} are based on a set of different elements $i$ and their abundances (total element abundance $\epsilon_\mathrm{tot}(i)$), the atmospheric pressure $p$, and the temperature $T$. As the atmospheric and condensate composition at equilibrium are calculated for each $(p,T)$ point individually, it is not possible to directly model the evolution of a primordial atmosphere which equilibrates with another source material, i.e., a planetary mantle of a given composition. However, a work-around can be undertaken by creating a set of total element abundances which are representative of this situation.}

{In order to investigate the contribution of non-H elements to the \ce{H2}-dominated atmosphere above a mantle with average carbonaceous chondritic composition, we create a set of total element abundances, which results in a hydrogen dominated atmosphere. The relative abundances of all elements but hydrogen are taken from the average CC \citep{Wasson1988}, while the H abundance is artificially increased. Increasing the total hydrogen abundance to 2\% mass fraction results in atmospheres of $\sim99\,$\%  molecular and atomic hydrogen for the parameter range studied in this work, i.e., in an atmosphere that is equivalent to an accreted primordial atmosphere. The ratio of \ce{H2/H} is dependent on the pressure and temperature.} Without this {hypothetical} increase in hydrogen abundance, {the equilibrium atmosphere for the average chondritic element abundance would result in a steam atmosphere, dominated by \ce{H2O} and \ce{CO2} but , in this work, we focus on a captured primordial \ce{H2}-atmospheres with additional outgassing of $^{40}$K.} The influence of changes to the total element abundance to the atmospheric composition is further discussed in Appendix \ref{App:Kfraction} and in \citet{Herbort2020,Herbort2021}.

\begin{figure}
\centering
\includegraphics[width = 1.0\linewidth, page=1]{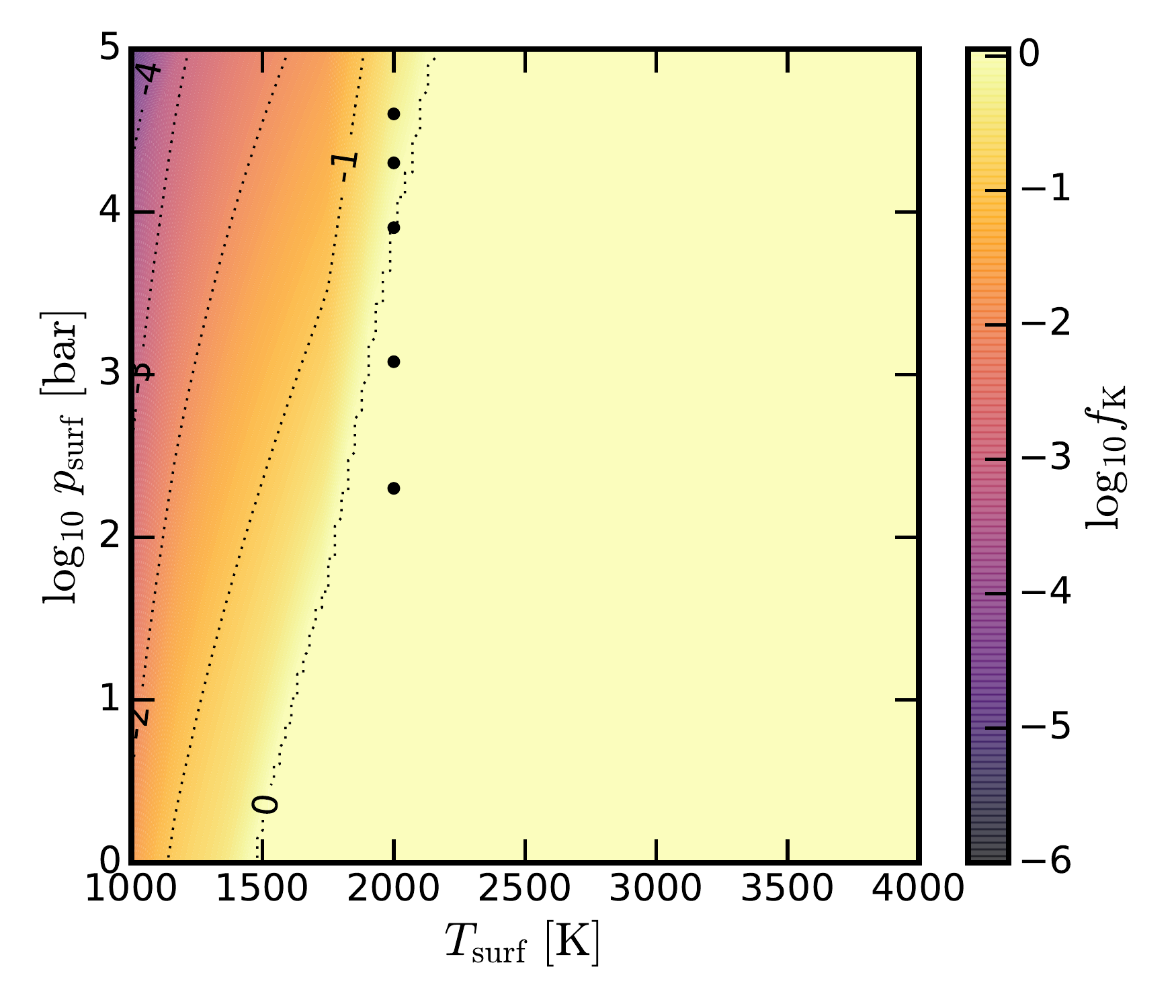}
\caption{K fraction in the gas phase dependent on $p_{\rm surf}$ of the H$_2$-envelope and assumed $T_{\rm surf}$ between 1000 - 4000 K of an underlying magma ocean. The equilibrium chemistry between the primordial H$_2$-dominated envelope and the magma ocean has been calculated with {\textsc{GGchem}} \citep{Woitke2018} for {an atmospheric} composition based on the averaged carbonaceous chondrite abundances {with increased hydrogen abundances} (see text). The numbers along the dotted lines indicate $\log(f_{\rm K})$.
{For $\log (f_\mathrm{K}) =0$ no K containing condensate is stable. The black dots for different pressure levels and temperatures of 2000\,K indicate the main ($p,T$) conditions investigated in this work.}}
\label{fig:Kfrac}
\end{figure}

In Fig.~\ref{fig:Kfrac}, the fractionation of K into gas and melt/solid is shown for the parameter range $T_{\rm surf}\in [\rm 1000\,K, 4000\,K]$ and $P_{\rm surf} \in [\rm 1\,bar , 10 000\,bar]$. The most stable K bearing condensate for the entire parameter range is $\rm KAlSi_2O_6[s]$. For $T\geq 2500$\,K no condensate bearing K is thermally stable.

We calculate the K gas fraction $f_{\rm K}$ as
\begin{equation}
f_{\rm K} = n_{\rm gas, K}/n_{\rm tot, K},
\end{equation}
where $n_{\rm gas,K}$ is the number density of all K bearing species in the gas phase, and $n_{\rm tot,K}$ the number density of all K bearing species in gas and condensate phase. This shows that for a large parameter range a high fraction of K can be expected to be in the gas phase. The reason for the small loss due to condensation of K in Fig.~\ref{fig:Kfrac} for $T_{\rm surf}{<}1300$\,K is a direct result of the high hydrogen abundance in the surrounding atmosphere.

Our results are also consistent with previous studies when we assume lower hydrogen abundances that correspond to silicate and steam atmospheres. \citet{Schaefer2010}, for instance, report that within their simulations between 0.7 – 11\% of K will be in the gas of a silicate atmosphere at a temperature of 1500\,K, and a pressure of 100\,bar. \citet{Fegley2016} found gas to melt fractions $d_{\rm K} = n_{\rm gas, K}/n_{\rm melt, K}$ for high pressure steam atmospheres between 270 - 1100\,bar, and 2000 - 3000\,K to be within $d_{\rm K}$ $\approx$ 0.0003 - 0.013. The direct comparison between the BSE and CC results from \citet{Fegley2016} and the ones presented here can be found in the Appendix~A, where we further investigate the K fraction in the gas under different atmospheric compositions.

Since we assume a pressure equilibrium between $M_{\rm ^{40}K,atm}(t)$, the amount of $^{40}$K in the atmosphere, and $M_{\rm ^{40}K,mo}(t)$, {the amount of $^{40}$K in the magma ocean}, it follows that the ratio
\begin{equation}\label{eq:dk}
f_{\rm K} = M_{\rm ^{40}K,atm}(t)/M_{\rm ^{40}K,tot}(t) = \rm const.,
\end{equation}
as long as potassium can be supplied from the underlying magma ocean or be delivered via impactors. {By way of example, for our scenario with an initial planetary mass of $M = 0.8\,M_{\rm Earth}$ and $f_{\rm K} = 0.1$, the fraction of $^{40}$K in the atmosphere will always be 10\% as long as $^{40}$K can be resupplied by the magma ocean or delivered by impactors. If potassium escapes from the atmosphere during one time step, it will be compensated through release of $^{40}$K from the magma ocean into the atmosphere, which implies that the abundance of $^{40}$K in the magma ocean decreases together with its abundance in the atmosphere to keep $f_{\rm K} = 0.1$ constant. If potassium is delivered by an impactor by a larger amount than escapes during the same time step, we assume that part of $^{40}$K will go into the magma ocean to again keep $f_{\rm K} = 0.1$ in the atmosphere constant, i.e., the abundance of potassium in both reservoirs will increase.} This also implies that $M_{\rm ^{40}K,atm}(t)$ is an ever changing mixture of potassium originally accreted from the nebula, {released from} the magma ocean, and delivered via impacts, that is continuously dragged partially into space by the escaping hydrogen atmosphere.

Time variations of the hydrogen and potassium are determined by the mass conservation equations, i.e.,
\begin{eqnarray}
\frac{d M_{\rm p}}{d t} = S_{\rm imp}, \\  \label{evol1}
\frac{d M_{\rm at}}{d t} = - L_{\rm H}, \\
\frac{d M_{\rm ^{40}K,neb}}{d t} = -L_{\rm H} M_{\rm ^{40}K,neb} \kappa_{K} /M_{\rm at}, \\
 \frac{d M_{\rm ^{40}K,mo}}{d t} = -f_{\rm K} L_{\rm H} M_{\rm ^{40}K,mo} \kappa_{\rm K} /M_{\rm at}, \\
 \frac{ d M_{\rm ^{40}K,imp}}{d t} = -f_{\rm  K} L_{\rm H} M_{\rm ^{40}K,imp} \kappa_{\rm K} / M_{\rm at} + S_{\rm imp} a_{\rm ^{40}K,imp},  \label{evol5}
\end{eqnarray}
where $L_{\rm H}$ is the loss rate,   $M_{\rm p}$ and  $M_{\rm at}$ are the masses of planet and atmosphere, $S_{\rm imp}$  is the source function of the mass impact, and $\kappa_{\rm K} = X_{\rm K\infty}/ X_{\rm K0}$, where $X_{\rm K\infty}$ and  $X_{\rm K0}$ are  the relative mass densities of potassium  at infinity and at the lower boundary, respectively.

\section{RESULTS AND DISCUSSION}\label{sec:results}
Here we discuss the escape of the captured H$_2$-dominated atmosphere envelopes from the studied protoplanet cases along the different stellar EUV-activity evolution tracks and then we present the evolution and loss of the embedded $^{40}$K isotopes that are dragged by the escaping hydrogen flow.

\subsection{Escape of the H$_2$-dominated primordial atmospheres}
We obtained quasi-stationary hydrodynamic solutions for a range of evolution times, i.e., $t = 5, 10, 15 ... 1000$ Myr. We start with the initial planetary and atmospheric masses/cases shown in Table~\ref{tab:params} and let all planets grow up to 1\,$M_{\rm Earth}$ through mass additions by {Moon or Mars-mass} impacts. This mass addition is a completely passive process that only changes the mass of the body and adds $^{40}$K and neglects any other effects such as the heating and/or potential escape of part of the atmosphere. {In reality, however, several percent of the atmosphere may be eroded by such impactors \citep[e.g.,][]{Schlichting2018,Lammer2020a}, thereby leading to faster erosion of the primordial atmosphere as assumed within our model.}

The EUV-driven mass loss of the respective primordial H$_2$-dominated atmospheres for the initially 0.55$M_{\rm Earth}$ to 1.0$M_{\rm Earth}$ protoplanetary masses at orbital locations of 1\,AU are shown in Fig.~\ref{fig:H-loss} for all of our three EUV-evolution tracks, i.e., for a slowly, moderately and fast rotating young G-type star. If we assume that the disk dissipated after 5\,Myr, one can see that the accumulated primordial atmospheres of protoplanetary bodies with masses $\leq$\,0.55\,$M_{\rm Earth}$ will be lost within $\leq$\,3 Myr after disk dissipation. Since the star's EUV flux stays saturated over this time frame for moderate and fast rotators, the loss rates for these two tracks stay identical, while the hydrogen atmosphere will last slightly longer for a slow rotator with $t_{\rm sat}\approx5$\,Myr. For planetary bodies with masses of 0.6\,$M_{\rm Earth}$, a more massive primordial atmosphere is captured, so that the loss of their H$_2$-envelopes take between $\approx$22--35\,Myr. Again, it will take longer for a slow rotator, while both, moderate and fast rotators are yet within the saturation phase until the entire atmosphere will have evaporated. The kinks in the lines correspond to the time when the growing protoplanets reached 1\,$M_{\rm Earth}$. One can also see that for initial protoplanetary masses of $\geq$\,$0.7\,M_{\rm Earth}$ it can take hundreds of Myr until the primordial atmosphere is lost; for slowly and moderately rotating G stars and planets with $1\,M_{\rm Earth}$ these captured atmospheres may even remain, which might render these planets uninhabitable.
\begin{figure}
\begin{center}
\includegraphics[width=1.0\columnwidth]{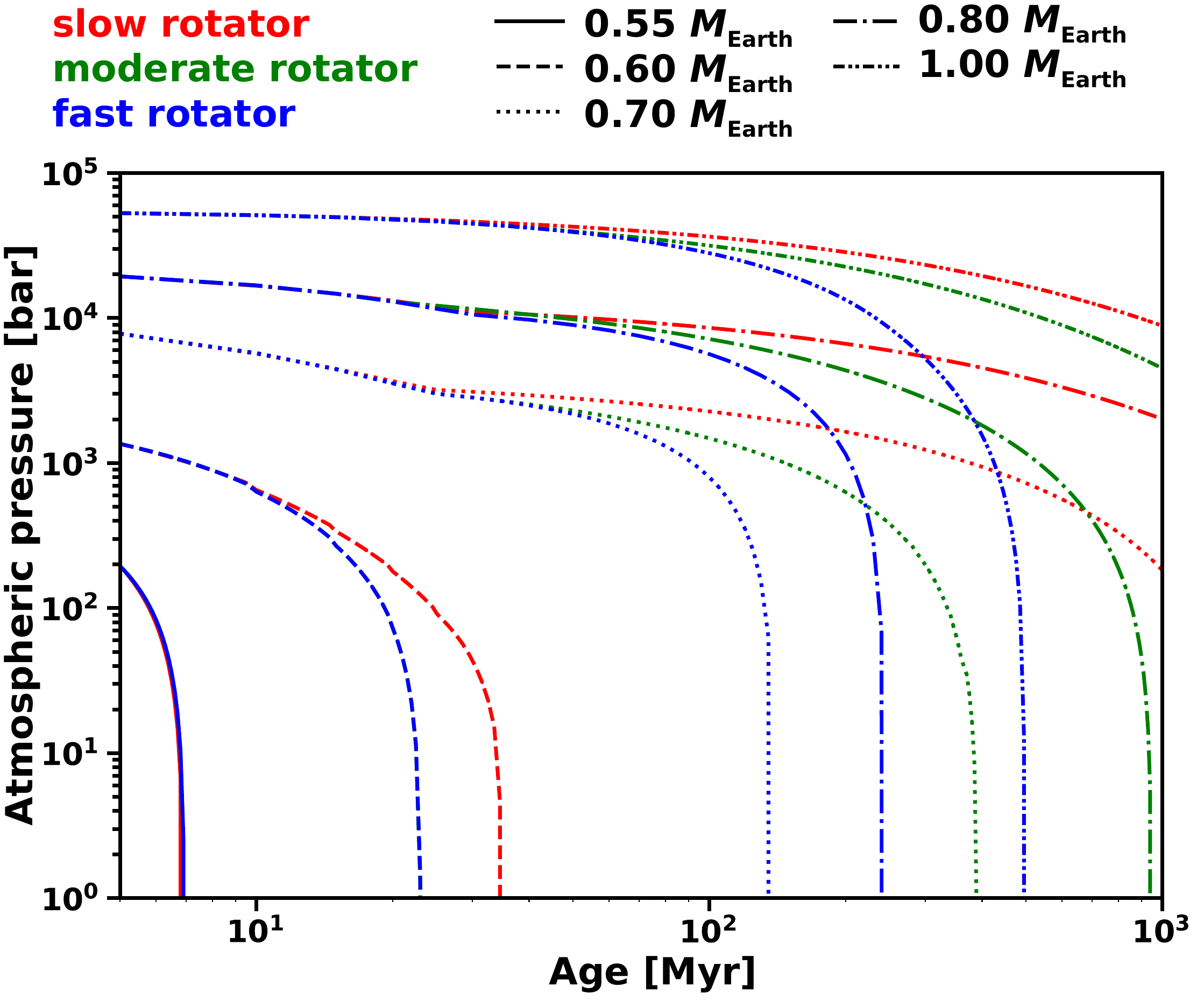}
\caption{The escape of the hydrogen atmospheres for our different cases and for slow (red), moderate (green), and fast rotating young G-stars (blue), respectively. The sudden changes in the loss rate, particularly visible for the 0.70\,$M_{\rm Earth}$ and 0.8\,$M_{\rm Earth}$ cases at $\approx$25 and 30\,Myr, respectively, are due to the assumed accumulated 0.1\,$M_{\rm Earth}$ impactor mass, which increase the gravity of the protoplanets, and consequently reduce the respective escape rates quite sudden. For a slow rotator, the hydrogen envelopes of the protoplanets with 0.8\,$M_{\rm Earth}$, and 1.0\,$M_{\rm Earth}$ cannot be lost entirely, and a relatively thick H$_2$-dominated atmosphere will remain; similarly for a moderate rotator and 1.0\,$M_{\rm Earth}$.}\label{fig:H-loss}
 \end{center}
\end{figure}

One should further note that these loss rates of primordial atmospheres should be considered as {lower} values because the erosion of atmospheric gas by large planetary embryos is neglected {\citep[see, e.g.,][]{Schlichting2018}}. However, in \citet{Lammer2020b}, SPH impact model runs indicate that, depending on impact velocity and angle, collisions with Moon-sized planetary embryos can remove between 1--10\,\% of atmospheric mass. {Planetesimals which have lower masses on the other hand will erode negligible amounts of the primordial atmosphere compared to} collisions with large planetary embryos. Therefore, one should keep in mind that the time scales for the atmospheric losses shown in Fig.~\ref{fig:H-loss} may be upper limits derived only by the thermal escape rates caused by the host star of the system. In the following section we investigate how much of the $^{40}$K might be lost together with the escaping H atoms, if one assumes that the outgassed $^{40}$K can reach the upper atmosphere.

\subsection{$^{40}$K loss and evolution}\label{sec:K-evo}

{As shown in Sect.~\ref{sec:comp} and Appendix~\ref{App:Kfraction}, the fraction of K remaining in the atmosphere for an atmosphere {at a given total pressure and temperature} is mainly depending on the hydrogen content of the atmosphere. For a hydrogen-rich element abundance, the resulting H$_2$-dominated atmosphere has not enough K-bearing molecules present to saturate the K-bearing condensate which results in them not being thermally stable so that K remains in the envelope resulting in $f_{\rm K}=1$. Sets of total element abundances with lower H abundances would result in steam atmospheres (which are not part of this study). For these, {the K-bearing condensate  KAlSi$_2$O$_6$ becomes} saturated and reduce the K-abundance in the gas phase to the vapour pressure of the condensates. This results in significantly lower values for $f_{\rm K}$. For low total pressures, the contribution of the vapour pressure limited gas phase of K becomes increasingly significant (see also Fig.~\ref{fig:Kfrac_all}).}

{Concerning steam atmosphere{, i.e., atmospheres dominated by water vapour,} it also has to be mentioned that in such an environment the dragging of K might generally be very limited. The steam atmosphere outgasses catastrophically when the magma ocean solidifies which can only happen when the frequency of impactors diminishes and the main accretion is over, and/or when most of the primordial atmosphere escaped to space since before that the greenhouse effect will be too strong to permit the solidification. That means that most of the remaining K will already have condensed and, depending on the scenario, the XUV flux could have already decreased to lower levels such that hydrodynamic escape may already be significantly weakened. Moreover, steam atmospheres are also populated by heavier elements such as oxygen and carbon species which influences the dragging of other trace elements such as K and, therefore, renders it insignificant, at least at heavier protoplanets as we study here \citep[see, e.g.,][]{Lichtenegger2016}.} Therefore, the assumption that K can be dragged away by escaping lighter molecules efficiently is justified especially for primordial H$_2$-dominated atmospheres.}

\begin{figure}
\begin{center}
\includegraphics[width=0.994\columnwidth]{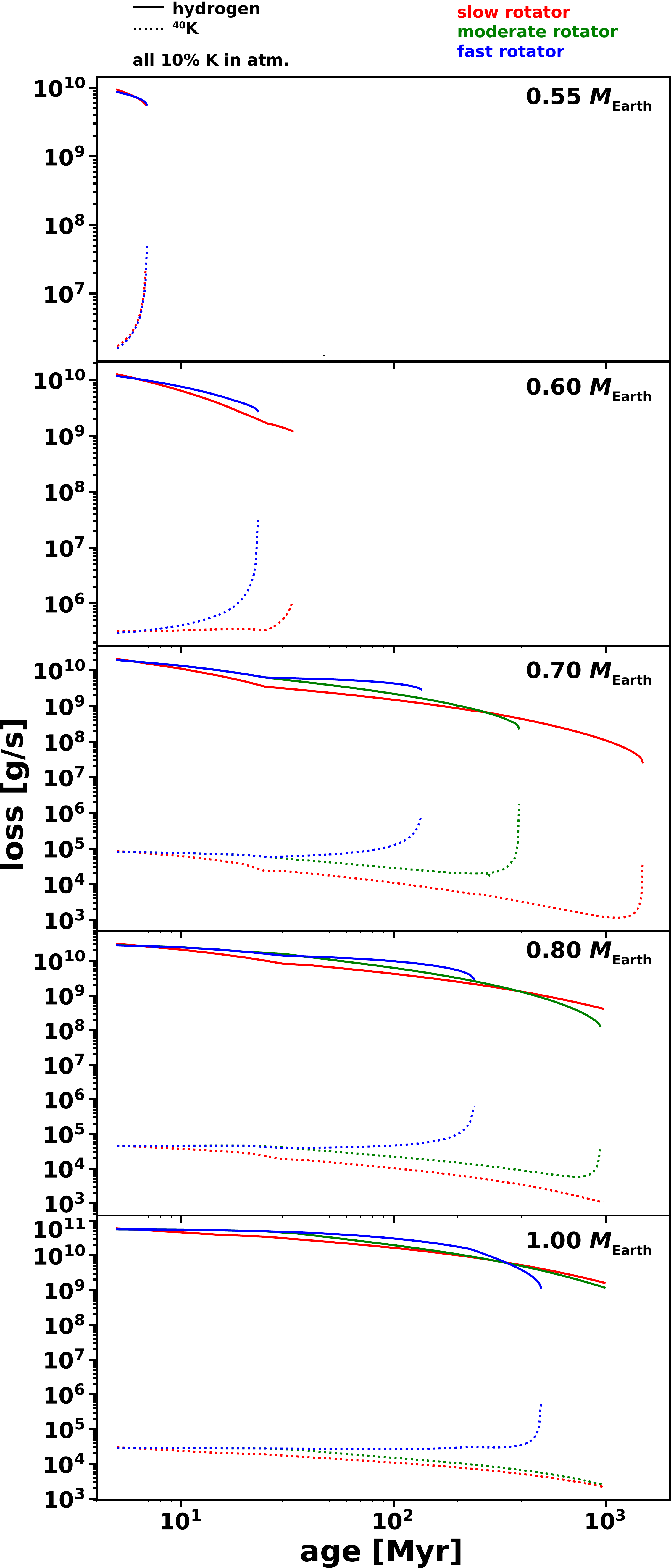}
\caption{The escape rates of hydrogen (solid lines) and of $^{40}$K that is mainly dragged by the escaping hydrogen (dotted lines) in $g/s$ for our 5 different cases, and for slow (red), moderate (green), and fast rotating young G-stars (blue), respectively. The loss rates of $^{40}$K decrease when the hydrogen loss rates increase, because the exobase level moves closer to the planet so that more of the heavier $^{40}$K is populated so that under this conditions more of the heavier elements can be dragged by the escaping H atoms.}\label{fig:escape}
 \end{center}
\end{figure}

Besides the corresponding hydrogen loss rates and the escape of the primordial atmosphere, we also interpolated $\kappa_{\rm K}$ as a function of time,
so that we can calculate the system of evolution equations (Eqs.~\ref{evol1}-\ref{evol5}) that are necessary for the calculation of the $^{40}$K loss rates. Fig.~\ref{fig:escape} shows the loss rates of hydrogen and the dragged $^{40}$K for the first 1\,Gyr after the gas disk evaporated for the 5 protoplanetary cases of Table~\ref{tab:params}. For this example we assumed that 10\,\% of the $^{40}$K is embedded in the escaping hydrogen dominated primordial atmospheres. For the cases in which the loss rates stop before 1\,Gyr, the hydrogen is completely lost and $^{40}K$ consequently stops to be dragged in to space. For the 0.8$M_{\rm Earth}$ slow rotator, and the 1.0$M_{\rm Earth}$ slow and moderate rotator cases, the primordial atmospheres did not escape during 1\,Gyr, and will also not entirely be lost subsequently. One can, therefore, also expect that $^{40}$K will condense and fall out to the surface, when surface/air temperatures will have decreased below the condensation temperature of K.

One can also see another interesting phenomenon, i.e., when the escape rate of hydrogen diminishes towards the end of the lifetime of the {respective} hydrogen atmospheres, the loss of $^{40}$K clearly increases. The reason for this behaviour can be found in the significantly decreased value for the exobase level, which increases the amount of the heavier $^{40}$K that is available to be dragged away at the respective height.

\begin{figure}
\begin{center}
\includegraphics[width=0.92\columnwidth]{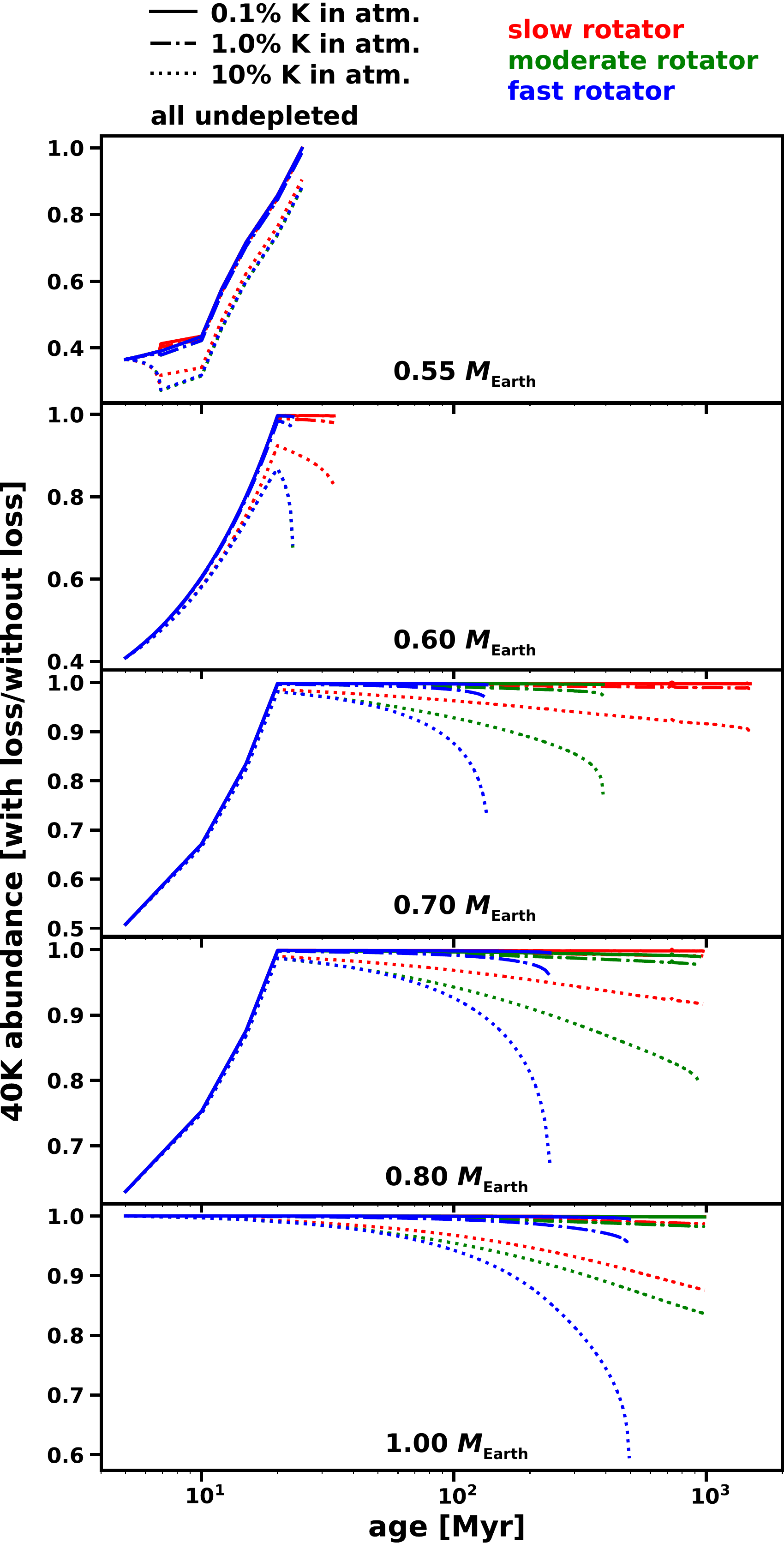}
\caption{The evolution $^{40}$K at the respective planets compared to the final abundance of $^{40}$K, if no atmospheric escape and the therewith connected dragging of the radioactive potassium isotope would be considered. For low values of $^{40}$K in the atmosphere, i.e., 0.1\% (solid) and 1.0\% (dash-dotted), respectively, almost no change in the abundance of $^{40}$K can be observed. Only for higher amounts (i.e., 10\% in this case; dotted), a significant effect is visible. Planets that lose their primordial atmosphere before they finish their accretion afterwards accumulate newly delivered $^{40}$K.Interestingly, massive planets generally lose more potassium, since the more massive hydrogen dominated primordial atmosphere can drag the heavy element over a longer time period. In addition, planets exposed to a fast rotator lose more than those orbiting moderate, or even slow rotators.}\label{fig:K-evo}
 \end{center}
\end{figure}

Fig.\ref{fig:K-evo} shows the evolution of the planetary $^{40}$K abundance through planetary growth and atmospheric loss normalized to a final $^{40}$K abundance that would not have suffered atmospheric escape {for $f_{\rm K}$\,=\,0.001,0.01, and 0.1}. Any mass additions to the planets in Fig.~\ref{fig:K-evo} are further assumed to be undepleted and to have {an average} carbonaceous chondritic composition. A different assumed composition, however, would not significantly alter our results, since we are only interested in the relative loss.

One can see that the loss of $^{40}$K is only efficient when $f_{\rm K}$, the ratio of $^{40}$K between the atmosphere and its total abundance, would be $\ge 0.01$. The effect of atmospheric escape at such massive bodies on the final abundance of $^{40}K$ for smaller values of $f_{\rm K}$, would, therefore, almost be negligible, and other loss processes will have to be considered to, e.g., explain the diminished K/U ratios of the terrestrial planets in the solar system compared to the chondritic values. However, for $f_{\rm K}> 0.01$ loss rates of the element can not be neglected anymore. Besides this, one can see that planets that start with larger masses at the time when the disk dissipated will likely lose more $^{40}$K in most of the cases, since these bodies accumulate more massive H$_2$-dominated primordial atmospheres that can drag $^{40}$K over a longer time period. Fast rotators further imply a greater loss of $^{40}$K than slowly, or even moderately rotating young G-stars do, simply due to the higher energy input into the upper atmosphere, which increases atmospheric escape of $^{40}$K. One can also see that for planets that have a mass of about 0.6$\,M_{\rm Earth}$ at the time when the gas disk evaporated, the loss of dragged  $^{40}$K isotopes is most efficient in all cases. This is due to a combination of accumulated primordial atmosphere mass, the gravity potential of the protoplanet at that time and the corresponding EUV-dependent hydrodynamic loss rates.

\begin{figure}
\begin{center}
\includegraphics[width=1.0\columnwidth]{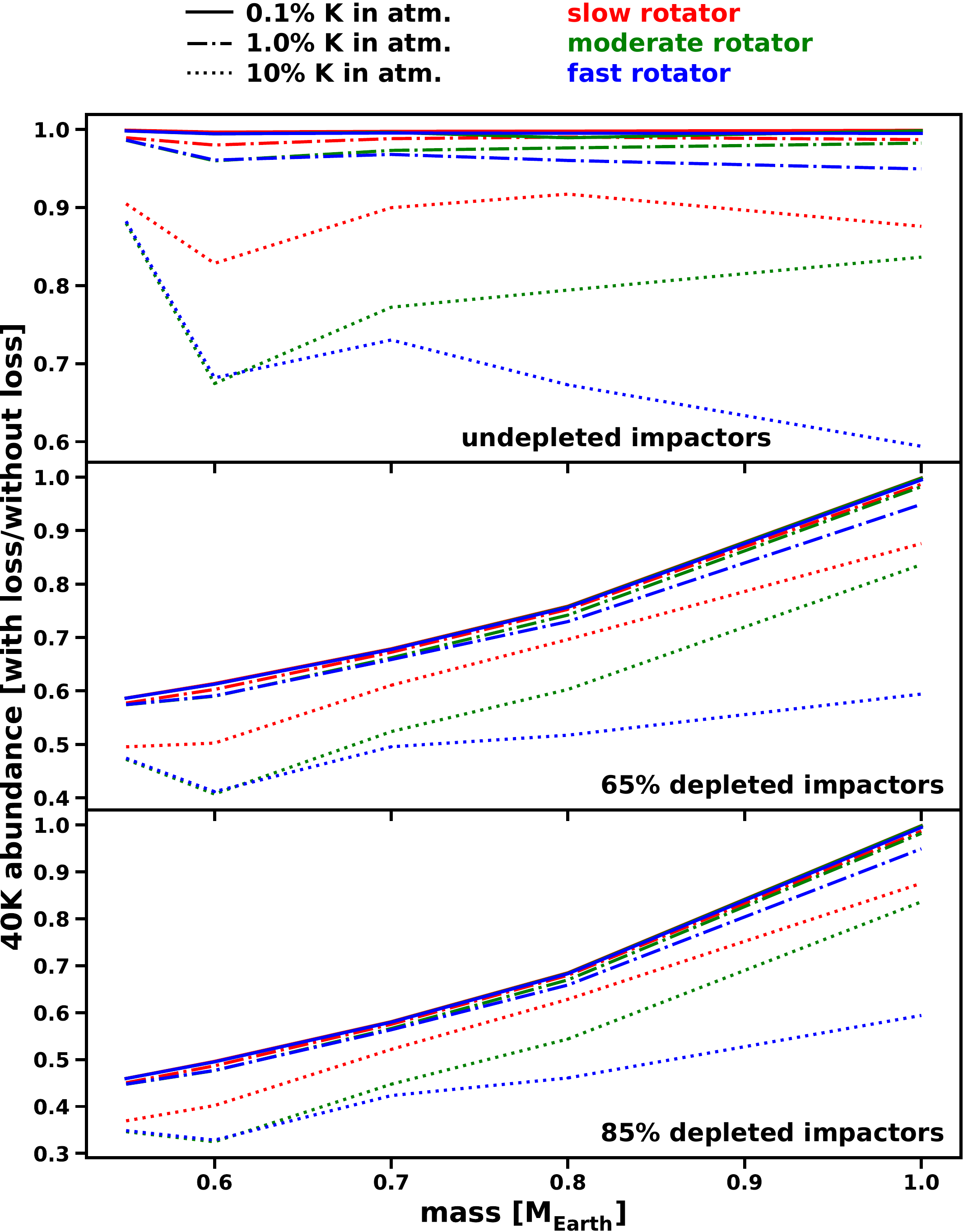}
\caption{The final outcome (i.e., corresponding to the end of the curves in Fig.~\ref{fig:K-evo}) of atmospheric escape of a primordial atmosphere that drags $^{40}$K on the final composition as compared to planets without atmospheric escape. {That is, it shows the depletion factor of $^{40}$K of each planet compared to the same planet but without escape (with 1.0 indicating no depletion).} Here, the upper panel shows our 5 planets ranging from 0.55\,$M_{\rm Earth}$ to 1.0\,$M_{\rm Earth}$ for slow (red), moderate (green), and fast rotating young G-stars (blue), in case that the impactors are not depleted in moderately volatile elements, and, therefore, show an undepleted carbonaceous chondritic composition. The middle and lower panels, however, show the same planets but with impactors being depleted in $^{40}$K by 65~\% and 85~\%, respectively. This illustrates that mostly depleted impactors can lead to a lower final abundance of moderately volatile elements.} \label{fig:K-final}
 \end{center}
\end{figure}

Fig.~\ref{fig:K-final} shows the end product of the planetary $^{40}$K abundance for $f_{\rm K}$\,=\,0.001,0.01, and 0.1, again normalized to the corresponding abundance value if one neglects loss to space. Here, we assumed that the planetary building blocks of the impactors onto the growing protoplanets were undepleted (upper panel), and, arbitrarily, 65\% (middle panel) and 85\%, respectively, depleted in $^{40}$K, to illustrate the effect of impactor depletion onto the evolution of the $^{40}$K abundance. Small planetary embryos with masses comparable to the moon or even smaller \citep{Hin2017,Young2019,Benedikt2020} can efficiently lose moderately volatile elements directly through outgassing from the magma ocean into space due to their low gravity, without the need of any additional atmosphere that can drag away K. This process could, therefore, be much more efficient than the loss process studied within this study, and might, therefore more strongly affect the final abundances of rocky exoplanets, as can be seen in Fig.~\ref{fig:K-final}. This figure illustrators that impactor depletion logically has a greater effect on initially smaller planetary embryos, since more material has to be delivered by depleted impactors. The initially fully grown planet, on the other hand, has no depleted impactors at all and ends with the same {K} abundance in all three panels. Due to the greater importance of depleted impactors for the initially smaller embryos, the relative $^{40}$K abundances in panels (b) and (c) rise towards higher masses. For panel (a), on the other hand, the abundance decreases for slow and fast rotators, since more $^{40}$K can be lost due to the more massive atmospheres as compared to the smaller protoplanets. Further research on this topic, however, will be needed in the future to better quantify the depletion of small planetary embryos.

\begin{figure}
\begin{center}
\includegraphics[width=1.0\columnwidth]{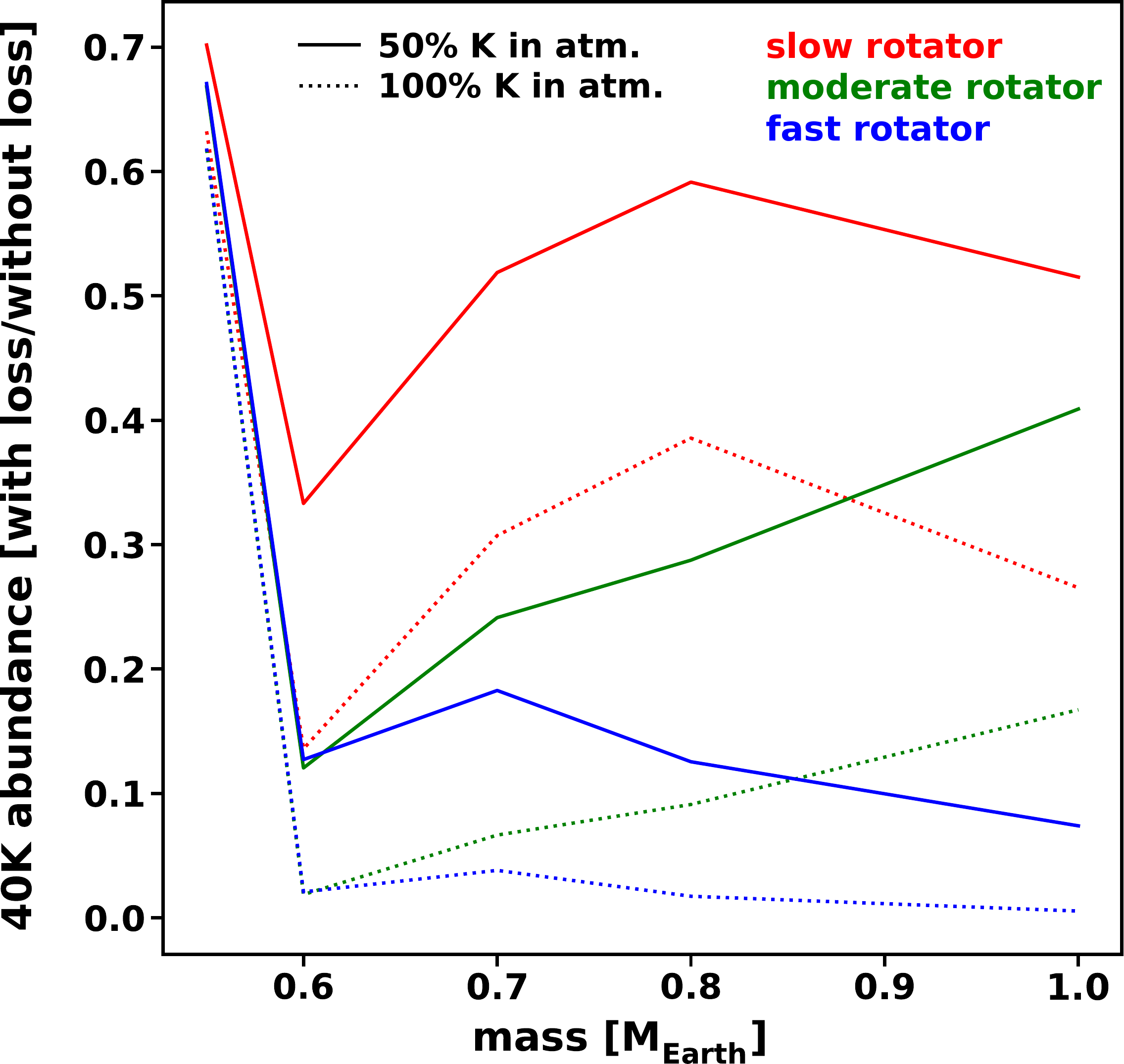}
\caption{The final outcome of atmospheric escape of a primordial atmosphere that drags $^{40}$K on the final composition as compared to planets without atmospheric escape for 50~\% and 100~\% of all $^{40}K$ within the atmosphere. Here, we assumed the impactors to be undepleted. Protoplanets with $\geq$~0.6\,$M_{\rm Earth}$ could lose almost all of their $^{40}$K, particularly for moderate and fast rotators. For depleted impactors, the final abundance of $^{40}$K would also be significantly lower for the slow rotator cases as well as for protoplanets with an initial mass of $\geq$~0.5\,$M_{\rm Earth}$.}\label{fig:K-final-50-100}
 \end{center}
\end{figure}

Finally, we also modeled all of our cases with $f_{\rm K}$\,=\,0.5, and 1.0, i.e., with 50\%, and 100\% of $^{40}$K in the atmosphere. This can be seen in Fig.~\ref{fig:K-final-50-100}, which represents the final outcome of these simulations with impactors assumed to be undepleted. In case that indeed $\geq$~50\% of all $^{40}$K within the magma ocean would be within the primordial H$_2$-dominated atmosphere, loss of the radioactive isotope would be significant, and could even lead to the entire escape of all $^{40}$K within the molten mantle. While the fast rotator cases for initial masses of $\geq0.6\,M_{\rm Earth}$ show the most pronounced losses within these simulations ranging from $\approx$~82\% to $>$~99~\% of all the $^{40}$K within the magma ocean being lost, smaller initial masses and all of the slow rotator scenarios (except 0.6\,$M_{\rm Earth}$ and $f_{\rm k}$=1.0), can retain from 27\% up to even 70~\% of all $^{40}$K. If one assumes depleted impactors, these values would be lower and closer to the fast rotator scenarios; the same would be the case for protoplanets that are irradiated by a moderately rotating young star.

However, we caution that it seems to be unlikely that indeed $\geq$~50~\% of all $^{40}$K that was initially within the magma ocean could be within the atmosphere at the same time, and/or can reach the upper atmosphere through diffusion or turbulence. Even though our geochemical model retrieved values for $f_{\rm k}$ of up to 1.0, we have to emphasize that within {\sc GGchem} only the upper part of the magma ocean will be equilibrated with the overlying atmosphere. That means, that the actual amount of $^{40}$K also depends on the convection time of the magma ocean which is estimated to be in the order of days \citep[][]{Solomatov2007} {but also on the diffusion of $^{40}$K into the upper atmosphere. Taking this into account, one can assume that a mixing ratio of $f_{\rm k}\sim 0.1$ may be a solid estimate. Since i) almost all of the $^{40}$K within the equilibrated part of the magma ocean will be in the atmosphere, and ii) the turnover timescale of the whole magma ocean is only in the order of days, values well below $f_{\rm k}< 0.1$ seem to be unlikely. Results for values well above $f_{\rm k}> 0.1$, on the other hand, may be similarly unlikely, at least for the upper atmosphere, due to the limited diffusion of $^{40}$K into the thermosphere from which the escape will take place.} However, the diffusion of $^{40}$K into the upper atmosphere, as well as simulating the turn-over time of the underlying magma ocean have to be modeled in more detail to quantify these processes properly which is beyond the scope of the present work.

In addition, we have to emphasize that we kept the surface temperature ($T_{\rm surf} = 2000\,\rm K$) as well as $f_{\rm K}$ constant for each simulation. This, however, is a strong simplification. $T_{\rm surf}$ in fact would likely decrease over time due to the decrease of surface pressure, which would in turn change the chemical equilibrium between magma ocean and atmosphere. As one can see in Fig.~\ref{fig:Kfrac} and in Appendix~\ref{App:Kfraction}, $f_{\rm K}$ changes in dependence of $T_{\rm surf}$ and surface pressure $p_{\rm surf}$, i.e., as pressure and temperature decrease over time due to the atmosphere's escape to space, $f_{\rm K}$ will decrease as well. Outgassing of $^{40}$K will even entirely halt at the time when the magma ocean solidifies which might take place before the whole atmosphere escaped (due to the pressure being too low for heating the surface efficiently), in case that frequent impactors will not prolong the magma ocean phase. The potential decrease of $f_{\rm K}$ over time additionally illustrates that setting $f_{\rm K}$ constant at 1.0 or 0.5 (see Fig.~\ref{fig:K-final-50-100}) will be an overestimation. These effects are important but coupling the different models for this research to account for these interconnections is beyond the current study which is mainly intended to quantitatively illustrate the versatile paths different planets can take through their evolution. However, we are planning to include these effects into future work.

From our study one can, therefore, conclude that terrestrial planets will end up with a wide diversity of radioactive heat producing elements, including $^{40}$K and, hence, we expect that many terrestrial planets will have initially different abundances than Earth. This is certainly a crucial key factor for whether these planets might later evolve to Earth-like habitats or not.

\subsection{Effect on planetary heat production}\label{sec:heat}
\begin{figure}
\begin{center}
\includegraphics[width=0.82\columnwidth]{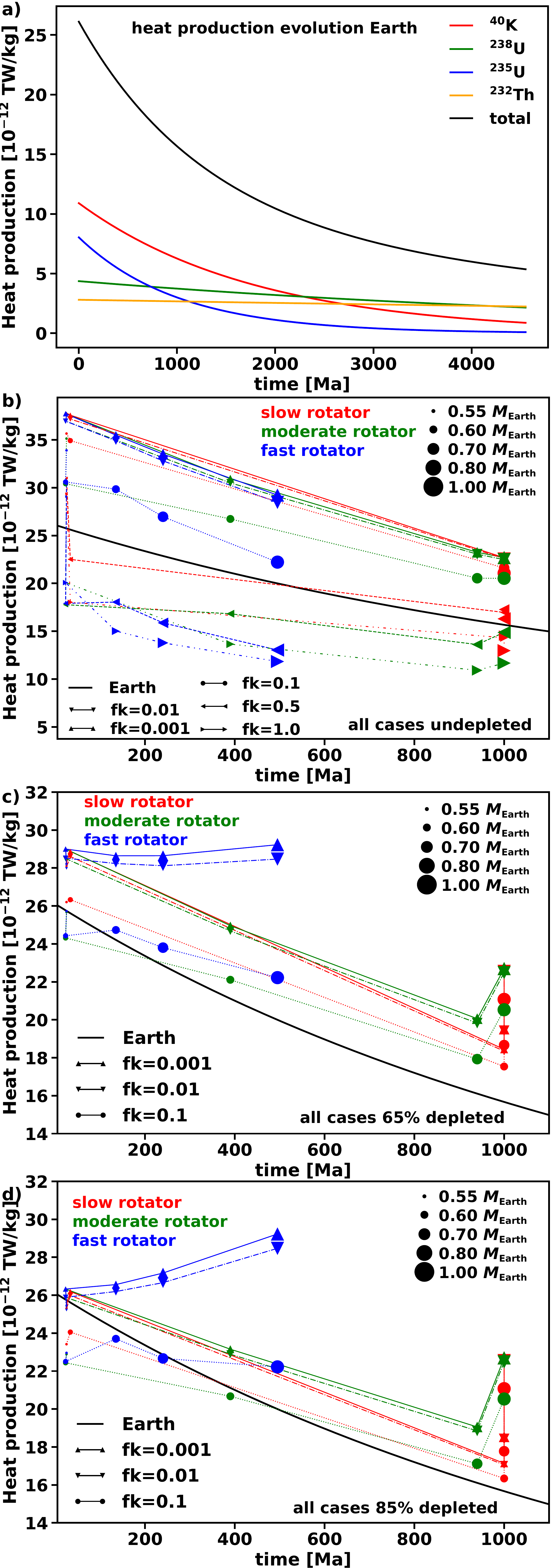}
\caption{Heat production of the Earth over time through radioactive decay of $^{40}$K, $^{235}$U, $^{238}$U, and $^{232}$Th (a), vs the end products of heat production of all of our cases at the time of final growth to 1.0\,$M_{\rm Earth}$ or, if happened later, of the loss of the primordial atmosphere for undepleted (b), 65\% depleted (c), and 85\% depleted impactors (d). In the latter 3 panels, the black line depicts the evolution of the Earth's total heat production as in panel (a). Note that the x-axis is different for (a) than for the other panels. Note also that the red, green, and blue lines connect the different initial protoplanetary masses (i.e., 0.55, 0.6, 0.7, 0.8, and 1.0\,M$_{\rm Earth}$) for the same conditions, i.e., same rotator and same $f_{rm K}$}.\label{fig:Earth-vs-Cases}
 \end{center}
\end{figure}

As discussed in Sect.~\ref{sec:intro}, the heat production in a planet's interior which is caused due to heat producing radioactive elements exerts a first order control on the bodies thermal evolution, tectonics, and hence on the planet's evolution to a potential Earth-like habitat \citep{Stein2004,ONeill2007,Jellinek2015,ONeill2020}. The process that we investigated within this study -- the loss of outgassed $^{40}$K due to the dragging of hydrodynamically escaping hydrogen atoms from a primordial atmosphere -- can, depending on the protoplanetary mass, $f_{\rm k}$, and the EUV flux evolution of its host star -- significantly alter the initial heat budget of such a planet, which might therefore have severe effects on its subsequent habitability.
Fig.~\ref{fig:Earth-vs-Cases} compares the evolution of the Earth's heat budget with the heat budget of our hypothetical planets after they evolved to their final state (i.e., after the planets reached their final mass, or, if happening later, the primordial atmosphere was either lost completely or stopped being lost hydrodynamically after $\sim$1\,Gyr).

Here, panel (a) shows the evolution of the Earth's heat production rate from its beginning to the present day including the radioactive isotopes $^{40}$K, $^{235}$U, $^{238}$U, and $^{232}$Th; $^{60}$Fe, and $^{26}$Al were neglected due to their extremely short half-lives. The heat production was calculated through the relation \citep{ONeill2020,Turcotte2002}
\begin{equation}\label{eq:heat-prod}
  H = \sum_i C_{0,i} H_i \mathrm{\exp}\left(\frac{t \mathrm{ln}2}{\tau_{1/2,i}}\right),
\end{equation}
where the summation $i$ is over the four radioactive nuclides $^{40}$K, $^{235}$U, $^{238}$U, and $^{232}$Th, $t$ is the time, $H_i$ the heating rates, $\tau_{1/2,i}$ the half-lifes of the radioactive isotopes, and  $C_{0,i}$ the respective present mean bulk silicate Earth concentrations. The natural abundances of the isotopes are factored into $C_{0,i}$, as described in \citet{ONeill2020}. Table~\ref{tab:heat-prod} lists all of the input parameters. One can clearly see that $^{40}$K dominates the heat budget for the first $\sim$2\,Gyr, afterwards the uranium isotopes get more important.

\begin{table*}
  \begin{center}
    \caption{Relevant data for heat producing elements for Earth \citep[see also,][]{ONeill2020}.}
    \label{tab:heat-prod}
\begin{tabular}{l|c|c|c|c}
  \hline
   Isotope & Half-life $\tau_{1/2}$ [Myr$^{a,b}$] & Natural abundance A [\%]$^{b}$  & Heating rate H [W/kg]$^{a}$ & Present mean BSE concentration C$_0$ [kg/kg]$^{c}$ \\
   \hline
  $^{40}$K & 1250 & 0.0117 & $28.761\times10^{-6}$  & $30.4\times10^{-9}$ \\
  $^{238}$U  & 4668 & 99.27 & $94.946\times10^{-6}$ & $22.7\times10^{-9}$ \\
 $^{235}$U & 703.8 & 0.72 & $568.402\times10^{-6}$ & $0.16\times10^{-9}$ \\
  $^{232}$Th & 14050 & 100 & $26.368\times10^{-6}$ &  $85\times10^{-9}$ \\
  \hline
\end{tabular}
\end{center}
\footnotesize
$^{a}$\citet{Ruedas2017}, $^{b}$\citet{ONeill2016}, $^{c}$\citet{ONeill2008}
\end{table*}

Panel (b) of Fig.~\ref{fig:Earth-vs-Cases} illustrates the total heat production for all of our final cases with undepleted impactors, i.e., for initial masses of 0.55\,--\,1.0\,$M_{\rm Earth}$, $f_{\rm k}=(0.001,0.01,0.1,0.5,1.0)$, and for slow, moderate, and fast rotating young G-stars. The only difference to the Earth lies within the varying abundance of $^{40}$K within the mantle due to its escape to space; everything else was assumed to be the same. As can be seen, for $f_{\rm k} = (0.001,0.01, 0.1)$ all of the different cases lie significantly above the Earth's total heat production evolution (black line). Only for the extreme, and potentially unlikely cases with $f_{\rm k} = (0.5, 1.0)$, most are below the Earth's evolution. This is an interesting outcome, which illustrates that, as long as not most or even all of $^{40}$K is within the primordial atmosphere where it can escape to space, other processes have to affect the final $^{40}$K abundance in addition.

As was shown by \citet{Lammer2020b}, the Earth's final K/U ratio could be explained by a mixture of different building blocks with various abundances of K, escape of K via the Earth's primordial atmosphere (same process as researched within this study), and by accretion of impactors that were already depleted in K. It is, therefore, no surprise that most of our cases in panel (b) show heat production rates that are higher than for the Earth, since these do neither include various initial building blocks of the protoplanets (but only K-rich carbonaceous chondrites), nor depleted impactors. Panels (c) and (d), on the other hand, show our simulations for $f_{\rm k} = (0.001,0.01,0.1)$ and for impactors that are 65\%, and 85\% depleted in $^{40}$K, respectively. For protoplanets with relatively small masses at the time of disk dissipation, i.e. ranging from $0.55 - 0.8\,M_{\rm Earth}$ in dependence of $f_{\rm k}$ and the initial rotation rate of the star, some of these cases already show heat production rates that are comparable or even below the Earth's case. This illustrates that atmospheric escape of $^{40}$K via the protoplanet's primordial atmosphere, together with depleted impactors could not only already explain the Earth's heat budget, but could even lead to planets with lower heat production rates. Interestingly, \citet{Lammer2020b} found that proto-Earth should have had a mass of $\approx0.55 - 0.6\,M_{\rm Earth}$ at the time the solar nebula disappeared, in case the Sun was approximately a slowly rotating young G-star. This fits relatively well with our 0.55 and 0.6\,$M_{\rm Earth}$ cases for a slow rotator and $f_{\rm k} = 0.1$ for an impactor depletion of 65\%, and $f_{\rm k}\approx 0.01$ for 85\%, respectively.

\begin{figure*}
\begin{center}
\includegraphics[width=1.95\columnwidth]{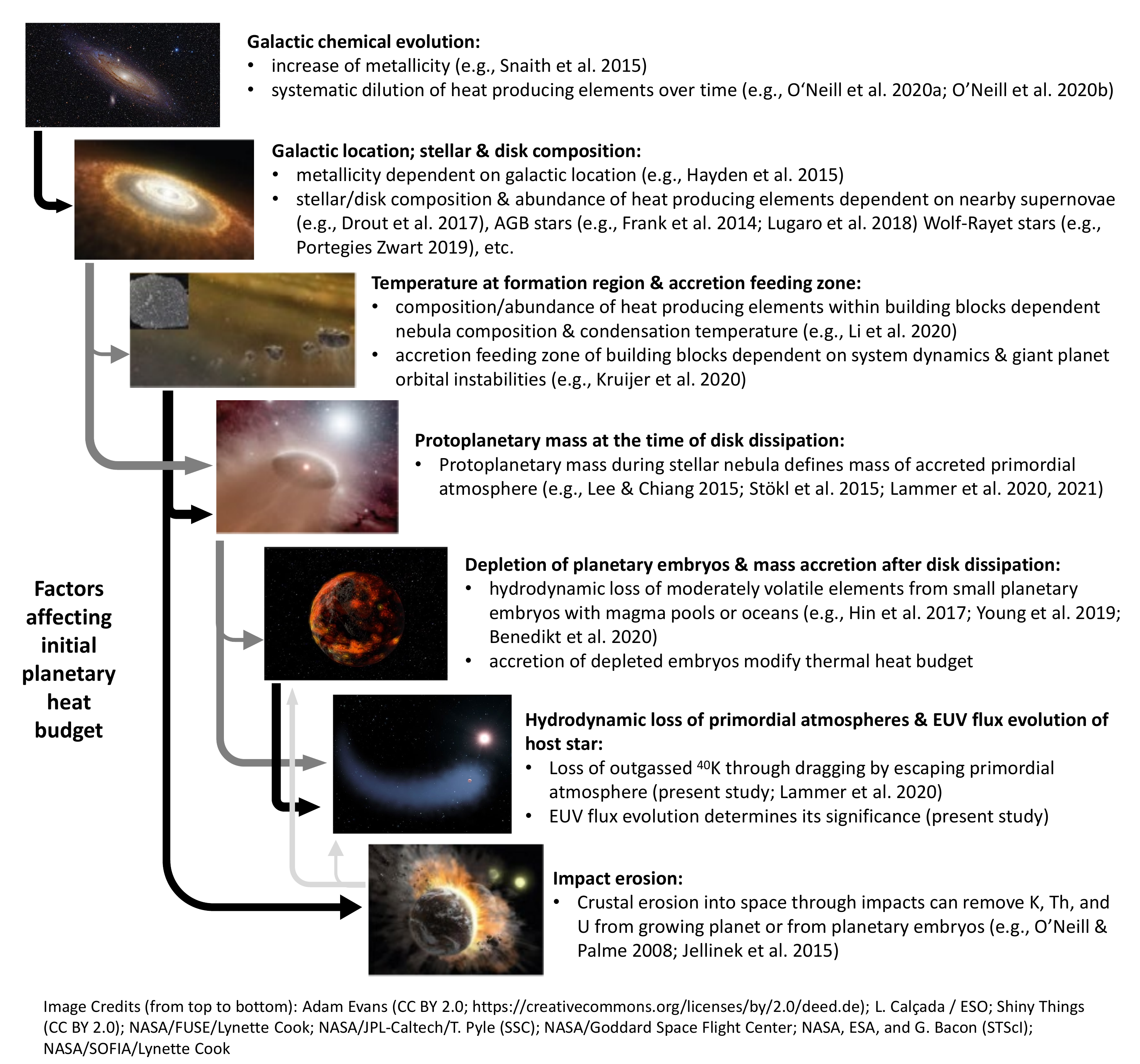}
\caption{Illustration of different factors that affect the final heat production budget of a rocky exoplanet.}\label{fig:illustration}
 \end{center}
\end{figure*}

However, besides impactor depletion and the initial building blocks, other factors might additionally affect the final heat budget of a rocky exoplanet, as depicted and described within the illustrative Fig.~\ref{fig:illustration}. We describe all of them below:

\begin{description}
  \item[{Galactic chemical evolution}] Over time, metallicity increases within the galaxy \citep[e.g.,][]{Snaith2015}. Furthermore, heat producing elements dilute over galactic time \citep[e.g.,][]{ONeill2020,ONeill2020b}. So, while there are on average more metals to form planets, the relative abundance of heat producing elements decreases. One can, therefore, expect that planets that formed early during galactic evolution might on average have ended up with a bigger heat budget than planets that formed more recently.
  \item[{Galactic location \& stellar composition}] Besides the evolution in metallicity, the abundance of metals is also dependent on the location within the galactic disk \citep[e.g.,][]{Hayden2015}. Its abundance within a certain location of the galaxy is further dependent on its immediate environment. That is, the initial stellar composition, including the composition of the stellar nebula, depends on the frequency of nearby supernovae that produce such elements \citep[e.g.,][]{Drout2017}. Similarly, nearby AGB \citep[e.g.,][]{Frank2014,Lugaro2018}, or Wolf-rayet stars \citep[e.g.,][]{PortegiesZwart2019} can modify the final heat budget of a planet.
  \item[{Temperature at formation region \& feeding zone}] The abundance of heat producing elements, particularly of the moderately volatile isotope $^{40}$K within the initial building blocks of a protoplanet is certainly also dependent on the temperature within the disk where a planet forms, and/or from where its building blocks came from \citep[see, e.g.,][]{Sossi2022}. If the respective temperature is close or above the condensation temperature $T_{\rm cond}$ of $^{40}$K, less potassium will {condense} onto the respective building blocks than for temperatures clearly below $T_{\rm cond}$. Moreover, the feeding zone for the accretion of a planet also depends on the system's dynamics such as potential giant planet instabilities that can lead to the scattering of K-rich building blocks from the outer to the inner stellar system, as it was the case for the early solar system \citep[e.g.,][]{Kruijer2020}.
  \item[{Protoplanetary mass at the time of disk dissipation}] The mass of the protoplanet at the end of the disk lifetime obviously affects the amount of hydrogen that can be accreted from the nebula \citep[e.g.,][]{Lee2015,Stoekl2015,Lammer2020b,Lammer2021}. As was shown within this study, different planetary masses at the time of disk dissipation might result in significant differences in $^{40}$K abundances.
  \item[{Depletion of planetary embryos \& mass accretion}] Moderately volatile elements can easily and substantially escape hydrodynamically from small planetary embryos through outgassing from magma pools or oceans -- even {without} the need of hydrogen dragging  \citep{Hin2017,Young2019,Benedikt2020}. This is the reason for the impactor depletion assumed within this study. If such depleted embryos are accreted onto the protoplanet, they are certainly modifying its heat budget.
  \item[{Primordial atmosphere loss \& EUV flux evolution}] As this study shows, $^{40}$K that will be outgassed from an underlying magma ocean into the primordial atmosphere can be lost together with the hydrodynamically escaping primordial atmosphere, thereby leading to significantly different planetary heat budgets. This process is also highly dependent on the evolution of the host star's EUV flux \citep[see also,][]{Lammer2020b}.
  \item[{Collisional erosion}] Impactors can erode the primordial crust of protoplanets by removing part of it into space. This process can, therefore, remove incompatible elements such as K, U, and Th from a planet, which would in turn alter its heat budget \citep[][]{ONeill2008,Jellinek2015}. It was already suggested that this processes guided the different evolutionary paths of the Earth and Venus; while collisional erosion removed part of the Earth's crust, reducing its heat budget, Venus did not experience this process as significantly as Earth did, thereby ending up with higher abundances of the radioactive isotopes \citep{Jellinek2015}.
\end{description}

\subsection{Implications for tectonics and habitability}\label{sec:implications}

Variable initial heat production can lead to large changes in tectonic history. For instance, in the case of Venus, if early collisional erosion affected Venus comparatively less than Earth, Venus's enhanced heat production may have resulted in it starting in a stagnant lid regime.  From this hot early state, simulations suggest a divergent tectonic evolution compared to Earth. The evolutionary scenarios of \citet{ONeill2013} demonstrate that an early hot stagnant Venus may have evolved into a current episodic-overturn state – highlighting the criticality of the initial temperatures and heat budget to a planet's long-term evolution.

The plausible variability of these initial conditions over the galaxy -- exemplified in Fig.~\ref{fig:Earth-vs-Cases} for different stellar rotation rates, magma ocean solubilities, and late impactor composition -- is large. A significant number of outcomes lay above Earth's anticipated early heat production. At face value, this suggests that hot stagnant or volcanic heat-pipe dominated planets may be common. However, the compositional dependence is strong; for low-solubility magma oceans ($f_{\rm K} \geq 0.5$), with high K-loss from the atmosphere, very cool planets are possible. At this end of the spectrum, early plate tectonics becomes plausible.

It has been suggested that plate tectonics may be a prerequisite for long-term habitability, and this predicate has influenced exosolar planet debate \citep{ONeill2020b}. However, Earth probably has evolved its tectonic state over its lifetime, maintaining its surface habitability through complex geological feedbacks \citep{ONeill2007}, including the association of surface tectonics and the geodynamo \citep{ONeill2020b}. These feedback systems are highly non-linear, and variable tectonic evolutionary paths --expected from the diversity of early heat production modelled here -- raise the possibility of a spectrum of habitability states in the galactic exoplanet catalogue.

\section{Conclusion}\label{sec:conclusion}
The radioactive heat producing isotope $^{40}$K abundance in accumulated H$_2$-dominated primordial atmospheres of growing protoplanets with initial masses between 0.55-1.0$M_{\rm Earth}$ at the time when the gas disk evaporated were estimated by applying equilibrium condensation models with GG$_{\rm CHEM}$, an equilibrium geochemistry code. In case that a planet accreted an H$_2$-dominant primordial atmosphere, the moderately volatile element $^{40}$K can be outgassed from the underlying hot magma ocean and, due to the lack of the formation of K-bearing condensates under such conditions, one can expect that $^{40}$K isotopes can populate primordial atmospheres to a great extend. By applying a multispecies hydrodynamic upper atmosphere evolution model, we found that $^{40}$K can be lost together with the escaping primordial atmosphere if the elements will indeed be embedded within the H atom flow. Our results, therefore, indicate that thermal escape of primordial atmospheres that drag away the embedded $^{40}$K isotopes can modify the heat budget of terrestrial planets. {However, this process cannot not explain the whole depletion of moderately volatile elements at the final planets and other processes have to be taken into account in additions such as the loss of $^{40}$K directly from the precursor material such as smaller planetary embryos with masses below $\sim$1\,M$_{\rm Mars}$ through steam atmospheres or through direct outgassing/vaporization from planetesimals/embryos with a mass smaller than the moon \citep[e.g.,][]{Hin2017,Sossi2019,Young2019,Benedikt2020}.} Our results nevertheless indicates that the final heat budget of terrestrial planets, depending on accretion history, and the EUV flux evolution of its host star, will be very diverse. This will lead to various thermal and tectonic pathways for terrestrial planets, which has a crucial impact on their habitability conditions.

\section*{Acknowledgements}
NVE acknowledges support from the Russian Science Foundation project 18-12-00080 in the frame of hydrodynamic simulation of the planetary upper atmosphere.
OH acknowledges the PhD stipend form the University of St Andrews' Centre for Exoplanet Science. ML and PO acknowledge the Austrian Science Fund (FWF): P30949-N36. We thank Bruce Fegley from the Department of Earth and Planetary Sciences, Washington University in St. Louis for discussions on magma ocean outgassed potassium in high pressure steam atmospheres. {Finally, we acknowledge Paolo Sossi whose referee comments significantly improved our article.}
\section*{DATA AVAILABILITY}
The data underlying this article will be shared on reasonable request to the corresponding author.



\bibliographystyle{mnras}
\typeout{}
\bibliography{references} 

\begin{thebibliography}{}
\makeatletter
\relax
\def\mn@urlcharsother{\let\do\@makeother \do\$\do\&\do\#\do\^\do\_\do\%\do\~}
\def\mn@doi{\begingroup\mn@urlcharsother \@ifnextchar [ {\mn@doi@}
  {\mn@doi@[]}}
\def\mn@doi@[#1]#2{\def\@tempa{#1}\ifx\@tempa\@empty \href
  {http://dx.doi.org/#2} {doi:#2}\else \href {http://dx.doi.org/#2} {#1}\fi
  \endgroup}
\def\mn@eprint#1#2{\mn@eprint@#1:#2::\@nil}
\def\mn@eprint@arXiv#1{\href {http://arxiv.org/abs/#1} {{\tt arXiv:#1}}}
\def\mn@eprint@dblp#1{\href {http://dblp.uni-trier.de/rec/bibtex/#1.xml}
  {dblp:#1}}

\bibitem[\protect\citeauthoryear{{Abdrakhimov} \& {Basilevsky}}{{Abdrakhimov}
  \& {Basilevsky}}{2002}]{Abdrakhimov2002}
{Abdrakhimov} A.~M.,  {Basilevsky} A.~T.,  2002, Solar System Research, \href
  {https://ui.adsabs.harvard.edu/abs/2002SoSyR..36..136A} {36, 136}

\bibitem[\protect\citeauthoryear{{Arevalo} \& {McDonough}}{{Arevalo} \&
  {McDonough}}{2010}]{Arevalo2010}
{Arevalo} Ricardo J.,  {McDonough} W.~F.,  2010, \mn@doi [Chemical Geology]
  {10.1016/j.chemgeo.2009.12.013}, \href
  {https://ui.adsabs.harvard.edu/abs/2010ChGeo.271...70A} {271, 70}

\bibitem[\protect\citeauthoryear{{Asplund}, {Grevesse}, {Sauval}  \&
  {Scott}}{{Asplund} et~al.}{2009}]{Asplund2009}
{Asplund} M.,  {Grevesse} N.,  {Sauval} A.~J.,   {Scott} P.,  2009, \mn@doi
  [\araa] {10.1146/annurev.astro.46.060407.145222}, \href
  {https://ui.adsabs.harvard.edu/abs/2009ARA&A..47..481A} {47, 481}

\bibitem[\protect\citeauthoryear{{Basilevsky}}{{Basilevsky}}{1997}]{Basilevsky1997a}
{Basilevsky} A.~T.,  1997, \mn@doi [\jgr] {10.1029/97JE00413}, \href
  {https://ui.adsabs.harvard.edu/abs/1997JGR...102.9257B} {102, 9257}

\bibitem[\protect\citeauthoryear{{Benedikt}, {Scherf}, {Lammer}, {Marcq},
  {Odert}, {Leitzinger}  \& {Erkaev}}{{Benedikt} et~al.}{2020}]{Benedikt2020}
{Benedikt} M.~R.,  {Scherf} M.,  {Lammer} H.,  {Marcq} E.,  {Odert} P.,
  {Leitzinger} M.,   {Erkaev} N.~V.,  2020, \mn@doi [\icarus]
  {10.1016/j.icarus.2020.113772}, \href
  {https://ui.adsabs.harvard.edu/abs/2020Icar..34713772B} {347, 113772}

\bibitem[\protect\citeauthoryear{Birnstiel, Fang  \& Johansen}{Birnstiel
  et~al.}{2016}]{Birnstiel2016}
Birnstiel T.,  Fang M.,   Johansen A.,  2016, \mn@doi [Space Science Reviews]
  {10.1007/s11214-016-0256-1}, 205, 41

\bibitem[\protect\citeauthoryear{{Bizzarro}, {Ulfbeck}, {Trinquier}, {Thrane},
  {Connelly}  \& {Meyer}}{{Bizzarro} et~al.}{2007}]{Bizzarro2007}
{Bizzarro} M.,  {Ulfbeck} D.,  {Trinquier} A.,  {Thrane} K.,  {Connelly} J.~N.,
    {Meyer} B.~S.,  2007, \mn@doi [Science] {10.1126/science.1141040}, \href
  {https://ui.adsabs.harvard.edu/abs/2007Sci...316.1178B} {316, 1178}

\bibitem[\protect\citeauthoryear{{Bouhifd} \& {Jephcoat}}{{Bouhifd} \&
  {Jephcoat}}{2011}]{Bouhifd2011}
{Bouhifd} M.~A.,  {Jephcoat} A.~P.,  2011, \mn@doi [Earth and Planetary Science
  Letters] {10.1016/j.epsl.2011.05.006}, \href
  {https://ui.adsabs.harvard.edu/abs/2011E&PSL.307..341B} {307, 341}

\bibitem[\protect\citeauthoryear{{Bower}, {Hakim}, {Sossi}  \& {Sanan}}{{Bower}
  et~al.}{2022}]{Bower2022}
{Bower} D.~J.,  {Hakim} K.,  {Sossi} P.~A.,   {Sanan} P.,  2022, \mn@doi [\psj]
  {10.3847/PSJ/ac5fb1}, \href
  {https://ui.adsabs.harvard.edu/abs/2022PSJ.....3...93B} {3, 93}

\bibitem[\protect\citeauthoryear{{Brasser}}{{Brasser}}{2013}]{Brasser2013}
{Brasser} R.,  2013, \mn@doi [\ssr] {10.1007/s11214-012-9904-2}, \href
  {https://ui.adsabs.harvard.edu/abs/2013SSRv..174...11B} {174, 11}

\bibitem[\protect\citeauthoryear{{Brown}}{{Brown}}{2001}]{Brown2001}
{Brown} T.~M.,  2001, \mn@doi [\apj] {10.1086/320950}, \href
  {https://ui.adsabs.harvard.edu/abs/2001ApJ...553.1006B} {553, 1006}

\bibitem[\protect\citeauthoryear{{Claire}, {Sheets}, {Cohen}, {Ribas},
  {Meadows}  \& {Catling}}{{Claire} et~al.}{2012}]{Claire2012}
{Claire} M.~W.,  {Sheets} J.,  {Cohen} M.,  {Ribas} I.,  {Meadows} V.~S.,
  {Catling} D.~C.,  2012, \mn@doi [\apj] {10.1088/0004-637X/757/1/95}, \href
  {https://ui.adsabs.harvard.edu/abs/2012ApJ...757...95C} {757, 95}

\bibitem[\protect\citeauthoryear{{Dauphas}}{{Dauphas}}{2017}]{Dauphas2017}
{Dauphas} N.,  2017, \mn@doi [\nat] {10.1038/nature20830}, \href
  {https://ui.adsabs.harvard.edu/abs/2017Natur.541..521D} {541, 521}

\bibitem[\protect\citeauthoryear{{Dorfi}, {Pikall}, {St{\"o}kl}  \&
  {Gautschy}}{{Dorfi} et~al.}{2006}]{Dorfi2006}
{Dorfi} E.~A.,  {Pikall} H.,  {St{\"o}kl} A.,   {Gautschy} A.,  2006, \mn@doi
  [Computer Physics Communications] {10.1016/j.cpc.2005.12.012}, \href
  {https://ui.adsabs.harvard.edu/abs/2006CoPhC.174..771D} {174, 771}

\bibitem[\protect\citeauthoryear{{Drout} et~al.,}{{Drout}
  et~al.}{2017}]{Drout2017}
{Drout} M.~R.,  et~al., 2017, \mn@doi [Science] {10.1126/science.aaq0049},
  \href {https://ui.adsabs.harvard.edu/abs/2017Sci...358.1570D} {358, 1570}

\bibitem[\protect\citeauthoryear{{Erkaev} et~al.,}{{Erkaev}
  et~al.}{2013}]{Erkaev2013}
{Erkaev} N.~V.,  et~al., 2013, \mn@doi [Astrobiology] {10.1089/ast.2012.0957},
  \href {https://ui.adsabs.harvard.edu/abs/2013AsBio..13.1011E} {13, 1011}

\bibitem[\protect\citeauthoryear{{Erkaev} et~al.,}{{Erkaev}
  et~al.}{2014}]{Erkaev2014}
{Erkaev} N.~V.,  et~al., 2014, \mn@doi [\planss] {10.1016/j.pss.2013.09.008},
  \href {https://ui.adsabs.harvard.edu/abs/2014P&SS...98..106E} {98, 106}

\bibitem[\protect\citeauthoryear{{Erkaev}, {Lammer}, {Odert}, {Kulikov}  \&
  {Kislyakova}}{{Erkaev} et~al.}{2015}]{Erkaev2015}
{Erkaev} N.~V.,  {Lammer} H.,  {Odert} P.,  {Kulikov} Y.~N.,   {Kislyakova}
  K.~G.,  2015, \mn@doi [\mnras] {10.1093/mnras/stv130}, \href
  {https://ui.adsabs.harvard.edu/abs/2015MNRAS.448.1916E} {448, 1916}

\bibitem[\protect\citeauthoryear{{Fegley}, {Jacobson}, {Williams}, {Plane},
  {Schaefer}  \& {Lodders}}{{Fegley} et~al.}{2016}]{Fegley2016}
{Fegley} Jr. B.,  {Jacobson} N.~S.,  {Williams} K.~B.,  {Plane} J.~M.~C.,
  {Schaefer} L.,   {Lodders} K.,  2016, \mn@doi [\apj]
  {10.3847/0004-637X/824/2/103}, \href
  {http://adsabs.harvard.edu/abs/2016ApJ...824..103F} {824, 103}

\bibitem[\protect\citeauthoryear{{Fegley}, {Lodders}  \& {Jacobson}}{{Fegley}
  et~al.}{2020}]{Fegley2020}
{Fegley} Bruce J.,  {Lodders} K.,   {Jacobson} N.~S.,  2020, \mn@doi [Chemie
  der Erde / Geochemistry] {10.1016/j.chemer.2019.125594}, \href
  {https://ui.adsabs.harvard.edu/abs/2020ChEG...80l5594F} {80, 125594}

\bibitem[\protect\citeauthoryear{{Fossati} et~al.,}{{Fossati}
  et~al.}{2017}]{Fossati2017}
{Fossati} L.,  et~al., 2017, \mn@doi [\aap] {10.1051/0004-6361/201629716},
  \href {https://ui.adsabs.harvard.edu/abs/2017A&A...598A..90F} {598, A90}

\bibitem[\protect\citeauthoryear{{Frank}, {Meyer}  \& {Mojzsis}}{{Frank}
  et~al.}{2014}]{Frank2014}
{Frank} E.~A.,  {Meyer} B.~S.,   {Mojzsis} S.~J.,  2014, \mn@doi [\icarus]
  {10.1016/j.icarus.2014.08.031}, \href
  {https://ui.adsabs.harvard.edu/abs/2014Icar..243..274F} {243, 274}

\bibitem[\protect\citeauthoryear{{Gaillard} et~al.,}{{Gaillard}
  et~al.}{2022}]{Gaillard2022}
{Gaillard} F.,  et~al., 2022, \mn@doi [Earth and Planetary Science Letters]
  {10.1016/j.epsl.2021.117255}, \href
  {https://ui.adsabs.harvard.edu/abs/2022E&PSL.57717255G} {577, 117255}

\bibitem[\protect\citeauthoryear{{G{\"u}del}}{{G{\"u}del}}{2007}]{Guedel2007}
{G{\"u}del} M.,  2007, \mn@doi [Living Reviews in Solar Physics]
  {10.12942/lrsp-2007-3}, \href
  {https://ui.adsabs.harvard.edu/abs/2007LRSP....4....3G} {4, 3}

\bibitem[\protect\citeauthoryear{{Harper} \& {Jacobsen}}{{Harper} \&
  {Jacobsen}}{1996a}]{Harper1996a}
{Harper} C.~L.,  {Jacobsen} S.~B.,  1996a, \mn@doi [\gca]
  {10.1016/0016-7037(96)00027-0}, \href
  {https://ui.adsabs.harvard.edu/abs/1996GeCoA..60.1131H} {60, 1131}

\bibitem[\protect\citeauthoryear{{Harper} \& {Jacobsen}}{{Harper} \&
  {Jacobsen}}{1996b}]{Harper1996b}
{Harper} Charles~L. J.,  {Jacobsen} S.~B.,  1996b, \mn@doi [Science]
  {10.1126/science.273.5283.1814}, \href
  {https://ui.adsabs.harvard.edu/abs/1996Sci...273.1814H} {273, 1814}

\bibitem[\protect\citeauthoryear{{Hayden} et~al.,}{{Hayden}
  et~al.}{2015}]{Hayden2015}
{Hayden} M.~R.,  et~al., 2015, \mn@doi [\apj] {10.1088/0004-637X/808/2/132},
  \href {https://ui.adsabs.harvard.edu/abs/2015ApJ...808..132H} {808, 132}

\bibitem[\protect\citeauthoryear{{Herbort}, {Woitke}, {Helling}  \&
  {Zerkle}}{{Herbort} et~al.}{2020}]{Herbort2020}
{Herbort} O.,  {Woitke} P.,  {Helling} C.,   {Zerkle} A.,  2020, \mn@doi [\aap]
  {10.1051/0004-6361/201936614}, \href
  {https://ui.adsabs.harvard.edu/abs/2020A&A...636A..71H} {636, A71}

\bibitem[\protect\citeauthoryear{{Herbort}, {Woitke}, {Helling}  \&
  {Zerkle}}{{Herbort} et~al.}{2022}]{Herbort2021}
{Herbort} O.,  {Woitke} P.,  {Helling} C.,   {Zerkle} A.~L.,  2022, \mn@doi
  [\aap] {10.1051/0004-6361/202141636}, \href
  {https://ui.adsabs.harvard.edu/abs/2022A&A...658A.180H} {658, A180}

\bibitem[\protect\citeauthoryear{{Hin} et~al.,}{{Hin} et~al.}{2017}]{Hin2017}
{Hin} R.~C.,  et~al., 2017, \mn@doi [\nat] {10.1038/nature23899}, \href
  {https://ui.adsabs.harvard.edu/abs/2017Natur.549..511H} {549, 511}

\bibitem[\protect\citeauthoryear{{Ikoma} \& {Genda}}{{Ikoma} \&
  {Genda}}{2006}]{Ikoma2006}
{Ikoma} M.,  {Genda} H.,  2006, \mn@doi [\apj] {10.1086/505780}, \href
  {https://ui.adsabs.harvard.edu/abs/2006ApJ...648..696I} {648, 696}

\bibitem[\protect\citeauthoryear{{Jacobsen}, {Ranen}, {Petaev}, {Remo},
  {O'Connell}  \& {Sasselov}}{{Jacobsen} et~al.}{2008}]{Jacobsen2008}
{Jacobsen} S.~B.,  {Ranen} M.~C.,  {Petaev} M.~I.,  {Remo} J.~L.,  {O'Connell}
  R.~J.,   {Sasselov} D.~D.,  2008, \mn@doi [Philosophical Transactions of the
  Royal Society of London Series A] {10.1098/rsta.2008.0174}, \href
  {https://ui.adsabs.harvard.edu/abs/2008RSPTA.366.4129J} {366, 4129}

\bibitem[\protect\citeauthoryear{{Jellinek} \& {Jackson}}{{Jellinek} \&
  {Jackson}}{2015}]{Jellinek2015}
{Jellinek} A.~M.,  {Jackson} M.~G.,  2015, \mn@doi [Nature Geoscience]
  {10.1038/ngeo2488}, \href
  {https://ui.adsabs.harvard.edu/abs/2015NatGe...8..587J} {8, 587}

\bibitem[\protect\citeauthoryear{{Johansen}, {Mac Low}, {Lacerda}  \&
  {Bizzarro}}{{Johansen} et~al.}{2015}]{Johansen2015}
{Johansen} A.,  {Mac Low} M.-M.,  {Lacerda} P.,   {Bizzarro} M.,  2015, \mn@doi
  [Science Advances] {10.1126/sciadv.1500109}, \href
  {https://ui.adsabs.harvard.edu/abs/2015SciA....1E0109J} {1, 1500109}

\bibitem[\protect\citeauthoryear{{Johansen}, {Ronnet}, {Bizzarro}, {Schiller},
  {Lambrechts}, {Nordlund}  \& {Lammer}}{{Johansen}
  et~al.}{2021}]{Johansen2021}
{Johansen} A.,  {Ronnet} T.,  {Bizzarro} M.,  {Schiller} M.,  {Lambrechts} M.,
  {Nordlund} {\r{A}}.,   {Lammer} H.,  2021, \mn@doi [Science Advances]
  {10.1126/sciadv.abc0444}, \href
  {https://ui.adsabs.harvard.edu/abs/2021SciA....7..444J} {7, eabc0444}

\bibitem[\protect\citeauthoryear{{Johnson} \& {Goldblatt}}{{Johnson} \&
  {Goldblatt}}{2018}]{Johnson2018}
{Johnson} B.~W.,  {Goldblatt} C.,  2018, \mn@doi [Geochemistry, Geophysics,
  Geosystems] {10.1029/2017GC007392}, \href
  {https://ui.adsabs.harvard.edu/abs/2018GGG....19.2516J} {19, 2516}

\bibitem[\protect\citeauthoryear{{Johnstone}, {Zhilkin}, {Pilat-Lohinger},
  {Bisikalo}, {G{\"u}del}  \& {Eggl}}{{Johnstone} et~al.}{2015}]{Johnstone2015}
{Johnstone} C.~P.,  {Zhilkin} A.,  {Pilat-Lohinger} E.,  {Bisikalo} D.,
  {G{\"u}del} M.,   {Eggl} S.,  2015, \mn@doi [\aap]
  {10.1051/0004-6361/201425134}, \href
  {https://ui.adsabs.harvard.edu/abs/2015A&A...577A.122J} {577, A122}

\bibitem[\protect\citeauthoryear{{Johnstone}, {Lammer}, {Kislyakova}, {Scherf}
  \& {G{\"u}del}}{{Johnstone} et~al.}{2021}]{Johnstone2021}
{Johnstone} C.~P.,  {Lammer} H.,  {Kislyakova} K.~G.,  {Scherf} M.,
  {G{\"u}del} M.,  2021, \mn@doi [Earth and Planetary Science Letters]
  {10.1016/j.epsl.2021.117197}, \href
  {https://ui.adsabs.harvard.edu/abs/2021E&PSL.57617197J} {576, 117197}

\bibitem[\protect\citeauthoryear{{Kasting} \& {Pollack}}{{Kasting} \&
  {Pollack}}{1983}]{Kasting1983}
{Kasting} J.~F.,  {Pollack} J.~B.,  1983, \mn@doi [\icarus]
  {10.1016/0019-1035(83)90212-9}, \href
  {https://ui.adsabs.harvard.edu/abs/1983Icar...53..479K} {53, 479}

\bibitem[\protect\citeauthoryear{{Khan}, {Sossi}, {Liebske}, {Rivoldini}  \&
  {Giardini}}{{Khan} et~al.}{2022}]{Khan2022}
{Khan} A.,  {Sossi} P.~A.,  {Liebske} C.,  {Rivoldini} A.,   {Giardini} D.,
  2022, \mn@doi [Earth and Planetary Science Letters]
  {10.1016/j.epsl.2021.117330}, \href
  {https://ui.adsabs.harvard.edu/abs/2022E&PSL.57817330K} {578, 117330}

\bibitem[\protect\citeauthoryear{{Kruijer}, {Kleine}  \& {Borg}}{{Kruijer}
  et~al.}{2020}]{Kruijer2020}
{Kruijer} T.~S.,  {Kleine} T.,   {Borg} L.~E.,  2020, \mn@doi [Nature
  Astronomy] {10.1038/s41550-019-0959-9}, \href
  {https://ui.adsabs.harvard.edu/abs/2020NatAs...4...32K} {4, 32}

\bibitem[\protect\citeauthoryear{{Kubyshkina}, {Lendl}, {Fossati}, {Cubillos},
  {Lammer}, {Erkaev}  \& {Johnstone}}{{Kubyshkina}
  et~al.}{2018a}]{Kubyshkina2018b}
{Kubyshkina} D.,  {Lendl} M.,  {Fossati} L.,  {Cubillos} P.~E.,  {Lammer} H.,
  {Erkaev} N.~V.,   {Johnstone} C.~P.,  2018a, \mn@doi [\aap]
  {10.1051/0004-6361/201731816}, \href
  {https://ui.adsabs.harvard.edu/abs/2018A&A...612A..25K} {612, A25}

\bibitem[\protect\citeauthoryear{{Kubyshkina} et~al.,}{{Kubyshkina}
  et~al.}{2018b}]{Kubyshkina2018a}
{Kubyshkina} D.,  et~al., 2018b, \mn@doi [\apjl] {10.3847/2041-8213/aae586},
  \href {https://ui.adsabs.harvard.edu/abs/2018ApJ...866L..18K} {866, L18}

\bibitem[\protect\citeauthoryear{{Labrosse}, {Hernlund}  \&
  {Coltice}}{{Labrosse} et~al.}{2007}]{Labrosse2007}
{Labrosse} S.,  {Hernlund} J.~W.,   {Coltice} N.,  2007, \mn@doi [\nat]
  {10.1038/nature06355}, \href
  {https://ui.adsabs.harvard.edu/abs/2007Natur.450..866L} {450, 866}

\bibitem[\protect\citeauthoryear{{Lambrechts} \& {Johansen}}{{Lambrechts} \&
  {Johansen}}{2012}]{lambrechts2012}
{Lambrechts} M.,  {Johansen} A.,  2012, \mn@doi [\aap]
  {10.1051/0004-6361/201219127}, \href
  {https://ui.adsabs.harvard.edu/abs/2012A&A...544A..32L} {544, A32}

\bibitem[\protect\citeauthoryear{{Lammer}, {Erkaev}, {Odert}, {Kislyakova},
  {Leitzinger}  \& {Khodachenko}}{{Lammer} et~al.}{2013}]{Lammer2013}
{Lammer} H.,  {Erkaev} N.~V.,  {Odert} P.,  {Kislyakova} K.~G.,  {Leitzinger}
  M.,   {Khodachenko} M.~L.,  2013, \mn@doi [\mnras] {10.1093/mnras/sts705},
  \href {https://ui.adsabs.harvard.edu/abs/2013MNRAS.430.1247L} {430, 1247}

\bibitem[\protect\citeauthoryear{{Lammer} et~al.,}{{Lammer}
  et~al.}{2014}]{Lammer2014}
{Lammer} H.,  et~al., 2014, \mn@doi [\mnras] {10.1093/mnras/stu085}, \href
  {https://ui.adsabs.harvard.edu/abs/2014MNRAS.439.3225L} {439, 3225}

\bibitem[\protect\citeauthoryear{{Lammer} et~al.,}{{Lammer}
  et~al.}{2016}]{Lammer2016}
{Lammer} H.,  et~al., 2016, \mn@doi [\mnras] {10.1093/mnrasl/slw095}, \href
  {https://ui.adsabs.harvard.edu/abs/2016MNRAS.461L..62L} {461, L62}

\bibitem[\protect\citeauthoryear{{Lammer} et~al.,}{{Lammer}
  et~al.}{2018}]{Lammer2018}
{Lammer} H.,  et~al., 2018, \mn@doi [\aapr] {10.1007/s00159-018-0108-y}, \href
  {https://ui.adsabs.harvard.edu/abs/2018A&ARv..26....2L} {26, 2}

\bibitem[\protect\citeauthoryear{{Lammer} et~al.,}{{Lammer}
  et~al.}{2020a}]{Lammer2020a}
{Lammer} H.,  et~al., 2020a, \mn@doi [\ssr] {10.1007/s11214-020-00701-x}, \href
  {https://ui.adsabs.harvard.edu/abs/2020SSRv..216...74L} {216, 74}

\bibitem[\protect\citeauthoryear{{Lammer} et~al.,}{{Lammer}
  et~al.}{2020b}]{Lammer2020b}
{Lammer} H.,  et~al., 2020b, \mn@doi [\icarus] {10.1016/j.icarus.2019.113551},
  \href {https://ui.adsabs.harvard.edu/abs/2020Icar..33913551L} {339, 113551}

\bibitem[\protect\citeauthoryear{{Lammer}, {Brasser}, {Johansen}, {Scherf}  \&
  {Leitzinger}}{{Lammer} et~al.}{2021}]{Lammer2021}
{Lammer} H.,  {Brasser} R.,  {Johansen} A.,  {Scherf} M.,   {Leitzinger} M.,
  2021, \mn@doi [\ssr] {10.1007/s11214-020-00778-4}, \href
  {https://ui.adsabs.harvard.edu/abs/2021SSRv..217....7L} {217, 7}

\bibitem[\protect\citeauthoryear{{Lee} \& {Chiang}}{{Lee} \&
  {Chiang}}{2015}]{Lee2015}
{Lee} E.~J.,  {Chiang} E.,  2015, \mn@doi [\apj] {10.1088/0004-637X/811/1/41},
  \href {https://ui.adsabs.harvard.edu/abs/2015ApJ...811...41L} {811, 41}

\bibitem[\protect\citeauthoryear{{Lichtenegger} et~al.,}{{Lichtenegger}
  et~al.}{2016}]{Lichtenegger2016}
{Lichtenegger} H.~I.~M.,  et~al., 2016, \mn@doi [Journal of Geophysical
  Research (Space Physics)] {10.1002/2015JA022226}, \href
  {http://adsabs.harvard.edu/abs/2016JGRA..121.4718L} {121, 4718}

\bibitem[\protect\citeauthoryear{{Lodders}, {Palme}  \& {Gail}}{{Lodders}
  et~al.}{2009}]{Lodders2009}
{Lodders} K.,  {Palme} H.,   {Gail} H.~P.,  2009, \mn@doi [Landolt
  B\&ouml;rnstein] {10.1007/978-3-540-88055-4\_34}, \href
  {https://ui.adsabs.harvard.edu/abs/2009LanB...4B..712L} {4B, 712}

\bibitem[\protect\citeauthoryear{{Lopez} \& {Fortney}}{{Lopez} \&
  {Fortney}}{2014}]{Lopez2014}
{Lopez} E.~D.,  {Fortney} J.~J.,  2014, \mn@doi [\apj]
  {10.1088/0004-637X/792/1/1}, \href
  {https://ui.adsabs.harvard.edu/abs/2014ApJ...792....1L} {792, 1}

\bibitem[\protect\citeauthoryear{{Lugaro}, {Ott}  \& {Kereszturi}}{{Lugaro}
  et~al.}{2018}]{Lugaro2018}
{Lugaro} M.,  {Ott} U.,   {Kereszturi} {\'A}.,  2018, \mn@doi [Progress in
  Particle and Nuclear Physics] {10.1016/j.ppnp.2018.05.002}, \href
  {https://ui.adsabs.harvard.edu/abs/2018PrPNP.102....1L} {102, 1}

\bibitem[\protect\citeauthoryear{{Makide}, {Nagashima}, {Krot}, {Huss},
  {Ciesla}, {Hellebrand}, {Gaidos}  \& {Yang}}{{Makide}
  et~al.}{2011}]{Makide2011}
{Makide} K.,  {Nagashima} K.,  {Krot} A.~N.,  {Huss} G.~R.,  {Ciesla} F.~J.,
  {Hellebrand} E.,  {Gaidos} E.,   {Yang} L.,  2011, \mn@doi [\apjl]
  {10.1088/2041-8205/733/2/L31}, \href
  {https://ui.adsabs.harvard.edu/abs/2011ApJ...733L..31M} {733, L31}

\bibitem[\protect\citeauthoryear{{Mamajek}}{{Mamajek}}{2009}]{Mamajek2009}
{Mamajek} E.~E.,  2009, in {Usuda} T.,  {Tamura} M.,   {Ishii} M.,  eds,
  American Institute of Physics Conference Series Vol. 1158, Exoplanets and
  Disks: Their Formation and Diversity. pp 3--10

\bibitem[\protect\citeauthoryear{{Marty}}{{Marty}}{2012}]{Marty2012}
{Marty} B.,  2012, \mn@doi [Earth and Planetary Science Letters]
  {10.1016/j.epsl.2011.10.040}, \href
  {http://adsabs.harvard.edu/abs/2012E%26PSL.313...56M} {313, 56}

\bibitem[\protect\citeauthoryear{{Mordasini}, {Molli{\`e}re}, {Dittkrist},
  {Jin}  \& {Alibert}}{{Mordasini} et~al.}{2015}]{Mordasini2015}
{Mordasini} C.,  {Molli{\`e}re} P.,  {Dittkrist} K.~M.,  {Jin} S.,   {Alibert}
  Y.,  2015, \mn@doi [International Journal of Astrobiology]
  {10.1017/S1473550414000263}, \href
  {https://ui.adsabs.harvard.edu/abs/2015IJAsB..14..201M} {14, 201}

\bibitem[\protect\citeauthoryear{{Murray-Clay}, {Chiang}  \&
  {Murray}}{{Murray-Clay} et~al.}{2009}]{Murray-Clay2009}
{Murray-Clay} R.~A.,  {Chiang} E.~I.,   {Murray} N.,  2009, \mn@doi [\apj]
  {10.1088/0004-637X/693/1/23}, \href
  {https://ui.adsabs.harvard.edu/abs/2009ApJ...693...23M} {693, 23}

\bibitem[\protect\citeauthoryear{{Murthy}, {van Westrenen}  \& {Fei}}{{Murthy}
  et~al.}{2003}]{Murthy2003}
{Murthy} V.~R.,  {van Westrenen} W.,   {Fei} Y.,  2003, \mn@doi [\nat]
  {10.1038/nature01560}, \href
  {https://ui.adsabs.harvard.edu/abs/2003Natur.423..163M} {423, 163}

\bibitem[\protect\citeauthoryear{{Nimmo} \& {Kleine}}{{Nimmo} \&
  {Kleine}}{2015}]{Nimmo2015}
{Nimmo} F.,  {Kleine} T.,  2015, \mn@doi [Washington DC American Geophysical
  Union Geophysical Monograph Series] {10.1002/9781118860359.ch5}, \href
  {https://ui.adsabs.harvard.edu/abs/2015GMS...212...83N} {212, 83}

\bibitem[\protect\citeauthoryear{O'Neill}{O'Neill}{2016}]{ONeill2016}
O'Neill H.~S.,  2016, Heat-Producing Elements (HPEs).
Springer International Publishing, Cham, pp~1--6,
  \mn@doi{10.1007/978-3-319-39193-9_265-1}, \url
  {https://doi.org/10.1007/978-3-319-39193-9_265-1}

\bibitem[\protect\citeauthoryear{{O'Neill} \& {Palme}}{{O'Neill} \&
  {Palme}}{2008}]{ONeill2008}
{O'Neill} H.~S.~C.,  {Palme} H.,  2008, \mn@doi [Philosophical Transactions of
  the Royal Society of London Series A] {10.1098/rsta.2008.0111}, \href
  {http://adsabs.harvard.edu/abs/2008RSPTA.366.4205O} {366, 4205}

\bibitem[\protect\citeauthoryear{O'Neill, Lenardic, Moresi, Torsvik  \&
  Lee}{O'Neill et~al.}{2007}]{ONeill2007}
O'Neill C.,  Lenardic A.,  Moresi L.,  Torsvik T.,   Lee C.,  2007, \mn@doi
  [Earth and Planetary Science Letters] {10.1016/j.epsl.2007.04.056}, 262, 552

\bibitem[\protect\citeauthoryear{{O'Neill}, {Lenardic}, {H{\"o}ink}  \&
  {Coltice}}{{O'Neill} et~al.}{2013}]{ONeill2013}
{O'Neill} C.,  {Lenardic} A.,  {H{\"o}ink} T.,   {Coltice} N.,  2013, {Mantle
  Convection and Outgassing on Terrestrial Planets}.
pp 473--486, \mn@doi{10.2458/azu\_uapress\_9780816530595-ch019}

\bibitem[\protect\citeauthoryear{{O'Neill}, {O'Neill}  \& {Jellinek}}{{O'Neill}
  et~al.}{2020a}]{ONeill2020}
{O'Neill} C.,  {O'Neill} H.~S.~C.,   {Jellinek} A.~M.,  2020a, \mn@doi [\ssr]
  {10.1007/s11214-020-00656-z}, \href
  {https://ui.adsabs.harvard.edu/abs/2020SSRv..216...37O} {216, 37}

\bibitem[\protect\citeauthoryear{{O'Neill}, {Lowman}  \& {Wasiliev}}{{O'Neill}
  et~al.}{2020b}]{ONeill2020b}
{O'Neill} C.,  {Lowman} J.,   {Wasiliev} J.,  2020b, \mn@doi [\icarus]
  {10.1016/j.icarus.2020.114025}, \href
  {https://ui.adsabs.harvard.edu/abs/2020Icar..35214025O} {352, 114025}

\bibitem[\protect\citeauthoryear{{Odert} et~al.,}{{Odert}
  et~al.}{2018}]{Odert2018}
{Odert} P.,  et~al., 2018, \mn@doi [\icarus] {10.1016/j.icarus.2017.10.031},
  \href {https://ui.adsabs.harvard.edu/abs/2018Icar..307..327O} {307, 327}

\bibitem[\protect\citeauthoryear{{Owen} \& {Wu}}{{Owen} \&
  {Wu}}{2016}]{Owen2016}
{Owen} J.~E.,  {Wu} Y.,  2016, \mn@doi [\apj] {10.3847/0004-637X/817/2/107},
  \href {https://ui.adsabs.harvard.edu/abs/2016ApJ...817..107O} {817, 107}

\bibitem[\protect\citeauthoryear{{Owen}, {Shaikhislamov}, {Lammer}, {Fossati}
  \& {Khodachenko}}{{Owen} et~al.}{2020}]{Owen2020}
{Owen} J.~E.,  {Shaikhislamov} I.~F.,  {Lammer} H.,  {Fossati} L.,
  {Khodachenko} M.~L.,  2020, \mn@doi [\ssr] {10.1007/s11214-020-00756-w},
  \href {https://ui.adsabs.harvard.edu/abs/2020SSRv..216..129O} {216, 129}

\bibitem[\protect\citeauthoryear{{Padhi}, {Korenaga}  \& {Ozima}}{{Padhi}
  et~al.}{2012}]{Padhi2012}
{Padhi} C.~M.,  {Korenaga} J.,   {Ozima} M.,  2012, \mn@doi [Earth and
  Planetary Science Letters] {10.1016/j.epsl.2012.06.013}, \href
  {https://ui.adsabs.harvard.edu/abs/2012E&PSL.341....1P} {341, 1}

\bibitem[\protect\citeauthoryear{{Portegies Zwart}}{{Portegies
  Zwart}}{2019}]{PortegiesZwart2019}
{Portegies Zwart} S.,  2019, \mn@doi [\aap] {10.1051/0004-6361/201833974},
  \href {https://ui.adsabs.harvard.edu/abs/2019A&A...622A..69P} {622, A69}

\bibitem[\protect\citeauthoryear{{Ribas}, {Guinan}, {G{\"u}del}  \&
  {Audard}}{{Ribas} et~al.}{2005}]{Ribas2005}
{Ribas} I.,  {Guinan} E.~F.,  {G{\"u}del} M.,   {Audard} M.,  2005, \mn@doi
  [\apj] {10.1086/427977}, \href
  {https://ui.adsabs.harvard.edu/abs/2005ApJ...622..680R} {622, 680}

\bibitem[\protect\citeauthoryear{{Ruedas}}{{Ruedas}}{2017}]{Ruedas2017}
{Ruedas} T.,  2017, \mn@doi [Geochemistry, Geophysics, Geosystems]
  {10.1002/2017GC006997}, \href
  {https://ui.adsabs.harvard.edu/abs/2017GGG....18.3530R} {18, 3530}

\bibitem[\protect\citeauthoryear{Sackmann \& Boothroyd}{Sackmann \&
  Boothroyd}{2003}]{Sackmann2003}
Sackmann I.-J.,  Boothroyd A.~I.,  2003, \mn@doi [The Astrophysical Journal]
  {10.1086/345408}, 583, 1024

\bibitem[\protect\citeauthoryear{{Savage}, {Moynier}  \& {Boyet}}{{Savage}
  et~al.}{2022}]{Savage2022}
{Savage} P.~S.,  {Moynier} F.,   {Boyet} M.,  2022, \mn@doi [\icarus]
  {10.1016/j.icarus.2022.115172}, \href
  {https://ui.adsabs.harvard.edu/abs/2022Icar..38615172S} {386, 115172}

\bibitem[\protect\citeauthoryear{{Schaefer} \& {Fegley}}{{Schaefer} \&
  {Fegley}}{2007}]{Schaefer2007}
{Schaefer} L.,  {Fegley} B.,  2007, \mn@doi [\icarus]
  {10.1016/j.icarus.2006.09.002}, \href
  {http://adsabs.harvard.edu/abs/2007Icar..186..462S} {186, 462}

\bibitem[\protect\citeauthoryear{{Schaefer} \& {Fegley}}{{Schaefer} \&
  {Fegley}}{2010}]{Schaefer2010}
{Schaefer} L.,  {Fegley} B.,  2010, \mn@doi [Icarus]
  {10.1016/j.icarus.2010.01.026}, \href
  {http://adsabs.harvard.edu/abs/2010Icar..208..438S} {208, 438}

\bibitem[\protect\citeauthoryear{{Schaefer}, {Lodders}  \& {Fegley}}{{Schaefer}
  et~al.}{2012}]{Schaefer2012}
{Schaefer} L.,  {Lodders} K.,   {Fegley} B.,  2012, \mn@doi [\apj]
  {10.1088/0004-637X/755/1/41}, \href
  {https://ui.adsabs.harvard.edu/abs/2012ApJ...755...41S} {755, 41}

\bibitem[\protect\citeauthoryear{{Schiller}, {Connelly}, {Glad}, {Mikouchi}  \&
  {Bizzarro}}{{Schiller} et~al.}{2015}]{Schiller2015}
{Schiller} M.,  {Connelly} J.~N.,  {Glad} A.~C.,  {Mikouchi} T.,   {Bizzarro}
  M.,  2015, \mn@doi [Earth and Planetary Science Letters]
  {10.1016/j.epsl.2015.03.028}, \href
  {https://ui.adsabs.harvard.edu/abs/2015E&PSL.420...45S} {420, 45}

\bibitem[\protect\citeauthoryear{{Schiller}, {Bizzarro}  \&
  {Fernandes}}{{Schiller} et~al.}{2018}]{Schiller2018}
{Schiller} M.,  {Bizzarro} M.,   {Fernandes} V.~A.,  2018, \mn@doi [\nat]
  {10.1038/nature25990}, \href
  {https://ui.adsabs.harvard.edu/abs/2018Natur.555..507S} {555, 507}

\bibitem[\protect\citeauthoryear{{Schlichting} \& {Mukhopadhyay}}{{Schlichting}
  \& {Mukhopadhyay}}{2018}]{Schlichting2018}
{Schlichting} H.~E.,  {Mukhopadhyay} S.,  2018, \mn@doi [\ssr]
  {10.1007/s11214-018-0471-z}, \href
  {https://ui.adsabs.harvard.edu/abs/2018SSRv..214...34S} {214, 34}

\bibitem[\protect\citeauthoryear{{Sekiya}, {Nakazawa}  \& {Hayashi}}{{Sekiya}
  et~al.}{1980}]{Sekiya1980}
{Sekiya} M.,  {Nakazawa} K.,   {Hayashi} C.,  1980, \mn@doi [Progress of
  Theoretical Physics] {10.1143/PTP.64.1968}, \href
  {https://ui.adsabs.harvard.edu/abs/1980PThPh..64.1968S} {64, 1968}

\bibitem[\protect\citeauthoryear{{Shematovich}, {Ionov}  \&
  {Lammer}}{{Shematovich} et~al.}{2014}]{Shematovich2014}
{Shematovich} V.~I.,  {Ionov} D.~E.,   {Lammer} H.,  2014, \mn@doi [\aap]
  {10.1051/0004-6361/201423573}, \href
  {https://ui.adsabs.harvard.edu/abs/2014A&A...571A..94S} {571, A94}

\bibitem[\protect\citeauthoryear{{Snaith}, {Haywood}, {Di Matteo}, {Lehnert},
  {Combes}, {Katz}  \& {G{\'o}mez}}{{Snaith} et~al.}{2015}]{Snaith2015}
{Snaith} O.,  {Haywood} M.,  {Di Matteo} P.,  {Lehnert} M.~D.,  {Combes} F.,
  {Katz} D.,   {G{\'o}mez} A.,  2015, \mn@doi [\aap]
  {10.1051/0004-6361/201424281}, \href
  {https://ui.adsabs.harvard.edu/abs/2015A&A...578A..87S} {578, A87}

\bibitem[\protect\citeauthoryear{{Solomatov} \& {Barr}}{{Solomatov} \&
  {Barr}}{2007}]{Solomatov2007}
{Solomatov} V.~S.,  {Barr} A.~C.,  2007, \mn@doi [Physics of the Earth and
  Planetary Interiors] {10.1016/j.pepi.2007.06.007}, \href
  {https://ui.adsabs.harvard.edu/abs/2007PEPI..165....1S} {165, 1}

\bibitem[\protect\citeauthoryear{{Sossi}, {Klemme}, {O'Neill}, {Berndt}  \&
  {Moynier}}{{Sossi} et~al.}{2019}]{Sossi2019}
{Sossi} P.~A.,  {Klemme} S.,  {O'Neill} H. S.~C.,  {Berndt} J.,   {Moynier} F.,
   2019, \mn@doi [\gca] {10.1016/j.gca.2019.06.021}, \href
  {https://ui.adsabs.harvard.edu/abs/2019GeCoA.260..204S} {260, 204}

\bibitem[\protect\citeauthoryear{{Sossi}, {Burnham}, {Badro}, {Lanzirotti},
  {Newville}  \& {O'Neill}}{{Sossi} et~al.}{2020}]{Sossi2020}
{Sossi} P.~A.,  {Burnham} A.~D.,  {Badro} J.,  {Lanzirotti} A.,  {Newville} M.,
    {O'Neill} H. S.~C.,  2020, \mn@doi [Science Advances]
  {10.1126/sciadv.abd1387}, \href
  {https://ui.adsabs.harvard.edu/abs/2020SciA....6.1387S} {6, eabd1387}

\bibitem[\protect\citeauthoryear{{Sossi}, {Stotz}, {Jacobson}, {Morbidelli}  \&
  {O'Neill}}{{Sossi} et~al.}{2022}]{Sossi2022}
{Sossi} P.~A.,  {Stotz} I.~L.,  {Jacobson} S.~A.,  {Morbidelli} A.,   {O'Neill}
  H. S.~C.,  2022, \mn@doi [Nature Astronomy] {10.1038/s41550-022-01702-2},
  \href {https://ui.adsabs.harvard.edu/abs/2022NatAs...6..951S} {6, 951}

\bibitem[\protect\citeauthoryear{{Stein}, {Schmalzl}  \& {Hansen}}{{Stein}
  et~al.}{2004}]{Stein2004}
{Stein} C.,  {Schmalzl} J.,   {Hansen} U.,  2004, \mn@doi [Physics of the Earth
  and Planetary Interiors] {10.1016/j.pepi.2004.01.006}, \href
  {https://ui.adsabs.harvard.edu/abs/2004PEPI..142..225S} {142, 225}

\bibitem[\protect\citeauthoryear{{Steller}, {Burkhardt}, {Yang}  \&
  {Kleine}}{{Steller} et~al.}{2022}]{Steller2022}
{Steller} T.,  {Burkhardt} C.,  {Yang} C.,   {Kleine} T.,  2022, \mn@doi
  [\icarus] {10.1016/j.icarus.2022.115171}, \href
  {https://ui.adsabs.harvard.edu/abs/2022Icar..38615171S} {386, 115171}

\bibitem[\protect\citeauthoryear{{St{\"o}kl}}{{St{\"o}kl}}{2008}]{Stoekl2008}
{St{\"o}kl} A.,  2008, \mn@doi [\aap] {10.1051/0004-6361:200810144}, \href
  {https://ui.adsabs.harvard.edu/abs/2008A&A...490.1181S} {490, 1181}

\bibitem[\protect\citeauthoryear{{St{\"o}kl} \& {Dorfi}}{{St{\"o}kl} \&
  {Dorfi}}{2007}]{Stoekl2007}
{St{\"o}kl} A.,  {Dorfi} E.~A.,  2007, \mn@doi [Computer Physics
  Communications] {10.1016/j.cpc.2007.06.012}, \href
  {https://ui.adsabs.harvard.edu/abs/2007CoPhC.177..815S} {177, 815}

\bibitem[\protect\citeauthoryear{{St{\"o}kl}, {Dorfi}  \& {Lammer}}{{St{\"o}kl}
  et~al.}{2015}]{Stoekl2015}
{St{\"o}kl} A.,  {Dorfi} E.,   {Lammer} H.,  2015, \mn@doi [\aap]
  {10.1051/0004-6361/201423638}, \href
  {https://ui.adsabs.harvard.edu/abs/2015A&A...576A..87S} {576, A87}

\bibitem[\protect\citeauthoryear{{St{\"o}kl}, {Dorfi}, {Johnstone}  \&
  {Lammer}}{{St{\"o}kl} et~al.}{2016}]{Stoekl2016}
{St{\"o}kl} A.,  {Dorfi} E.~A.,  {Johnstone} C.~P.,   {Lammer} H.,  2016,
  \mn@doi [\apj] {10.3847/0004-637X/825/2/86}, \href
  {https://ui.adsabs.harvard.edu/abs/2016ApJ...825...86S} {825, 86}

\bibitem[\protect\citeauthoryear{{Storey} \& {Hummer}}{{Storey} \&
  {Hummer}}{1995}]{Storey1995}
{Storey} P.~J.,  {Hummer} D.~G.,  1995, \mn@doi [\mnras]
  {10.1093/mnras/272.1.41}, \href
  {https://ui.adsabs.harvard.edu/abs/1995MNRAS.272...41S} {272, 41}

\bibitem[\protect\citeauthoryear{{Tu}, {Johnstone}, {G{\"u}del}  \&
  {Lammer}}{{Tu} et~al.}{2015}]{Tu2015}
{Tu} L.,  {Johnstone} C.~P.,  {G{\"u}del} M.,   {Lammer} H.,  2015, \mn@doi
  [\aap] {10.1051/0004-6361/201526146}, \href
  {https://ui.adsabs.harvard.edu/abs/2015A&A...577L...3T} {577, L3}

\bibitem[\protect\citeauthoryear{{Turcotte} \& {Schubert}}{{Turcotte} \&
  {Schubert}}{2002}]{Turcotte2002}
{Turcotte} D.~L.,  {Schubert} G.,  2002, {Geodynamics - 2nd Edition},
  \mn@doi{10.2277/0521661862.
}

\bibitem[\protect\citeauthoryear{{Wasson} \& {Kallemeyn}}{{Wasson} \&
  {Kallemeyn}}{1988}]{Wasson1988}
{Wasson} J.~T.,  {Kallemeyn} G.~W.,  1988, \mn@doi [Philosophical Transactions
  of the Royal Society of London Series A] {10.1098/rsta.1988.0066}, \href
  {https://ui.adsabs.harvard.edu/abs/1988RSPTA.325..535W} {325, 535}

\bibitem[\protect\citeauthoryear{{Watson}, {Donahue}  \& {Walker}}{{Watson}
  et~al.}{1981}]{Watson1981}
{Watson} A.~J.,  {Donahue} T.~M.,   {Walker} J.~C.~G.,  1981, \mn@doi [\icarus]
  {10.1016/0019-1035(81)90101-9}, \href
  {https://ui.adsabs.harvard.edu/abs/1981Icar...48..150W} {48, 150}

\bibitem[\protect\citeauthoryear{{Williams} \& {Mukhopadhyay}}{{Williams} \&
  {Mukhopadhyay}}{2019}]{Williams2019}
{Williams} C.~D.,  {Mukhopadhyay} S.,  2019, \mn@doi [\nat]
  {10.1038/s41586-018-0771-1}, \href
  {https://ui.adsabs.harvard.edu/abs/2019Natur.565...78W} {565, 78}

\bibitem[\protect\citeauthoryear{{Woitke}, {Helling}, {Hunter}, {Millard},
  {Turner}, {Worters}, {Blecic}  \& {Stock}}{{Woitke}
  et~al.}{2018}]{Woitke2018}
{Woitke} P.,  {Helling} C.,  {Hunter} G.~H.,  {Millard} J.~D.,  {Turner} G.~E.,
   {Worters} M.,  {Blecic} J.,   {Stock} J.~W.,  2018, \mn@doi [\aap]
  {10.1051/0004-6361/201732193}, \href
  {https://ui.adsabs.harvard.edu/abs/2018A&A...614A...1W} {614, A1}

\bibitem[\protect\citeauthoryear{{Wood}, {Smythe}  \& {Harrison}}{{Wood}
  et~al.}{2019}]{Wood2019}
{Wood} B.~J.,  {Smythe} D.~J.,   {Harrison} T.,  2019, \mn@doi [American
  Mineralogist] {10.2138/am-2019-6852CCBY}, \href
  {https://ui.adsabs.harvard.edu/abs/2019AmMin.104..844W} {104, 844}

\bibitem[\protect\citeauthoryear{{Yelle}}{{Yelle}}{2004}]{Yelle2004}
{Yelle} R.~V.,  2004, \mn@doi [\icarus] {10.1016/j.icarus.2004.02.008}, \href
  {https://ui.adsabs.harvard.edu/abs/2004Icar..170..167Y} {170, 167}

\bibitem[\protect\citeauthoryear{{Young}, {Shahar}, {Nimmo}, {Schlichting},
  {Schauble}, {Tang}  \& {Labidi}}{{Young} et~al.}{2019}]{Young2019}
{Young} E.~D.,  {Shahar} A.,  {Nimmo} F.,  {Schlichting} H.~E.,  {Schauble}
  E.~A.,  {Tang} H.,   {Labidi} J.,  2019, \mn@doi [\icarus]
  {10.1016/j.icarus.2019.01.012}, \href
  {https://ui.adsabs.harvard.edu/abs/2019Icar..323....1Y} {323, 1}

\bibitem[\protect\citeauthoryear{{Zahnle} \& {Kasting}}{{Zahnle} \&
  {Kasting}}{1986}]{Zahnle1986}
{Zahnle} K.~J.,  {Kasting} J.~F.,  1986, \mn@doi [\icarus]
  {10.1016/0019-1035(86)90051-5}, \href
  {https://ui.adsabs.harvard.edu/abs/1986Icar...68..462Z} {68, 462}

\makeatother
\end{thebibliography}




\begin{appendix}
\onecolumn

\section{Further Details on Upper Atmosphere Model}\label{App:Model}
For computational convenience we introduce normalized quantities as follows:

\begin{eqnarray}
X_{\rm K} = \rho_{\rm K}/\rho , \quad    \tilde V = V/V_{T0}, \quad   V_{T0} =\sqrt{k T_0/m_{\rm H}},  \quad  \tilde T = T/T_0,
\\ \tilde P = P/ (\rho_0 V_{T0}^2 ), \quad   \tilde \rho =\rho/\rho_0, \quad u= (\tilde V_{\rm K}- \tilde V), \\
 m = m_{\rm K}/ m_{\rm H},
\quad \tilde r = r/ R_0, \quad \tilde t = t V_{T0}/R_0 , \quad \tilde \nu = \nu_c / \nu_{c0}, \\
\varepsilon = V_{T0}/(\nu_{c0} R_{0}), \quad \lambda = G m_{\rm H} M_{\rm pl}/(r_0 k_{\rm B} T_0), \quad \Gamma = L_{\rm H}/(\rho_0 V_{T_0} R_0^2).
\end{eqnarray}
Here, subscript $0$ is related to the lower boundary.

Hereafter, we will leave out $\tilde{}$~ for simplicity. Using these normalized parameters we obtain the following dimensionless equations

\begin{eqnarray}
 \frac{d V}{d t}  + \frac{1}{\rho}\nabla P= \lambda g +
 X_K \nu\frac{u}{(m+1)}   , \label{V_norm}\\
 X_K \frac{d  (V +\varepsilon u)}{d t}  + \frac{1}{\rho}\nabla P_K= \lambda g X_K-
 X_K \nu \frac{u}{(m +1)}    .      \label{V_K_nor}
\end{eqnarray}
Here, the partial pressure is
\begin{equation}
P_{\rm K} = X_K \xi P /m, \xi = 1/(1+X_{\rm H+}- 0.5 X_{\rm H2}), \label{P_K}
\end{equation}
where $X_{\rm H+}$ = $n_{\rm H+} m_{\rm H}/\rho$, $X_{\rm H2}$ = $n_{\rm H2} m_{\rm H2}/\rho$ .

Neglecting terms $~\varepsilon$ in equation (\ref{V_K_nor})
and substituting $d V/dt$ from equation (\ref{V_norm}),
we obtain
\begin{eqnarray}
\rho X_{\rm K} u = \frac{(m +1)}{\nu(1+X_{\rm K})}
\left[ \frac{d P}{d r} X_{\rm K}  - \frac{d P_{\rm K}}{d r} \right]  .  \label{uX_k}
\end{eqnarray}
Equation (\ref{Rho_K}) can be transformed as follows:
\begin{eqnarray}
\rho r^2\frac{\partial X_{\rm K}}{\partial t} + \rho V r^2 \frac{\partial  X_{\rm K}}{\partial r} +
\frac{\partial \left(r^2 \rho \varepsilon X_{\rm K}  u \right)}{\partial r}= 0. \label{Rho_K1}
\end{eqnarray}
Substituting  (\ref{uX_k})  to the mass conservation equation (\ref{Rho_K1}), we derive finally equation in a stationary case
\begin{eqnarray}
\rho V r^2 \frac{\partial X_k}{\partial r} +
\frac{d}{dr}\left \{\varepsilon r^2\frac{(m +1)}{\nu (1+ X_{\rm K})} \left[ \frac{dP}{dr}X_{\rm K}  -\frac{d P_{\rm K}}{dr}   \right]   \right \} =0.   \label{X_eq}
\end{eqnarray}
Equation (\ref{X_eq}) contains both advection and diffusion parts. In the particular case of equilibrium this equation yields the Boltzmann distribution.
In case of stationary flow and condition $X_{\rm K}  \ll $1 equation (\ref{X_eq})
can be solved analytically. In case of a stationary radial gas flow, the mass flux is constant $\rho V r^2 $ = $\Gamma$ = $const$, and thus equation (\ref{X_eq}) can be integrated
\begin{eqnarray}
\Gamma (X_{\rm K} - X_{\rm K\infty}) + (1+m)\nu^{-1} r^2\varepsilon
\left( X_{\rm K}\frac{d P}{d r} - \frac{d P_{\rm K}}{d r} \right) = 0.  \label{X_eq2}
\end{eqnarray}
 This equation can be rearranged as
\begin{equation}
 \varepsilon\frac{d X_{\rm K}}{d r}   -X_{\rm K} \frac{(B\varepsilon +1)}{A}  + \frac{X_{\rm K\infty}}{A} =0,  \label{X_eq3}
\end{equation}
where
\begin{eqnarray}
 A=   \frac{(1+ m) r^2  \xi P}{m \Gamma \nu },
 \quad B=   \frac{(1+ m) r^2 }{ m \Gamma \nu }\frac{d[( m -\xi) P]}{d  r}.
 \end{eqnarray}
The solution of  equation  (\ref{X_eq3}) can be written in the following integral form
\begin{equation}
X_{\rm K} = X_{\rm K\infty} \int_{\psi(r)}^{\psi_\infty} {exp(\psi -\psi')\frac{1}{(1+B\varepsilon)}d \psi'}, \label{sol}
\end{equation}
where
\begin{equation}
\psi = \frac{1}{\varepsilon}\int_1^{r}{\frac{(1+B\varepsilon)}{A} d r'}.
\end{equation}

Integrating (\ref{sol}) by parts, we get the expression
\begin{equation}
X_K = X_{\rm K\infty} \frac{1}{1+B\varepsilon} \left[ 1 - (1+B\varepsilon) \varepsilon^2
\int_\psi^{\psi_\infty}{(1+\varepsilon B)^{-3}\frac{d B}{d r} A \exp(\psi-\psi')d\psi'}\right ]. \label{Xk_int}
\end{equation}


Neglecting the second order term $\propto$  $\varepsilon$ in  (\ref{Xk_int}), we derive the analytical formula
\begin{eqnarray}
X_{\rm K\infty}  = X_{\rm K0} (1+\varepsilon B) = X_{\rm K0} \left \{1 + \varepsilon \frac{(1+m)}{m \Gamma}.
 \frac{d}{dr}\left [ P(m -\xi) \right ]  \right \} ,  \label{Xk_asym1}
\end{eqnarray}
In case of constant temperature, monatomic gas and Boltzmann distribution of the pressure, the
approximate formula (\ref{Xk_asym1}) is similar to that discussed by \citet{Zahnle1986}.

For the hydrogen constituents, we apply the finite difference numerical scheme
of MacCormack to integrate the system of equations in time.

Finite difference continuity equations for the total density, ions and atomic hydrogen are similar to each other
 \begin{eqnarray}
  \hat{n}_{i}^{k+1/2} =  \hat{n}_{k i}^k - \frac{\Delta t}{r_i^2}\frac{( \hat{\Gamma}_{i+1}^k  -
 \hat{\Gamma}_{i}^k  )}{(r_{i+1}-r_i)} + \hat{S}_{i}^k, \\
  \hat{n}_i^{k+1} = \frac{1}{2} (\hat {n}_i^{k+1/2}+\hat{n}_i^{k}) - \nonumber \\
 0.5\frac{\Delta t}{r_i^2}\frac{( \hat{\Gamma}_{i}^{k+1/2}  -
 \hat{\Gamma}_{i-1}^{k+1/2} )}{(r_i -r_{i-1})} + \hat{S}_{i}^{k+1/2} ,
\end{eqnarray}
where $\hat {n}$ = $ (\rho, n_N, n_{N+}, n_{N2+})$,  $\hat \Gamma = \hat{n} V r^2$, $\hat{S} = (0, S_N, S_{N+}, S_{N2+})$ .
The momentum finite difference equations can be written as
\begin{eqnarray}
{(\rho V)}_i^{k+1/2} = (\rho V)_i^k -
\frac{\Delta t}{r_i^2}\frac{( \Pi_{i+1}^k  -
 \Pi_i^k  )}{(r_{i+1}-r_i)} -  \nonumber \\
 0.5\frac{\Delta t}{r_i^2}(r_i^2\rho_i^k +r_{i+1}^2\rho_{i+1}^k)\frac{(U_{i+1}-U_i)}{(r_{i+1}-r_i)}  -   \nonumber \\
 0.5\frac{\Delta t}{r_i^2}{(r_i^2+r_{i+1}^2)} \frac{(P_{i+1}^k -P_{i}^k)}{(r_{i+1}-r_i)}  , \\
 {(\rho V)}_i^{k+1} = 0.5  [(\rho V)_i^{k+1/2} + (\rho V)_i^{k}] -\nonumber \\
  0.5\frac{\Delta t}{r_i^2}\frac{( \Pi_{i}^{k+1/2}  -
 \Pi_{i-1}^{k+1/2} )}{(r_i-r_{i-1})} +  \\
 -0.25\frac{\Delta t}{r_i^2}(r_i^2\rho_i^{k+1/2}+r_{i-1}^2\rho_{i-1}^{k+1/2}) \frac{(U_{i}-U_{i-1})}{(r_i - r_{i-1})}  -  \nonumber \\
 0.25\frac{\Delta t}{r_i^2}(r_i^2+r_{i-1}^2)\frac{( P_i^{k+1/2}-P_{i-1}^{k+1/2})}{(r_i - r_{i-1})}   ,
 \end{eqnarray}
where $\Pi = \rho V^2 r^2 $ .
And the energy conservation equation is approximated as follows:
\begin{eqnarray}
{W}_i^{k+1/2} = W_i^k - {\Delta t}\frac{( G_{i+1}^k  -
 G_i^k)}{(r_{i+1}-r_i)} +  {\Delta t} r_i^2 Q_{EUV i}^k   , \\
 {W}_i^{k+1} = 0.5(W_i^{k+1/2} + W_i^{k}) - 0.5{\Delta t}\frac{( G_{i}^{k+1/2}  -  G_{i-1}^{k+1/2})}{(r_i-r_{i-1})} + \nonumber\\
  0.5{\Delta t} r_i^2 Q_{EUV i}^{k+1/2} +
  \frac{\Delta t}{(r_{i+1}-r_{i-1}) }  \left[ r_{i+1/2}^2  \chi_{i+1/2} \frac{(T_{i+1}^{k+1}-T_{i}^{k+1})}{(r_{i+1}-r_i)} \right. \nonumber\\
  \left. - r_{i-1/2}^2 \chi_{i-1/2} \frac{(T_{i}^{k+1}-T_{i-1}^{k+1})}{(r_{i}-r_{i-1})} \right ] ,
\end{eqnarray}
where
$ W = r^2 (\rho V^2 /2 + \rho U + E )$, $G = V (W + P r^2)$.

This two-step MacCormack method is of second order approximation. It is suitable for smooth solutions without sharp discontinuities. A stationary radial distribution of the atmospheric quantities can be obtained as a result of the time relaxation.

The upper boundary is chosen at the exobase level ($r =R_{\rm exo}$) which is defined from the condition that the length scale  of the pressure variation $d = (d \ln(P)/ dr)^{-1}$ is equal to
the mean free path of particles between collisions $\ell_c = 1/(\sigma_c n) $, where $\sigma_c$ is the collision cross section.
This condition yields equation
\begin{eqnarray}
d \ln(P)/ dr = \sigma_{\rm c} n.
\end{eqnarray}

\section{K Fraction in the Atmosphere for Different Compositions}\label{App:Kfraction}
\begin{table}
\begin{center}
\caption{Different element abundances given in \% mass fractions used in this work.}
\label{tab:abundances}
{
\vspace*{-2mm}
\begin{tabular}{c|cccccccccc}
\hline
	&	BSE			&	CoCr	    &	MORB	&	CI	    		&chond. avg.		&	solar	\\
	&1				&2			&3			&4			&5								&6		\\ \hline
H	&	0.006		&	0.045	&	0.023	&	1.992	&0.9375$^\ast$&	98.4	\\
C	&	0.006		&	0.199	&	0.019	&	3.520	&1.6025&	0.316	\\
N	&	8.8E-05		&	0.006	&	5.5E-05	&	0.298	&0.07975&	0.092	\\
O	&	44.42		&	47.20	&	44.5		&	46.420	&40.8&	0.765	\\
F	&	0.002		&	0.053	&	0.017	&	0.0059	&0.0039&	0.000067\\
Na	&	0.29			&	2.36		&	2.012	&	0.505	&0.41&	0.0039	\\
Mg&	22.01		&	2.20		&	4.735	&	9.790	&12.6&	0.094	\\
Al	&	2.12			&	7.96		&	8.199	&	0.860	&1.305&	0.0074	\\
Si	&	21.61		&	28.80	&	23.62	&	10.820	&13.725&	0.089	\\
P	&	0.008		&	0.076	&	0.057	&	0.098	&0.09875&	0.00078	\\
S	&	0.027		&	0.070	&	0.110	&	5.411	&3.35&	0.041	\\
Cl	&	0.004		&	0.047	&	0.014	&	0.071	&0.03225&	0.0011	\\
K	&	0.02			&	2.14		&	0.152	&	0.055	&0.040375&	0.00041	\\
Ca	&	2.46			&	3.85		&	8.239	&	0.933	&1.4175&	0.0086	\\
Ti	&	0.12			&	0.401	&	0.851	&	0.046	&0.069&	0.00042	\\
Cr	&	0.29			&	0.013	&	0.033	&	0.268	&0.32125&	0.00222	\\
Mn&	0.11			&	0.072	&	0.132	&	0.195	&12.525&	0.0014	\\
Fe	&	6.27			&	4.32		&	7.278	&	18.710	&21.875	&	0.172	\\ \hline
sum&	99.77	&	99.81	&	99.99	&	100.00	&	&	100.00	\\ \hline
\end{tabular}}
\\*[1mm]$\ast$: For the enhanced \ce{H2}-rich model, the H abundance is increased to 2\% mass fraction, prior to re-normalisation of all elements to 100\% mass fraction\\
{(1)~Bulk Silicate Earth: \citet{Schaefer2012}; (2)~Continental Crust: \citet{Schaefer2012}; (3)~Mid Oceanic Ridge Basalt: \citet{Arevalo2010}; (4)~CI chondrite: \citet{Lodders2009}; (5)~Carbonaceous chondrite average: \citep{Wasson1988}; (6)~solar: \citet{Asplund2009}}
\end{center}
\end{table}

In the main body of the paper, the K fraction in the atmosphere is described in terms of $f_\mathrm{K}$, the amount of K in the atmosphere relative to the total amount of K in atmosphere and condensate (Eq.~\ref{eq:dk}). This $f_{\rm K}$ does not only depend on the assumed surface conditions (pressure and temperature), but also on the considered element composition, and the build up of the atmosphere itself. So far an H$_2$-dominated primordial atmosphere has been investigated. The total element abundances (i.e. gas + condensates) are based on a carbonaceous chondrite average from \citet{Wasson1988}, with an increased H abundance to resemble the hydrogen envelope (see Table~\ref{tab:abundances}). This results in a $\approx$ 99\% \ce{H2} atmosphere for the investigated parameter space. In this appendix, we want to set the assumptions in a broader context and further investigate the influence of different sets of total element abundances to the K fraction in atmospheres. This can also reflect changes in the total element abundances over time, when potentially a large part of the hydrogen envelope is lost to space.
Therefore we investigate various element abundances, ranging from Bulk Silicate Earth \citep[BSE][]{Schaefer2012}, Continental Crust \citep[CoCr][]{Schaefer2012}, Mid Oceanic Ridge Basalt \citep[MORB][]{Arevalo2010}, to chondritic abundances taken from a CI chondrite \citep{Lodders2009} as well as the unaltered averaged carbonaceous chondritic abundances from \citet{Wasson1988}. Furthermore we provide results for a solar abundance model \citep{Asplund2009}.
The mass fractions for these abundances are listed in Table~\ref{tab:abundances}.

\begin{figure*}
\centering
\includegraphics[width = .33\linewidth, page=1]{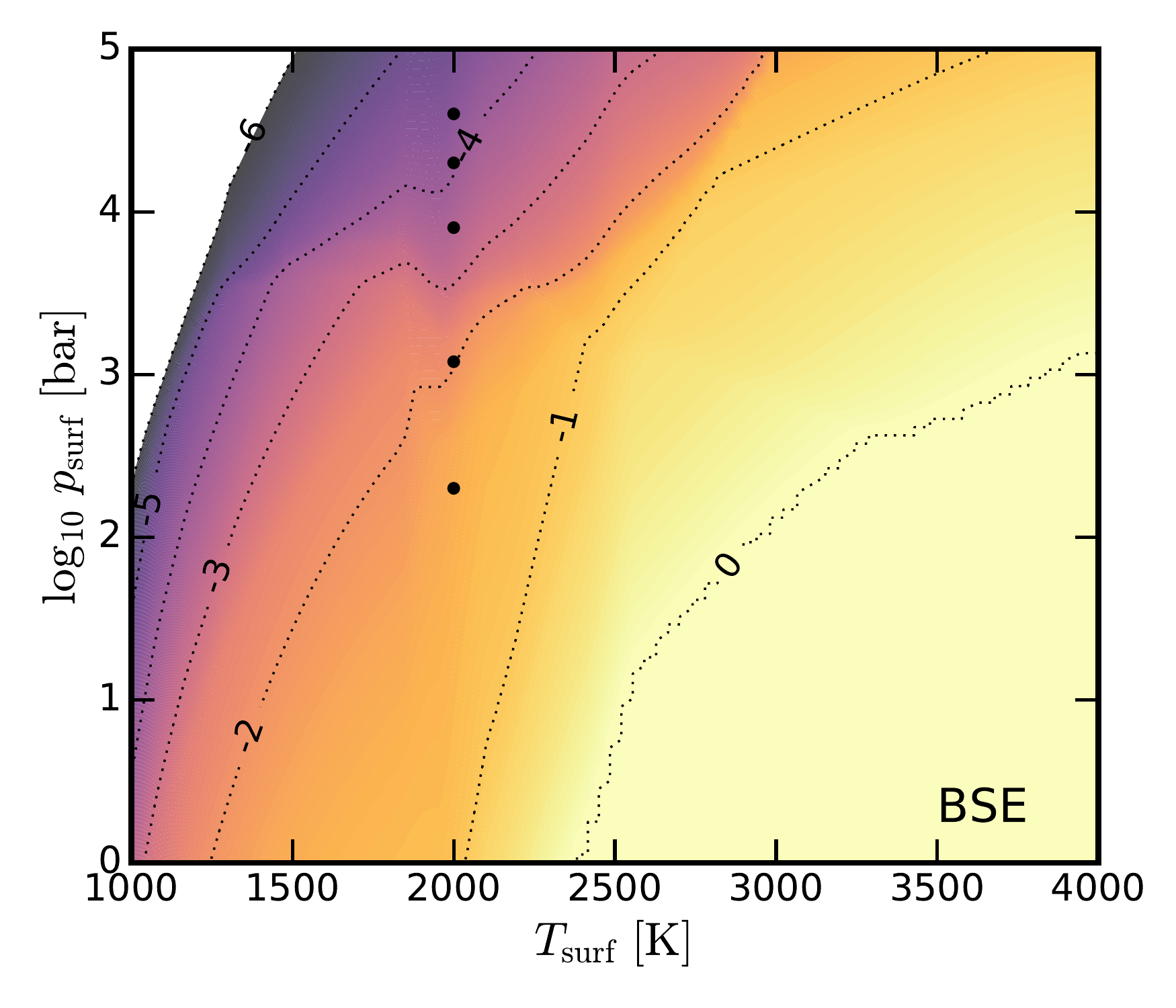}
\includegraphics[width = .33\linewidth, page=1]{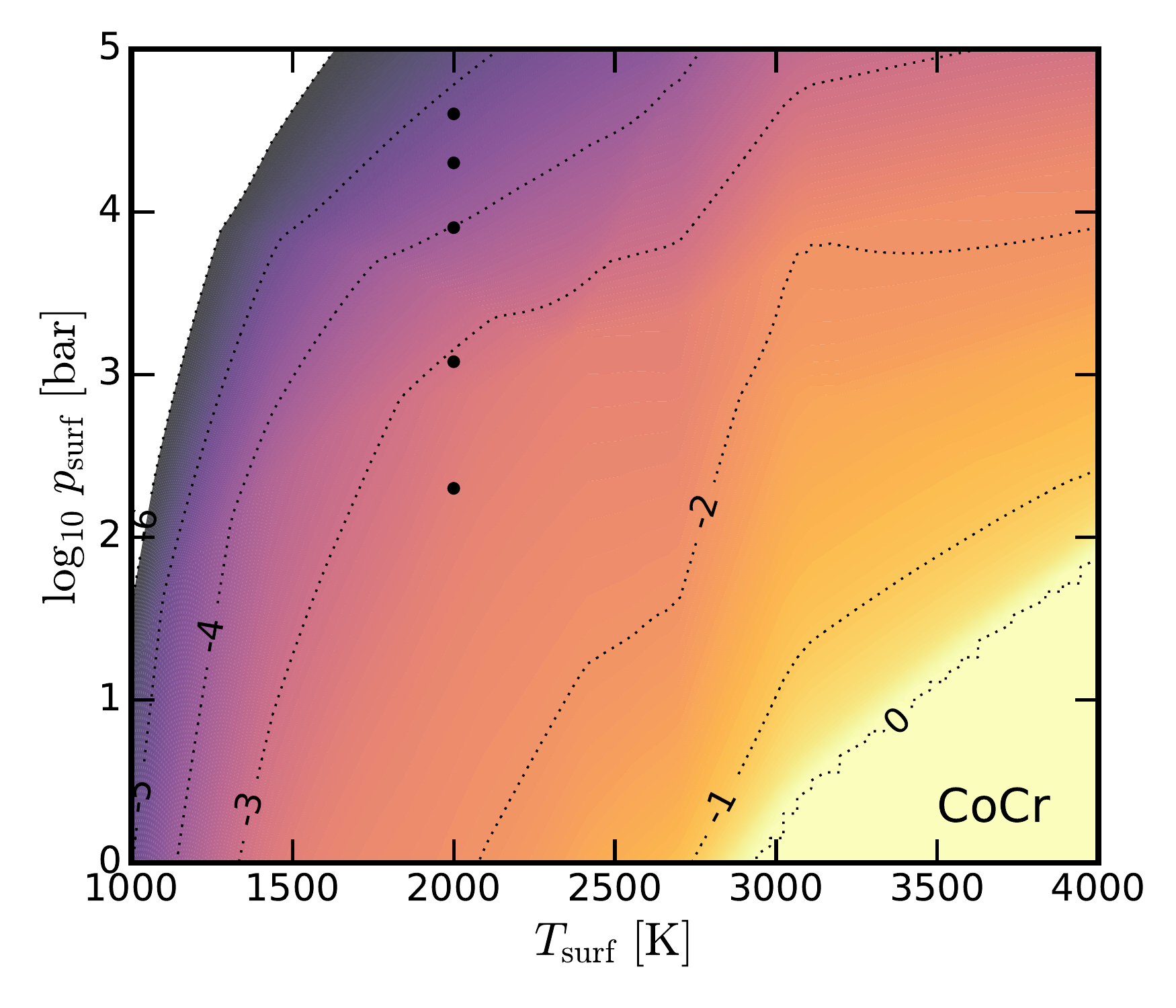}
\includegraphics[width = .33\linewidth, page=1]{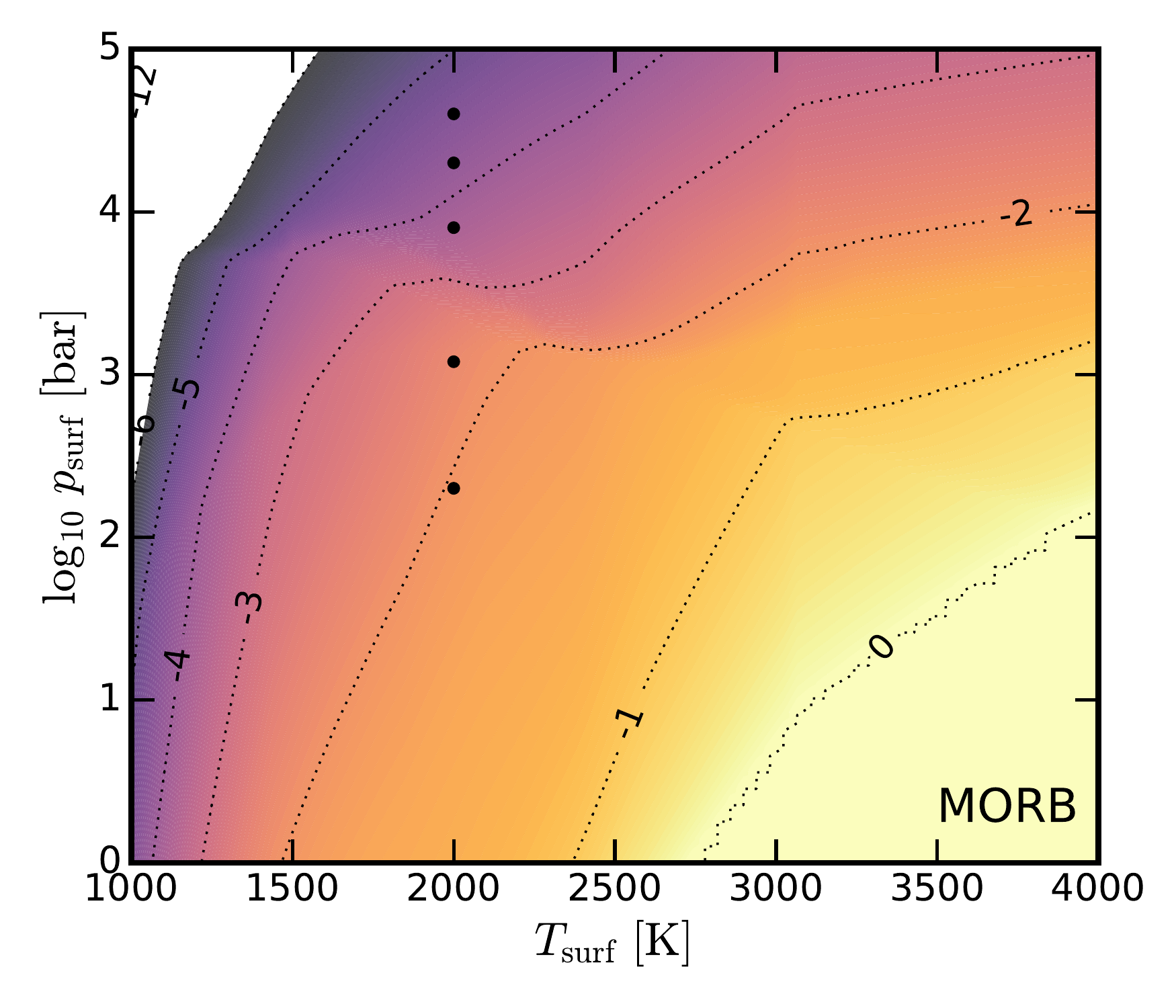}\\
\includegraphics[width = .33\linewidth, page=1]{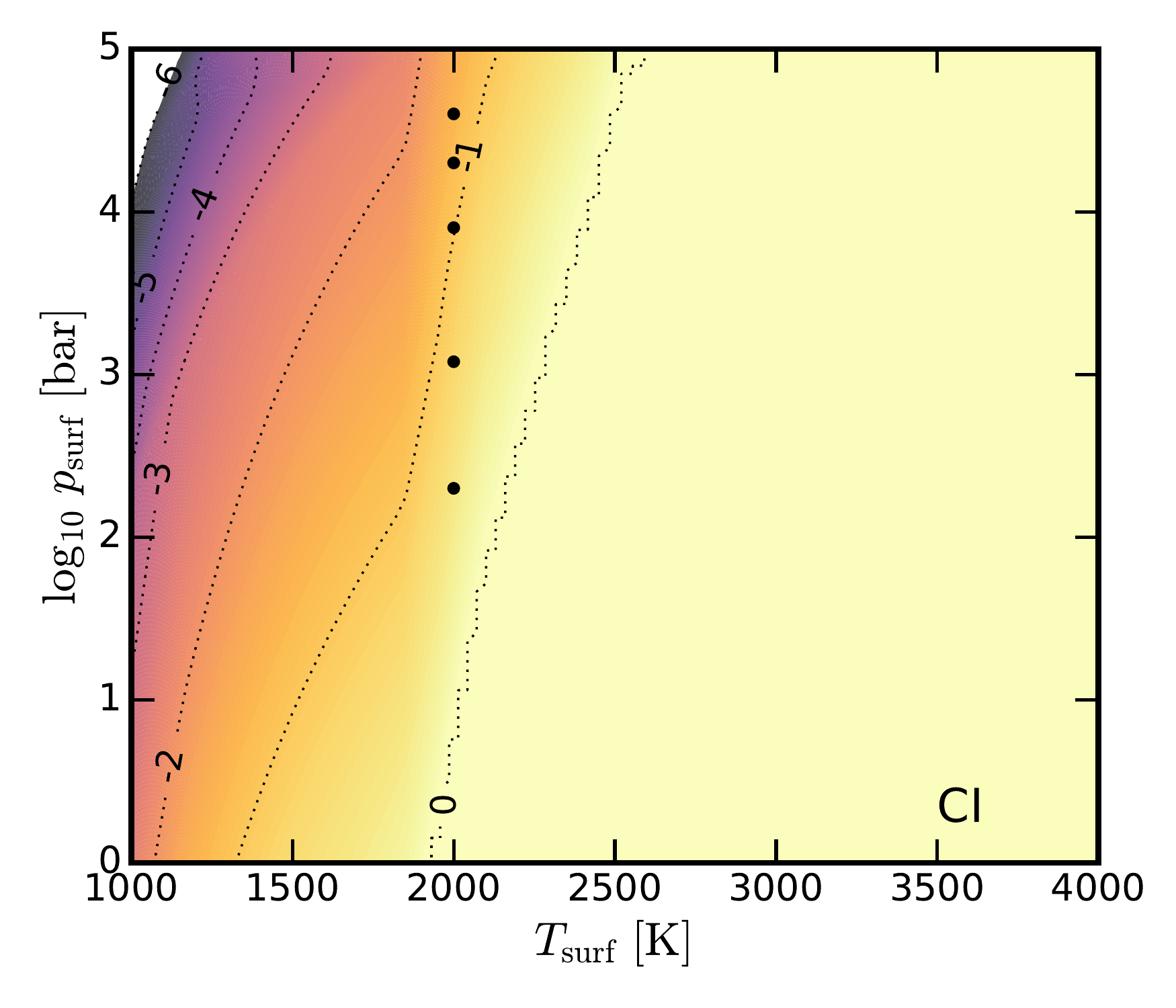}
\includegraphics[width = .33\linewidth, page=1]{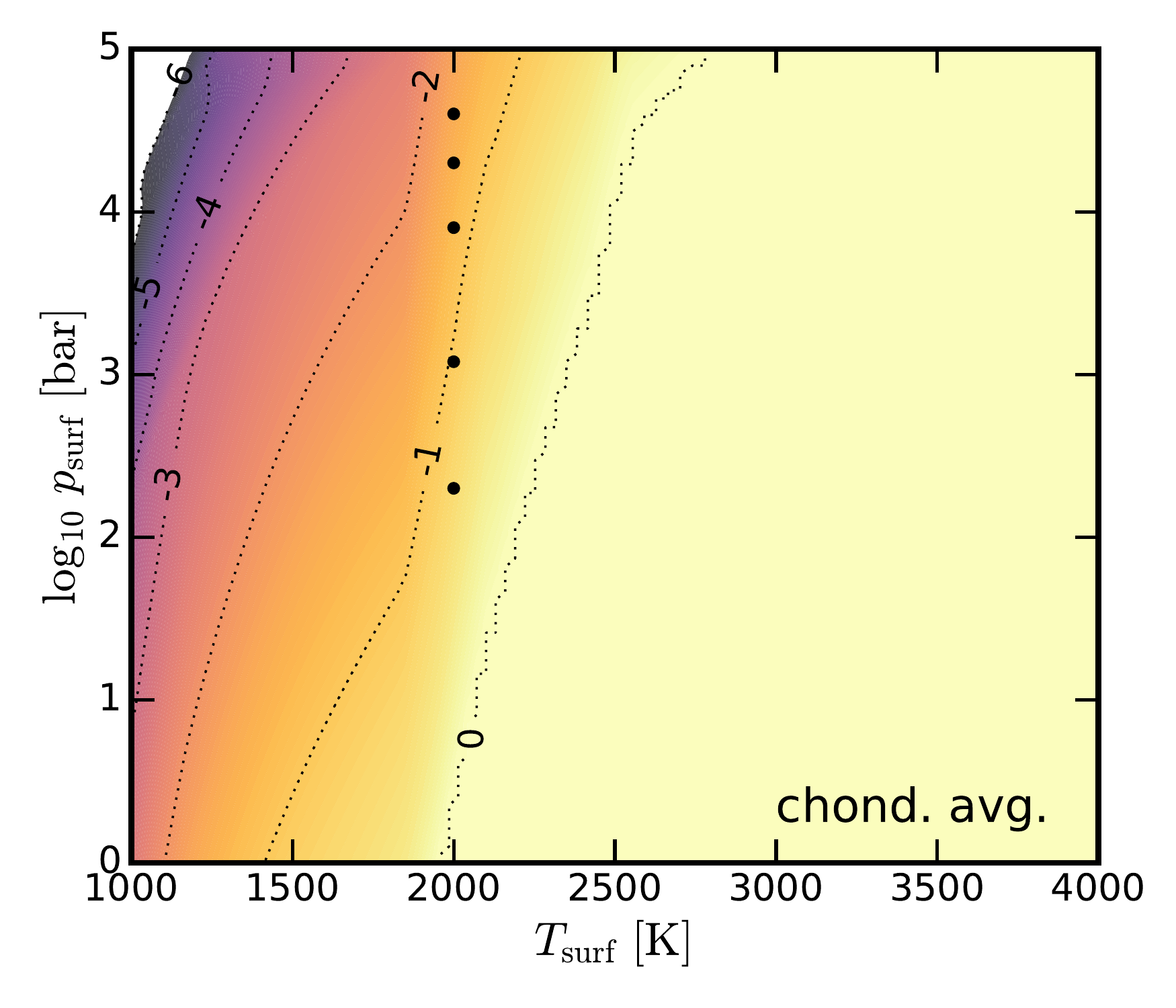}
\includegraphics[width = .33\linewidth, page=1]{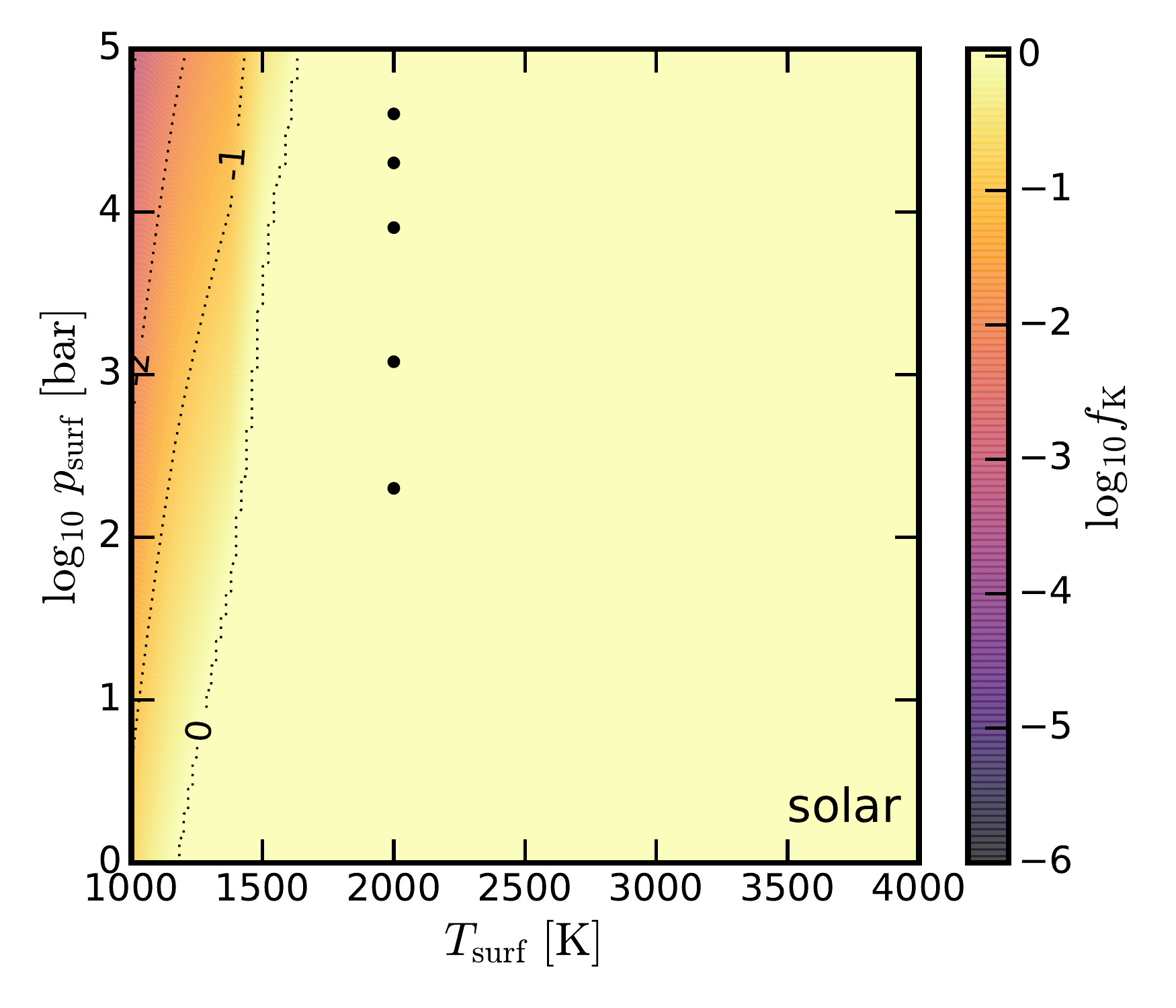}\\
\caption{Potassium fraction $f_\mathrm{K}$ for varying atmospheric temperatures and pressures for different sets of total element abunances. {Upper row:} BSE, CoCr, MORB. {lower row:} CI, Carb Cond average, solar.
Brighter colours refer to a higher $f_\mathrm{K}$. The dotted lines indicates lines of equal $\log(f_\mathrm{K})$.
For $\log (f_\mathrm{K}) =0$ no K containing condensate is stable. {The black dots for different pressure levels and temperatures of 2000\,K indicate the main ($p,T$) conditions investigated in this work.}}
\label{fig:Kfrac_all}
\end{figure*}

In Fig.~\ref{fig:Kfrac_all}, we show $f_\mathrm{K}$ for these various different total element abundances as calculated with the {\sc GGchem} code \citep[see][]{Herbort2021}. It should be noted, that for these temperatures, the atmosphere results to be a steam atmosphere where the main atmospheric constituents are H$_2$O, CO$_2$, and SO$_2$ and therefore resembles a steam atmosphere and not an H$_2$ dominated atmosphere, as assumed throughout the main body of this manuscript.
Only the solar abundance model and the H-increased chondritic average model, as used in the main body of the manuscript, show an H$_2$-dominated atmosphere.
For the atmospheres based on overall rock compositions $f_{\rm K}$ is about two orders of magnitude lower than for chondritic abundances, which have orders of magnitude more hydrogen available.
However, for high temperatures ($T>3000\,$K) and low pressures ($p<100\,$bar), $f_{\rm K}$ approaches 1 again.
These atmosphere may be more representative for rocky planets after they lost their hydrogen envelope and built up a secondary atmosphere.
The loss of heavier elements in such a scenario is, however, beyond the scope of this paper.

\begin{figure*}
\centering

\includegraphics[width = .33\linewidth, page=1]{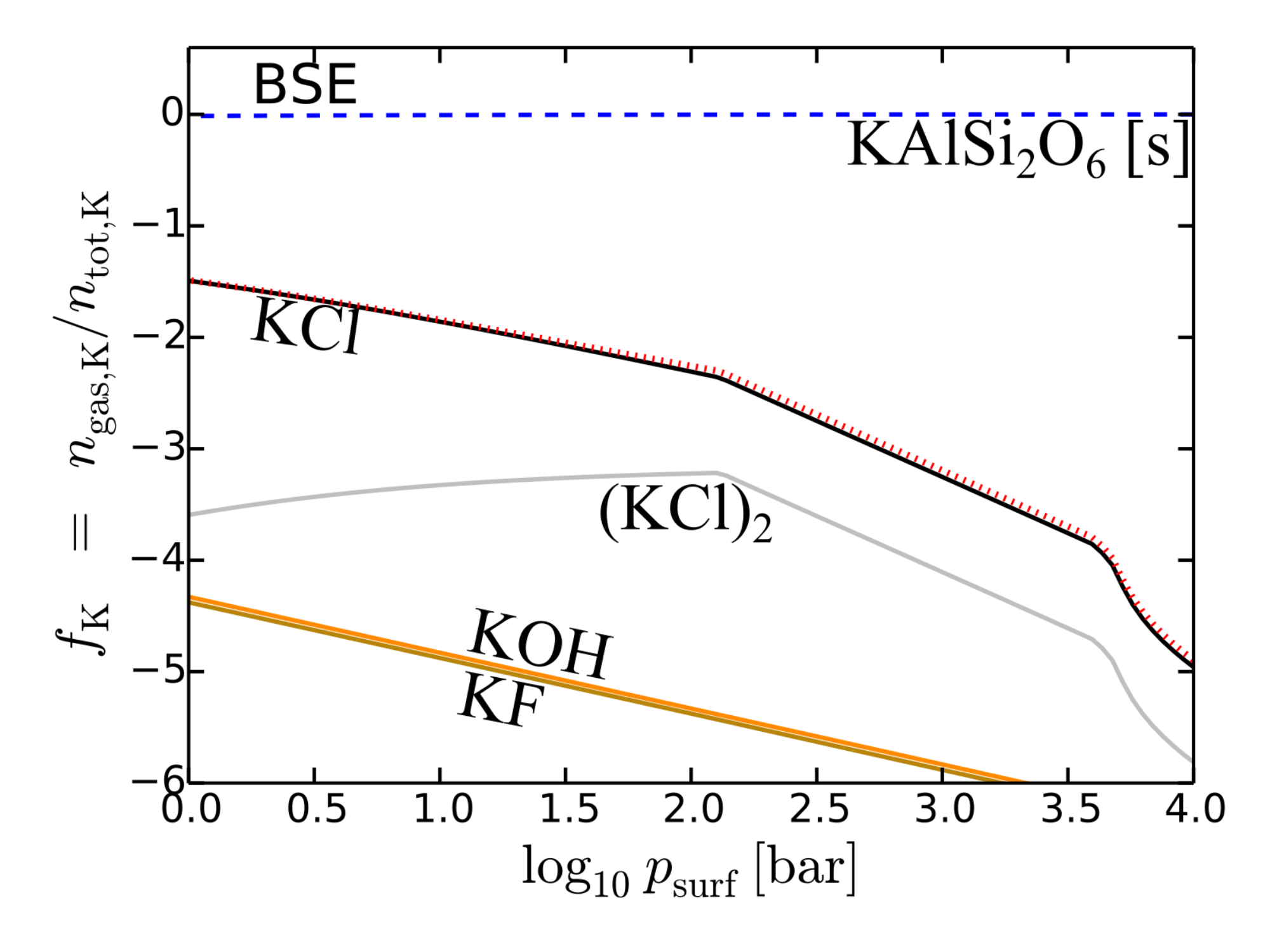}
\includegraphics[width = .33\linewidth, page=1]{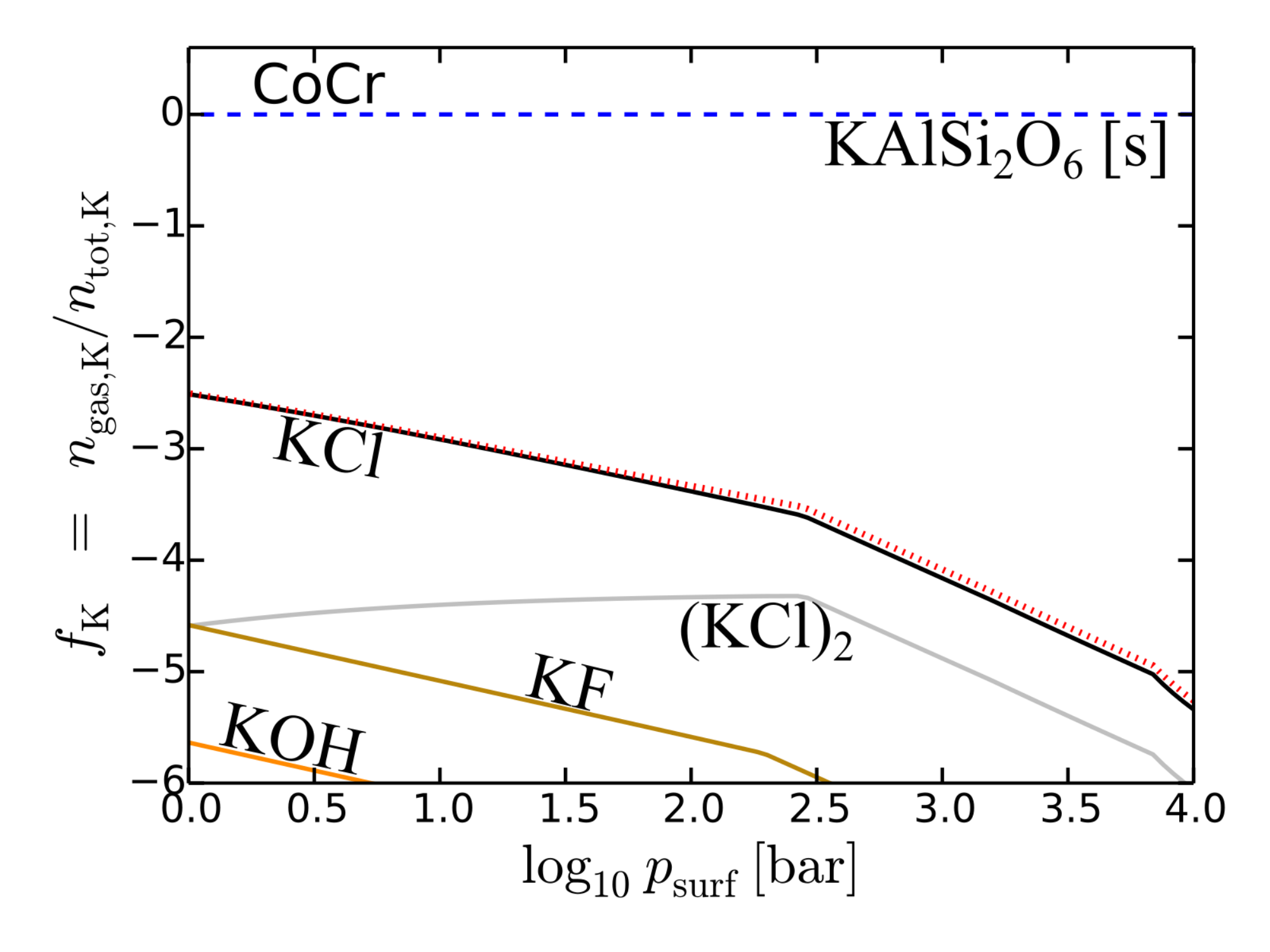}
\includegraphics[width = .33\linewidth, page=1]{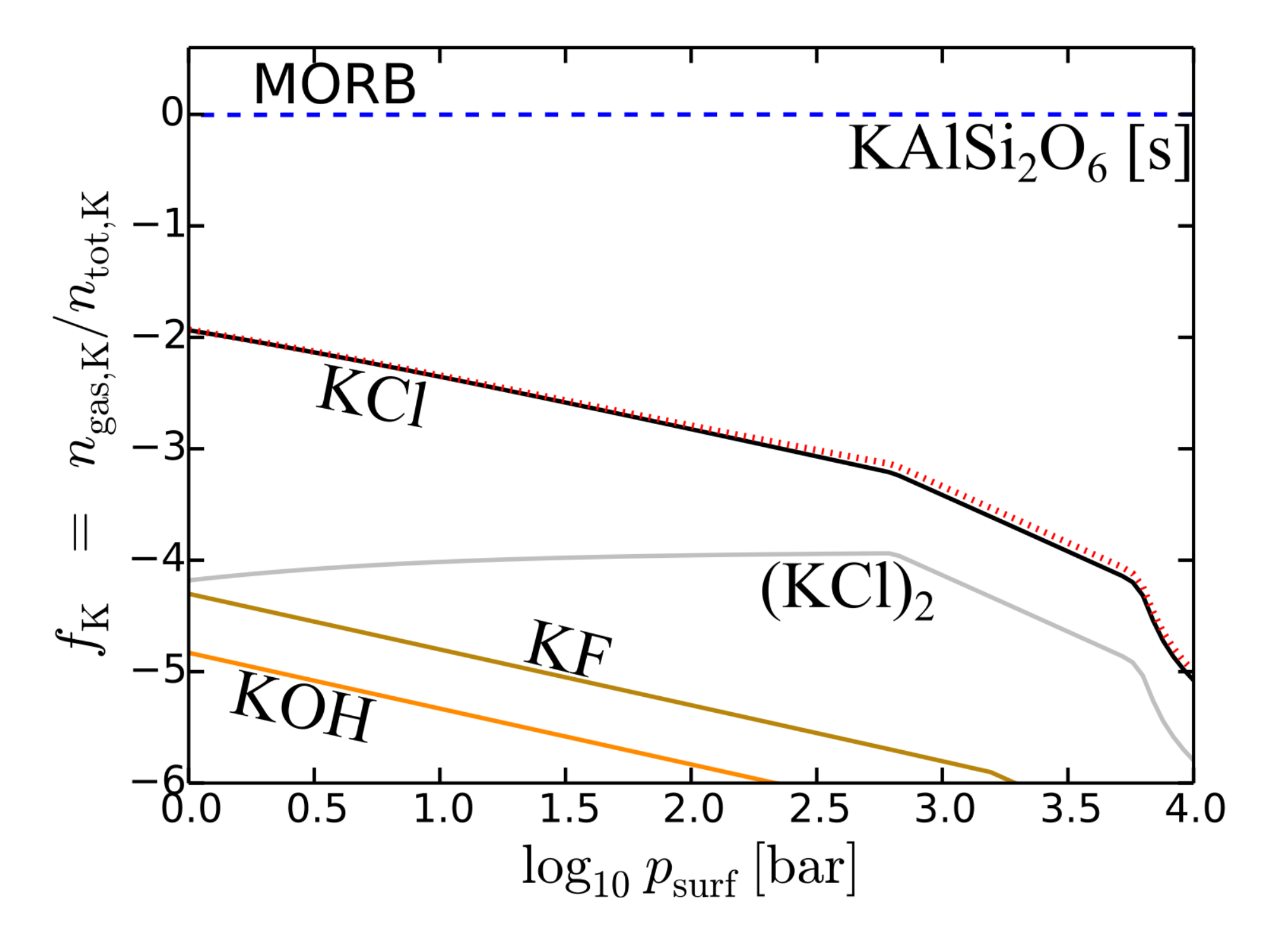}\\
\includegraphics[width = .33\linewidth, page=1]{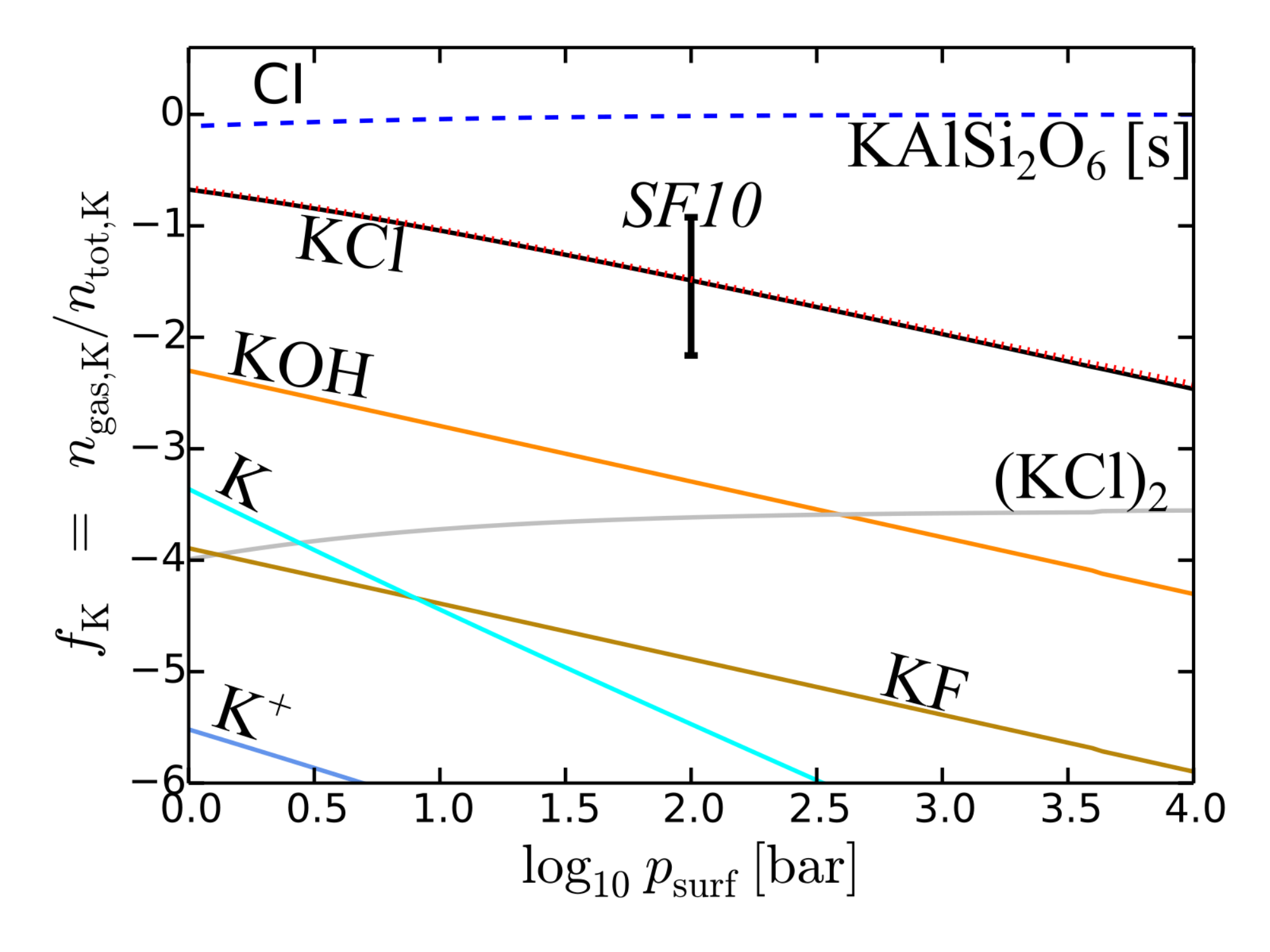}
\includegraphics[width = .33\linewidth, page=1]{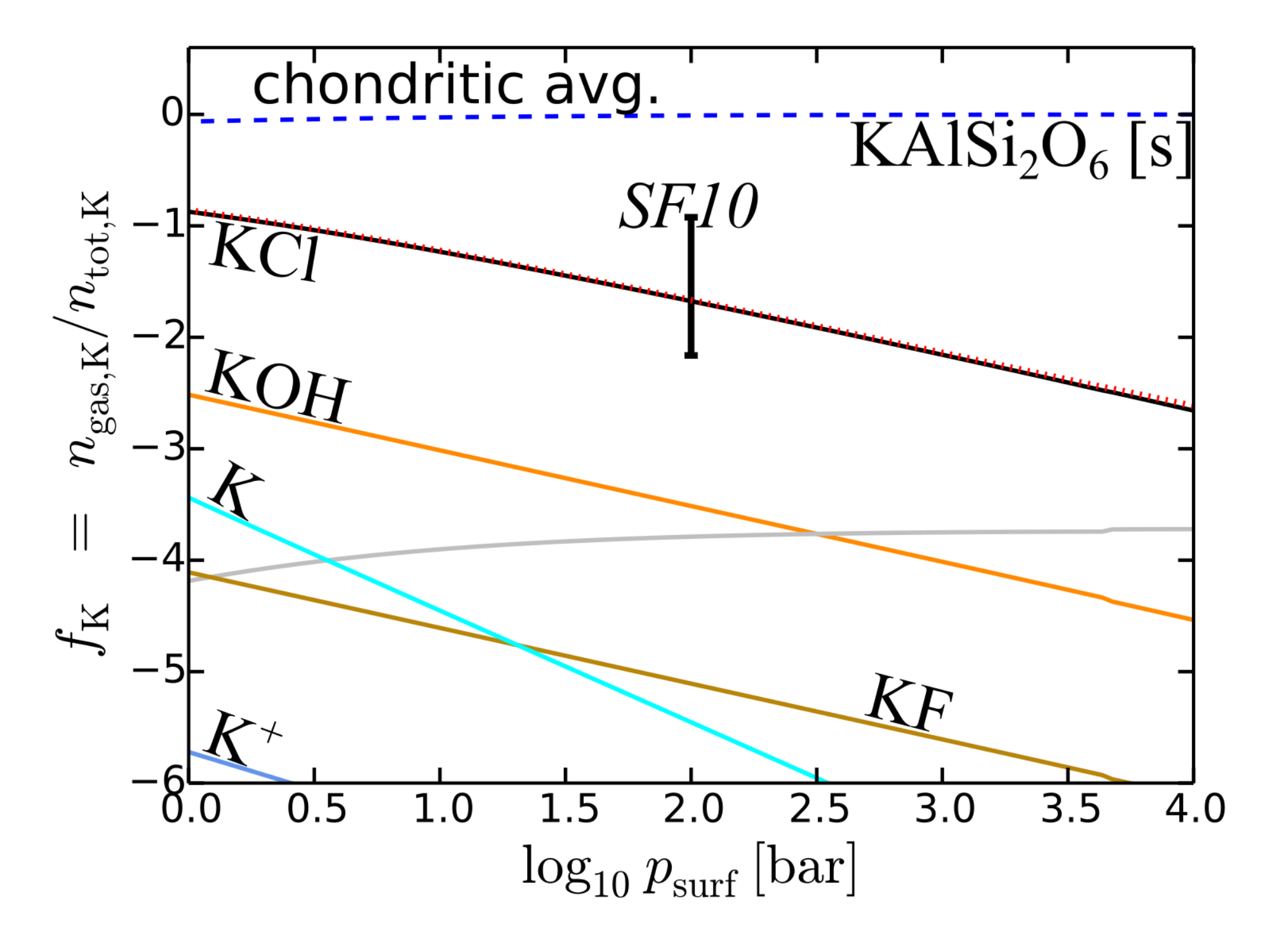}
\includegraphics[width = .33\linewidth, page=1]{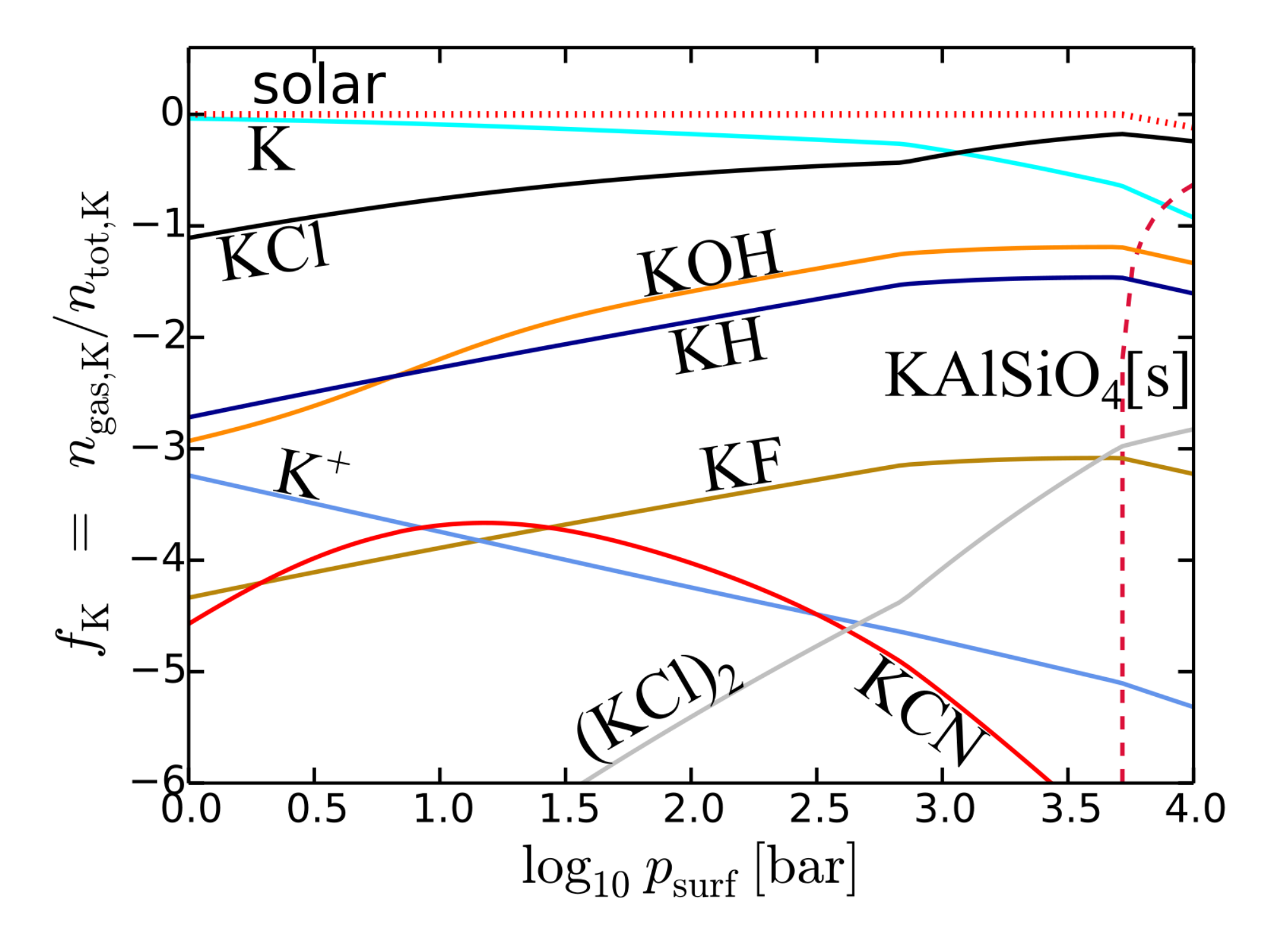}\\
%
\caption{For a constant temperature of $T=1500\,$K and a varying pressure, the different K components {of the gas and condensate phase} are visualised in comparison to the {total K abundance}.
The dashed lines represent condensate phase, while solid lines represent the gas phase.
The sum of all gas phase molecules is indicated with the dotted red line.
The sets of total element abundances are as in Fig.~\ref{fig:Kfrac_all}.
{The black range indicated with SF10 represents values taken from \citet{Schaefer2010} for comparison.}
}
\label{fig:Kfrac_1500K}
\end{figure*}

\begin{figure*}
\centering
\includegraphics[width = .33\linewidth, page=1]{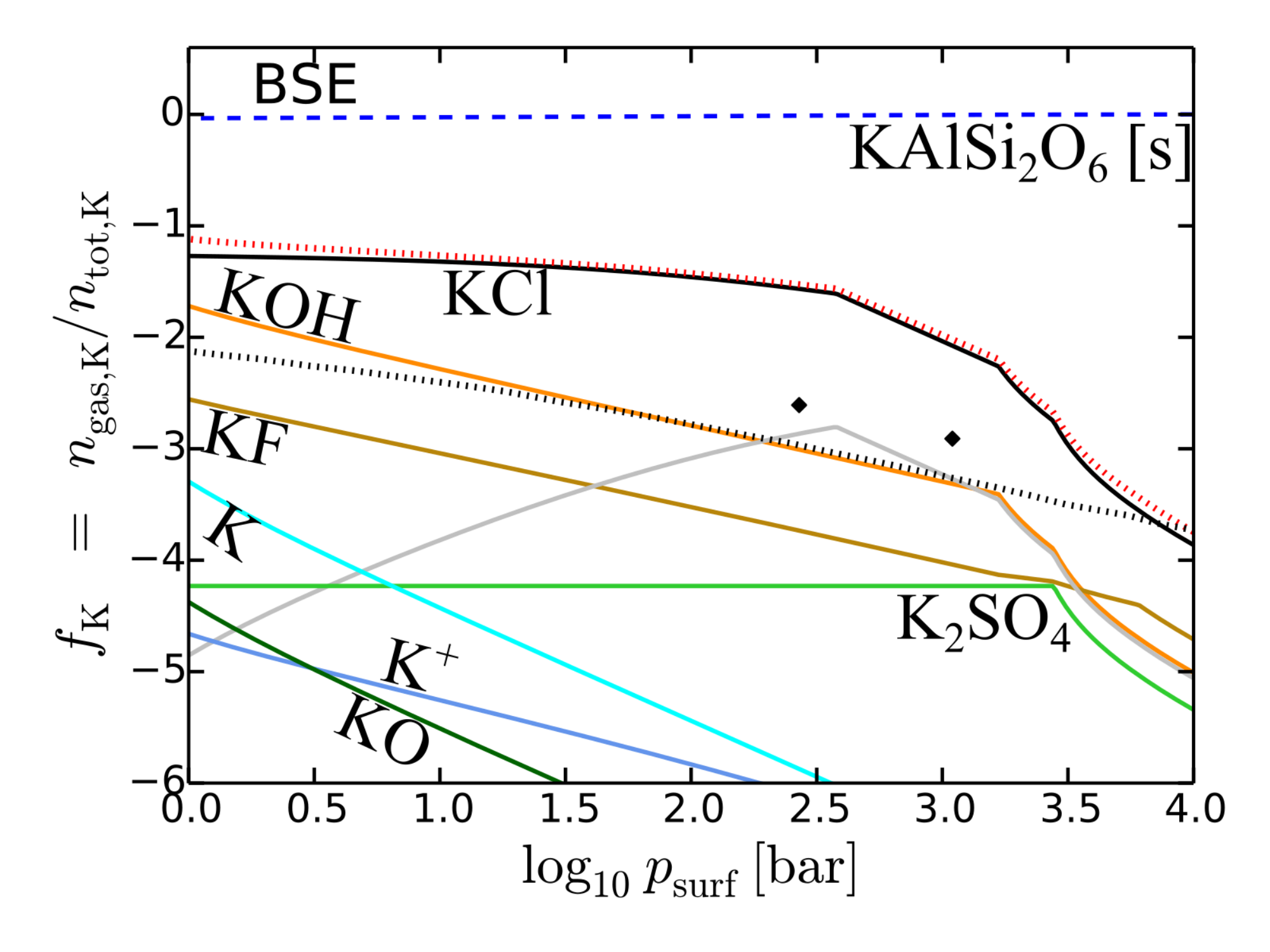}
\includegraphics[width = .33\linewidth, page=1]{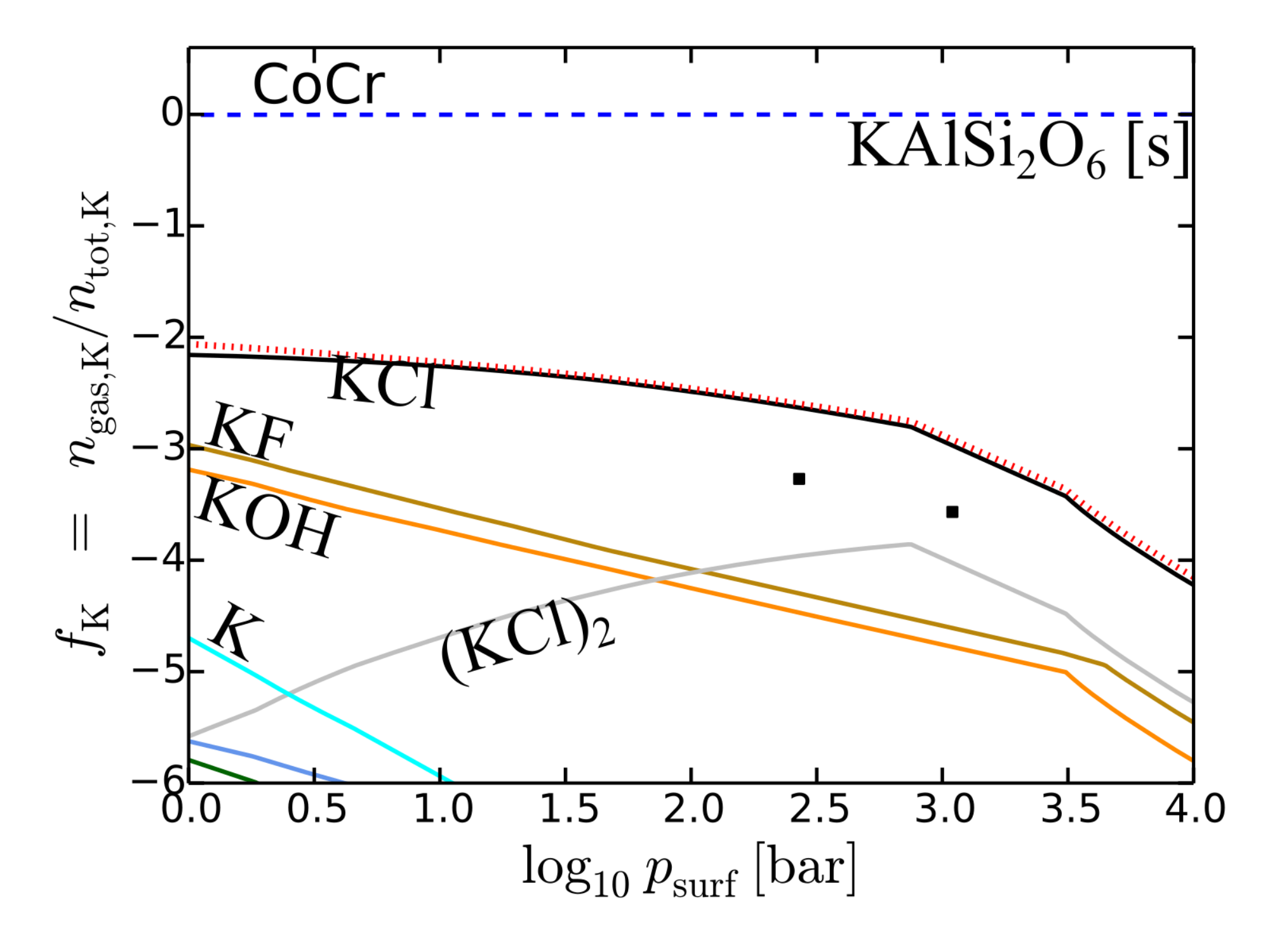}
\includegraphics[width = .33\linewidth, page=1]{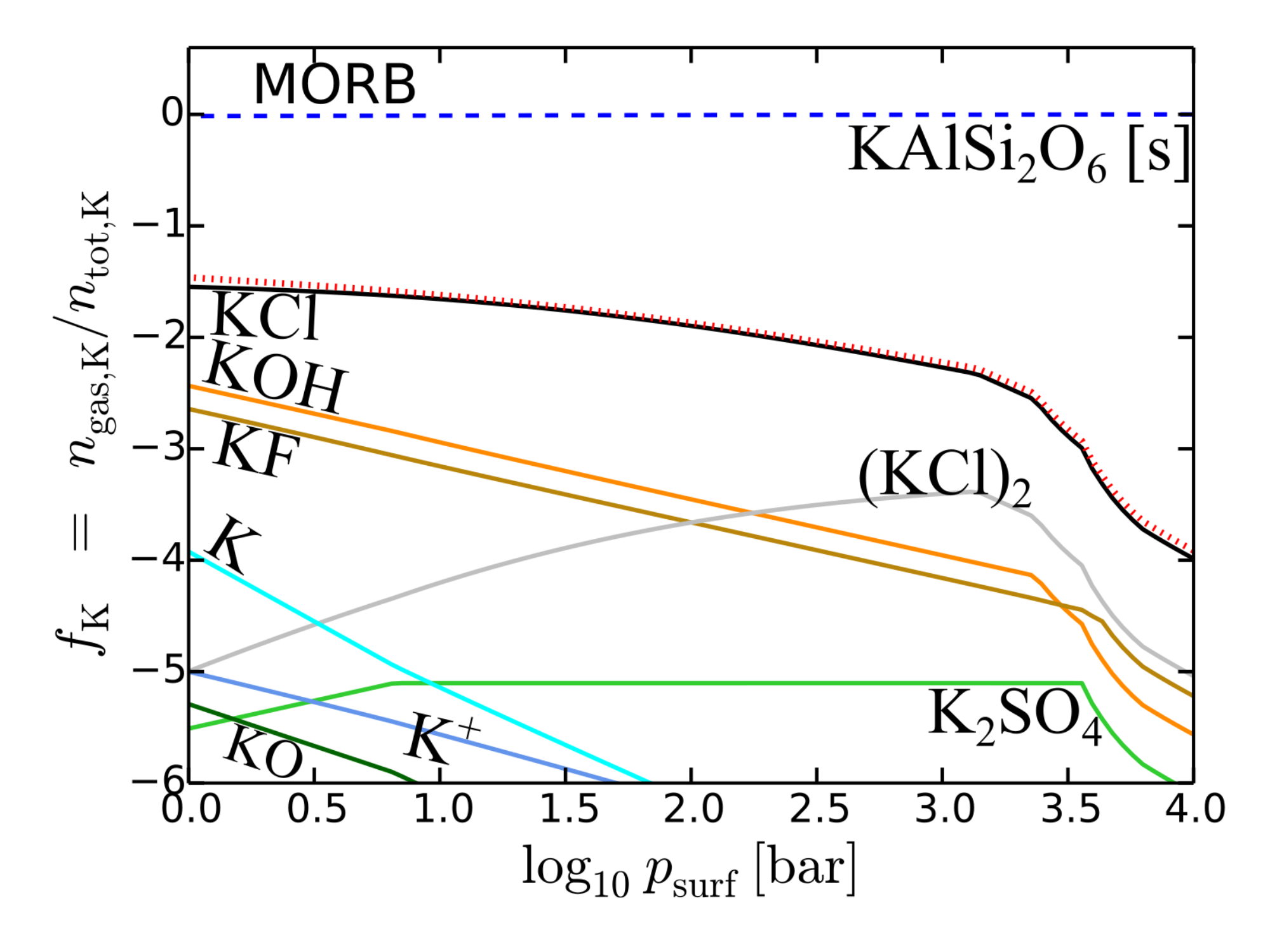}\\
\includegraphics[width = .33\linewidth, page=1]{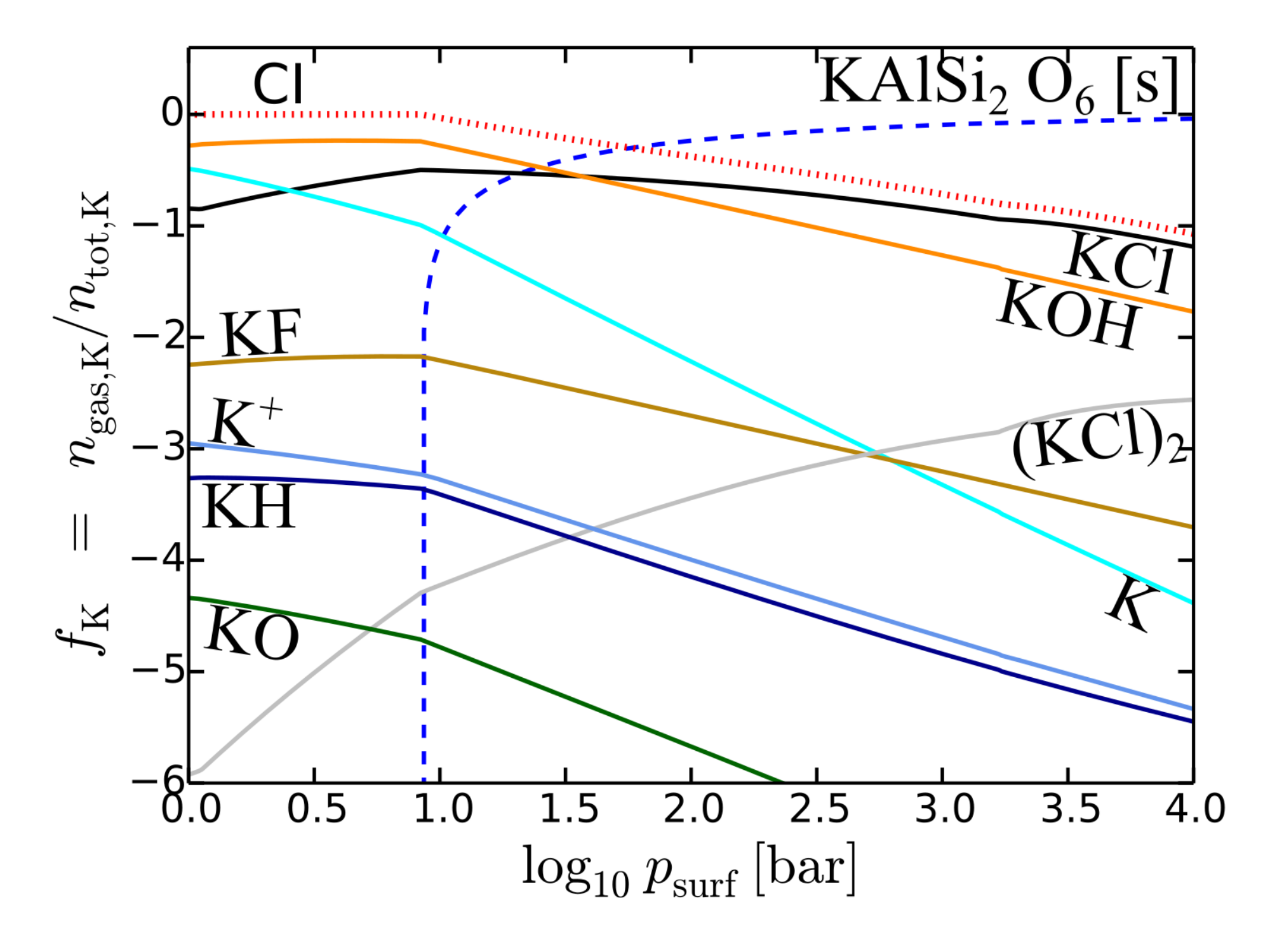}
\includegraphics[width = .33\linewidth, page=1]{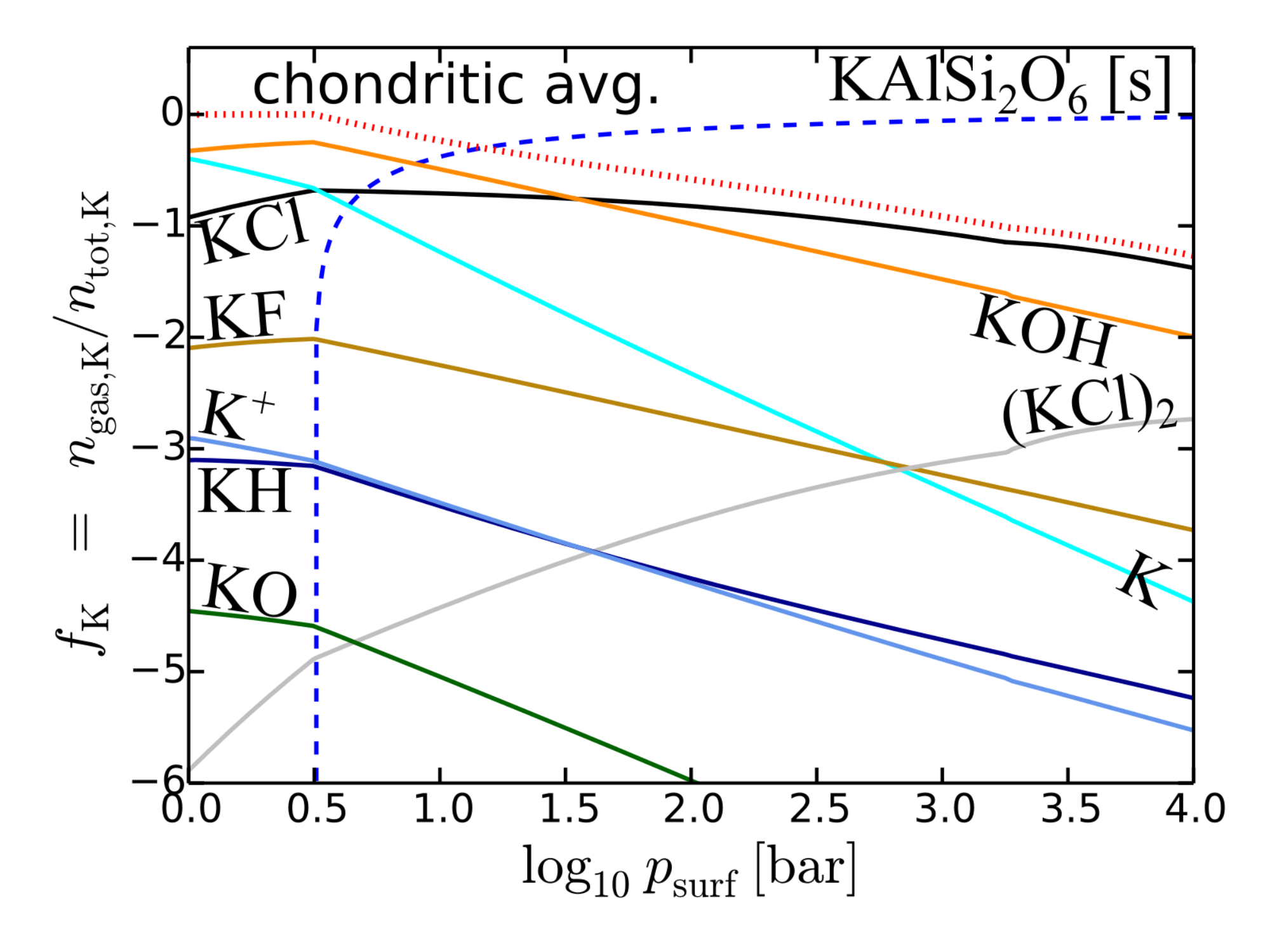}
\includegraphics[width = .33\linewidth, page=1]{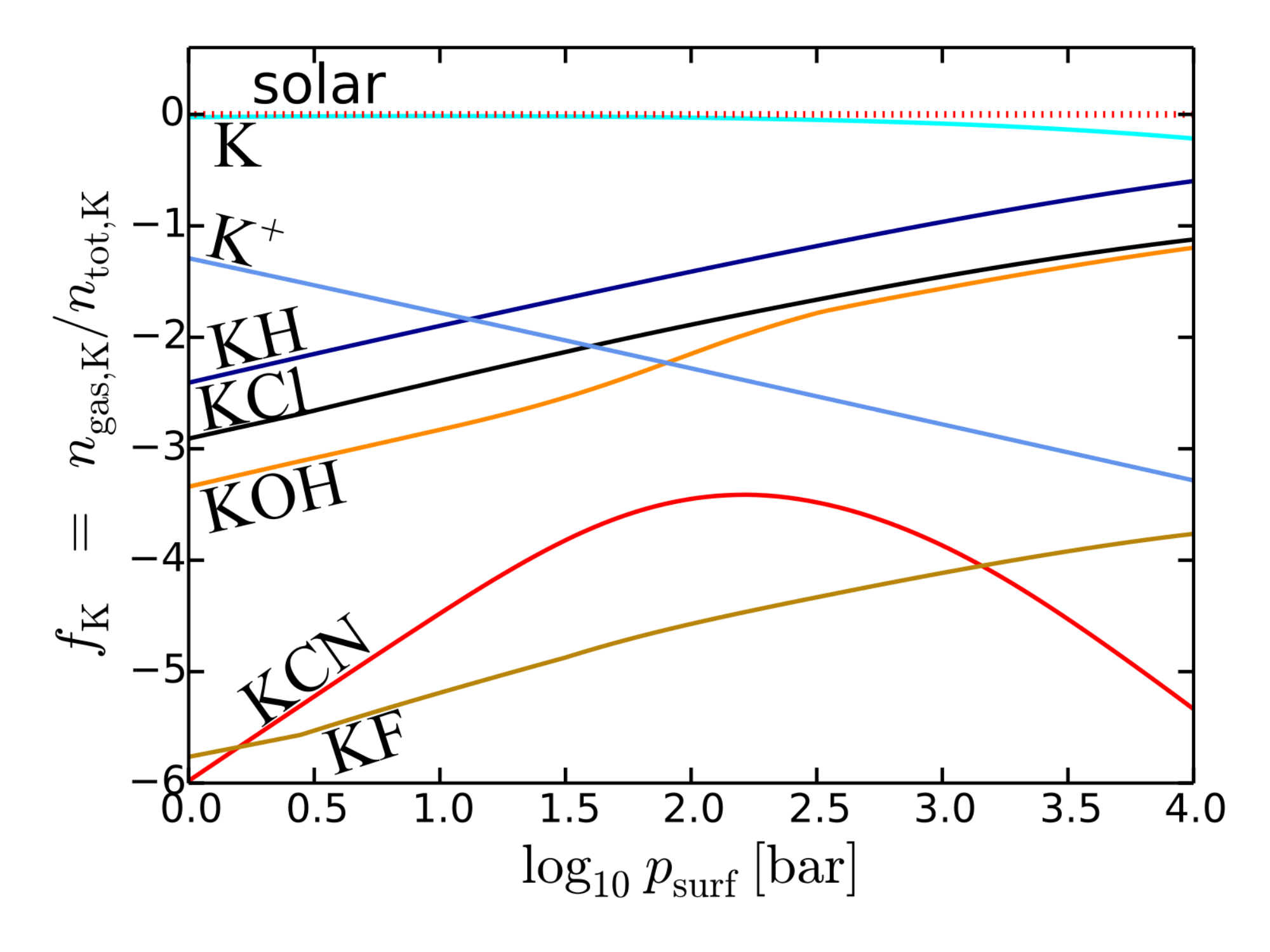}\\
\caption{As Fig~\ref{fig:Kfrac_1500K}, but for $T=2000\,$K. {The black dots for the BSE and CoCr abundance are taken from \citet{Fegley2016}, while the black dotted line for the BSE abundance is a recalculation by Fegley (private com.).}}
\label{fig:Kfrac_2000K}
\end{figure*}

In Figs.~\ref{fig:Kfrac_1500K} and \ref{fig:Kfrac_2000K} we show the distribution of different K bearing species for surface temperatures of 1500\,K and 2000\,K, respectively. We investigate different surface pressures and sets of total element abundances.
{Only one K bearing condensate is thermally stable for each set of total element abundances. These are \ce{KAlSiO4}[s] for solar abundances and \ce{KAlSi2O6}[s] for all other sets of total element abundances.
The gas-phase composition is more diverse with \ce{KCl}, \ce{KOH}, and \ce{K} being the most important K bearing gas species.}

Previous studies like \citet{Schaefer2010} report that within their simulations between 0.7-11\% of K will be in the gas of a silicate atmosphere at a temperature of 1500\,K, and a pressure of 100 bar.
{Our models based on carbonatious chondrites (CI and chondritic avg.) are in agreement with their results for chronditic compositions.}
{In general,} we see that for the models rich in H, more K is in the gas phase.

\citet{Fegley2016} calculated $f_{\rm K}$ for high pressure steam atmospheres between 270 - 1100\,bar, and 2000 - 3000\,K, resulting in $f_{\rm K}\approx$ 0.0003 - 0.013.
In Fig.~\ref{fig:Kfrac_2000K}, we show results of our various different element abundances for a temperature of $T=2000$\,K and a pressure range from 1\,bar to 10000\,bar.
For the corresponding sets of element abundances, we also include the values calculated by \citet{Fegley2016} for BSE and CoCr.
Furthermore, we show a recalculation of the K fractionation based on BSE abundances (Fegley, private com.).
The direct comparison shows an enhanced gas fraction {in our models} relative to those by \citet{Fegley2016}. {This is} likely a result of different elements {and condensate species} included in the respective models.
However, as in \citet{Fegley2016}, we see a similar qualitative difference between BSE and CoCr abundances, where more K is observed in the gas phase for CoCr than BSE.

\end{appendix}

\bsp	
\label{lastpage}

\end{document}